\documentclass[11pt,a4paper]{article}

\pdfoutput=1
\usepackage{amsmath,amssymb,authblk,nicefrac,hyperref,multirow,graphicx,cite}
\usepackage{color}
\usepackage[shortlabels]{enumitem} 

\graphicspath{{./}{./Figures/}}

\usepackage{geometry}
\usepackage{tikz}
\usetikzlibrary{positioning,arrows,patterns}
\usetikzlibrary{decorations.pathmorphing}
\usetikzlibrary{decorations.markings}
\usetikzlibrary{calc}
\tikzset{
    photon/.style={decorate, decoration={snake}, draw=black, thick},
    fermionnoarrow/.style={draw=black, postaction={decorate}, thick},
    scalar/.style={draw=black, postaction={decorate}, thick, dashed},
    fermion/.style={draw=black, postaction={decorate},decoration={markings,mark=at position .55 with {\arrow{>}}}, thick},
    gluon/.style={decorate, draw=black, decoration={coil,amplitude=4pt, segment length=5pt}, thick},
    vertex/.style={draw,shape=circle,fill=black,minimum size=3pt,inner sep=0pt} 
}

\newcommand{\hiddennote}[1]{}

\newcommand{\lag}{\mathcal{L}}
\newcommand{\amp}{\mathcal{M}}
\newcommand{\dsym}{\mathcal{D}}

\newcommand{\op}{\mathcal{O}}
\newcommand{\V}{\mathcal{V}}
\newcommand{\set}[1]{\mathbb{#1}}

\newcommand{\order}[1]{\mathcal{O}(#1)}
\newcommand{\abs}[1]{\lvert #1 \rvert}
\newcommand{\expect}[1]{\langle #1 \rangle}
\newcommand{\dlr}[1]{\overleftrightarrow{\partial_#1}}
\newcommand{\Dlr}[1]{\overleftrightarrow{D_#1}}
\newcommand{\brtilde}[1]{\breve{#1}}

\newcommand{\modeqref}[1]{Eq.~\eqref{#1}}
\newcommand{\modeqsref}[1]{Eqs.~\eqref{#1}}
\newcommand{\figref}[1]{Fig.~\ref{#1}}
\newcommand{\tabref}[1]{Table~\ref{#1}}
\newcommand{\tabsref}[1]{Tables~\ref{#1}}
\newcommand{\secref}[1]{Section~\ref{#1}}
\newcommand{\secsref}[1]{Sections~\ref{#1}}
\newcommand{\appref}[1]{Appendix~\ref{#1}}

\newcommand{\dk}{\chi}
\newcommand{\dm}{X}
\newcommand{\dkp}{\Psi}
\newcommand{\gfld}{\Phi}
\newcommand{\vis}{V}
\newcommand{\dms}{\phi}
\newcommand{\dmf}{\chi}
\newcommand{\dmfL}{\eta}
\newcommand{\dmfR}{\bar{\xi}^\dagger}
\newcommand{\dmfRbar}{\bar{\xi}}
\newcommand{\dps}{\omega}
\newcommand{\dpf}{\psi}
\newcommand{\dpfL}{\zeta}
\newcommand{\dpfR}{\bar{\upsilon}^\dagger}

\begin{document}

\title{A Systematic Effective Operator Analysis \\of Semi-Annihilating Dark Matter}
\author[1]{Yi Cai\thanks{\texttt{yi.cai@unimelb.edu.au}}} 
\author[2]{Andrew Spray\thanks{\texttt{a.spray.work@gmail.com}}}
\affil[1]{ARC Centre of Excellence for Particle Physics at the Terascale, School of Physics, 
The University of Melbourne, Victoria 3010, Australia}
\affil[2]{Center for Theoretical Physics of the Universe, Institute for Basic Science (IBS), Daejeon, 34051, Korea}

\maketitle

\begin{abstract}
  Semi-annihilation is a generic feature of dark matter theories stabilised by symmetries larger than a $\set{Z}_2$.  It contributes to thermal freeze out, but is irrelevant for direct and collider searches.  This allows semi-annihilating dark matter to avoid those limits in a natural way.  We use an effective operator approach to make the first model-independent study of the associated phenomenology.  We enumerate all possible operators that contribute to $2\to2$ semi-annihilation up to dimension 6, plus leading terms at dimension 7.  We find that when the only light states charged under the dark symmetry are dark matter, the model space is highly constrained.  Only fifteen operators exist, and just two for single-component dark sectors.  If there can be additional light, unstable ``dark partner'' states the possible phenomenology greatly increases, at the cost of additional model dependence in the dark partner decay modes.  We also derive the irreducible constraints on models with single-component dark matter from cosmic ray searches and astrophysical observations.  We find that for semi-annihilation to electrons and light quarks, the thermal relic cross sections can be excluded for dark matter masses up to 100~GeV.  However, significant model space for semi-annihilating dark matter remains.
\end{abstract}

\section{Introduction}\label{sec:intro}

The existence of dark matter (DM) is perhaps the strongest single piece of evidence for the necessity of new physics.  Cosmological measurements across a wide range of scales all point to the existence of a cold non-luminous component of matter in the Universe, with an energy density today about six times that of baryons.  However, despite several decades of strenuous experimental effort, no unambiguous non-gravitational signal has been found.  In the face of these null results, the microscopic properties of DM remain unknown.    

\begin{figure}
	\centering
	\begin{tikzpicture}[node distance=1cm and 1.75cm]
		\coordinate (v1);
		\coordinate[above left = of v1, label=above left:$\dk$] (i1);
		\coordinate[below left = of v1, label=below left:$\dk$] (i2);
		\coordinate[above right = of v1, label=above right:{$V$}] (o1);
		\coordinate[below right = of v1, label=below right:{$V$}] (o2);
		\draw[fermionnoarrow] (i1) -- (v1);
		\draw[fermionnoarrow] (v1) -- (i2);
		\draw[fermionnoarrow] (o2) -- (v1);
		\draw[fermionnoarrow] (v1) -- (o1);
		\draw[fill = white] (v1) circle (1);
		\fill[pattern = north west lines] (v1) circle (1);
	\end{tikzpicture}\qquad\qquad
	\begin{tikzpicture}[node distance=1cm and 1.75cm]
		\coordinate (v1);
		\coordinate[above left = of v1, label=above left:$\dk_1$] (i1);
		\coordinate[below left = of v1, label=below left:$\dk_2$] (i2);
		\coordinate[above right = of v1, label=above right:{$\dk_3$}] (o1);
		\coordinate[below right = of v1, label=below right:{$V$}] (o2);
		\draw[fermionnoarrow] (i1) -- (v1);
		\draw[fermionnoarrow] (v1) -- (i2);
		\draw[fermionnoarrow] (o2) -- (v1);
		\draw[fermionnoarrow] (v1) -- (o1);
		\draw[fill = white] (v1) circle (1);
		\fill[pattern = north west lines] (v1) circle (1);
	\end{tikzpicture}
	\caption{Two types of dark sector processes, where $\chi$ ($V$) is a dark (visible) sector field.  (Left): DM annihilation to/from, or scattering off, the SM; this is the only process possible when the dark matter is stabilised by a $\set{Z}_2$ symmetry.  (Right): Semi-annihilation, a non-decay process with an odd number of external visible particles, generically possible when the stabilising symmetry is larger than $\set{Z}_2$.}\label{fig:SA}
\end{figure}
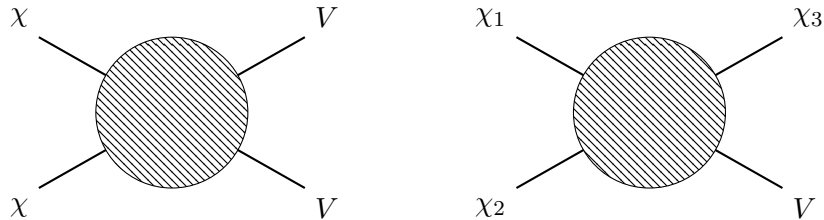

In particular, a combination of direct~\cite{1304.4279,1510.07754,1602.03489,1602.03781,lux2016talk,1607.07400}, collider~\cite{1604.07773,1611.03568}, and indirect (cosmic ray)~\cite{1604.00014,1609.08091,1611.03184} searches are seriously constraining the parameter space of one of the most commonly studied theoretical frameworks, thermal relic DM.   A key factor in the strength of these limits is the relation between rates at terrestrial experiments and the thermal relic density.  This derives from the common assumption that DM is stabilised by the existence of an unbroken $\set{Z}_2$ global symmetry, such as $R$-parity in supersymmetry, under which all Standard Model (SM) fields are even while the DM is odd.  Such a symmetry forbids all processes with an odd number of external DM fields.  If we restrict ourselves to $2\to 2$ processes, the only possible DM--SM connection is the well-known diagram shown on the left of \figref{fig:SA}.  Depending on the direction of time, this corresponds to DM annihilation to the SM (relevant for the relic density and indirect searches); DM scattering off the SM (relevant for direct searches); or DM production by SM annihilation (relevant for colliders).  With all these processes deriving from the same diagrams, their rates will be related.  The measured DM abundance lets us infer the annihilation cross section, and thereby a minimum sensitivity required to probe these models. 

However, this situation is not generic~\cite{0811.0172, 0907.1007, 0912.4496, 1003.5912}.  If the DM is charged under any global symmetry \emph{other} than $\set{Z}_2$, then processes with an odd number of dark sector fields can be allowed without inducing DM decay.  If we again restrict ourselves to $2\to 2$ processes, we have one new type of process, shown on the right of \figref{fig:SA}: $\dk_1\dk_2 \to \dk_3 V$, where $\dk_{1,2,3}$ are dark sector particles and $V$ an SM field.  This ``semi-annihilation'' (SA) changes dark sector number by one, compared to two for regular annihilation.  Most importantly, while SA can be very important for determining the relic density and for indirect detection experiments, it is essentially irrelevant for collider and direct searches.  The rates at these experiments can be substantially reduced while still obtaining the correct thermal abundance.  SA therefore represents a simple and generic extension of DM model space which has the potential to significantly weaken current bounds.

The natural question is then to what extent semi-annihilating dark matter (SADM) \emph{can} be constrained by current and near future searches.  To answer that we must first consider what the possible SA processes are, and what range of models can exist.  While several previous works have explored SADM phenomenology~\cite{err1202.2962,1210.7817,1211.1014,1402.6449,1403.4960,1405.5870,1410.2246,1507.08295,1508.03031,1509.08481,1510.02179,1511.09247,1601.06355,1607.01001,1611.09360}, these studies have typically been done in the context of specific models constructed for the purpose.  No systematic investigation has yet been attempted.  This is in contrast to conventional models of annihilating dark matter, where model-independent studies have a long history~\cite{Birkedal:2004xn,Feng:2005gj,Birkedal:2005ep}.  Our goal in this paper is to exploit these tools to make a first step in addressing the model space, general phenomenology, and constraints for SADM.

Modern model-independent studies of DM fall into two broad categories~\cite{1603.08002}: higher-dimension operators in an effective field theory (EFT)~\cite{0808.3384,0912.4511,1008.1783,1111.5331}, and renormalisable simplified models~\cite{1308.6799,1506.03116,1510.03434,1512.00476,1603.01366}.  The former has several advantages for our purposes.  The most immediate are its the relative simplicity and direct connection to the processes of interest.  Once we specify the external states, it is straightforward to construct all possible operators up to a given dimension.  Additionally, SADM is immune to the main phenomenological drawback, the breakdown of the EFT at high energies $E > \Lambda$, where $\Lambda$ is the UV cut-off.  This is a particular concern for collider studies, due to the high centre-of-mass energies involved~\cite{1112.5457}.  However, SA is only relevant for thermal freeze-out and indirect searches, which involve non-relativistic processes at approximately a single energy scale $\sqrt{s} \sim 2m$.  This makes it much more likely that the EFT will remain phenomenologically valid.  Finally, we note that $2\to 2$ SA as in \figref{fig:SA} will normally derive from four-field operators, three of which will be charged under $\dsym$.  Any renormalisable UV completion will involve intermediate states also charged under $\dsym$, which therefore must be more massive than the DM.  This also increases our belief in the validity of the EFT.  For all these reasons, we consider here higher-dimension operators and defer the construction of general renormalisable models to future work.

The irreducible constraints on SADM will come from indirect searches looking for the (decay products of the) visible state produced when two DM particles semi-annihilate today.  Once we have constructed the effective operators, we can use flux limits from current and near-future telescopes to place lower bounds on the UV scale $\Lambda$.  We can compare these limits to the scales needed for the same operator to give the correct thermal relic density; if we can exclude this latter value, additional (semi)-annihilation channels are needed.  These are expected to exist in a complete model; however, annihilation will bring additional limits while having multiple comparable semi-annihilation channels represents a model-building challenge.    As such, excluding the thermal relic cross section for a given SA operator disfavours it as a significant DM--SM coupling.  We show that this is already possible in some channels, but significant model space remains.

The outline of this paper is as follows.  We begin in \secref{sec:method} by defining our assumptions, notation and strategy for enumerating the possible effective operators that lead to SA.  We then proceed to construct them in \secref{sec:DMonly} for models where all light dark sector states are stable, and in \secref{sec:DP} when there can be additional unstable light states.  For simplicity we assume in these sections that the DM are total gauge singlets.  We also outline some generic constraints these models must obey.  Current and future bounds for these operators from indirect searches are given in \secref{sec:ID}, and we end by giving our conclusions in \secref{sec:conc}.  A few additional details are deferred to the appendices.  Some useful tensor symmetry properties are reviewed in \appref{app:TensSym}.  Additional effective operators are listed in \appref{app:HigherD} for potentially interesting terms at dimension 7 and 8, and in \appref{app:DMwithEW} for DM with SM gauge charges.  Last in \appref{app:pwave} we prove that scalar DM semi-annihilating to photons or gluons in our formalism is $p$-wave suppressed.

\section{Methodology and Notation}\label{sec:method}

Our framework for SADM is based on the following set of assumptions:
\begin{itemize}
	\item Dark matter is one or more stable electrically-neutral colour-singlet scalar and/or fermion particles, which together saturate the observed relic density.
	\item DM stability is enforced by a global symmetry $\dsym$ larger than $\set{Z}_2$, under which all SM fields are singlets.
	\item There may be additional light unstable scalar and/or fermion particles, charged under both $\dsym$ and the SM gauge group, relevant to SA processes.
	\item We assume that the SA interactions between the dark matter and visible sector can be well approximated by generally non-renormalizable effective contact operators.
\end{itemize}
We denote the fields charged under $\dsym$ as the dark sector, even if they have SM gauge charges.  We call unstable dark sector fields ``dark partners'', for reasons that will later become apparent.

The restriction to new particles of spin 0 and $\nicefrac{1}{2}$ is primarily done to avoid complications inherent in higher spin states.  Note, however, that as the spin of the dark sector fields increase, the dimension of the effective operators that couple to the SM will also increase, making them less relevant.  Taking the DM to be exactly stable avoids constraints from its decay; as discussed in the \nameref{sec:intro}, SA then requires $\dsym \neq \set{Z}_2$.  Exact DM stability can persist even in the presence of quantum gravity if $\dsym$ is the residual subgroup of a gauge symmetry broken at high energies~\cite{0905.0956,1402.6449,1407.5492,1407.6588}.

Our assumptions then lead us to a low-energy Lagrangian of the form
\begin{equation}
	\lag = \lag_{SM} + \lag_{Dark} + \sum_{d = 4}^\infty \sum_i \frac{c_d^i}{\Lambda^{d-4}} \, \op_d^i \,,
\end{equation}
where $\lag_{SM}$ is the usual SM Lagrangian; $\lag_{Dark}$ is the renormalisable dark-sector Lagrangian; $\Lambda$ is the UV cut-off; and $\op_d^i$ are dimension-$d$ operators connecting the two sectors with  Wilson coefficients $c_d^i$.  Aside from the $\op_d^i$, the only other possible connection between the two sectors are gauge couplings in $\lag_{Dark}$, \emph{e.g} for dark partners.  In general the $\op_d^i$ will include terms that lead to DM annihilation as well as SA.  We assume that these can be made subdominant, and discuss the consequences of that assumption where appropriate.  We also use $\op_{SA}$ to specifically refer to operators that lead to SA.

We do not specify the mechanism responsible for determining the DM relic density.  In particular, we will not impose that the SA effective operators generate the correct thermal relic density.  If the SA interactions are too weak, we can always posit additional (co-)annihilations to lower the relic density to the observed value.  If SA processes would \emph{under}-produce DM, then we must invoke some non-thermal process (such as late decays of more massive states~\cite{Lin:2000qq}) to regenerate the DM population.  In both cases, we see that the relic density is sensitive to details of the complete model not captured by our effective operator formalism.  However, when we show constraints in \secref{sec:ID}, we will show where in our parameter space SA processes alone generate the observed thermal abundance.

We allow for the existence of multiple DM states.  This is generically possible when $\dsym$ is sufficiently large; for example, if $\dsym = \set{Z}_4$ and the lightest and next-to-lightest states have charges 2 and 1 respectively, both are automatically stable.  However, while we will enumerate operators for arbitrary DM multiplicity, when we set limits we will focus on models where the DM is unique.  Limits on multicomponent dark sectors depend on (and are suppressed) by the fractional abundance of each state.  We would need to specify those abundances in some manner, which amounts to specifying how the DM abundance is set.

The set of possible effective operators coupling the SM to the dark sector is trivially infinite.  To make further progress, we make some further restrictions:
\begin{itemize}
	\item We focus on $2\to 2$ SA processes of the form $\dk_1\dk_2 \to \dk_3 \vis$, where $\dk_{1,2}$ are DM, $\dk_3$ either DM or a dark partner, and $\vis$ a SM particle.
	\item We give all operators up to dimension 6, plus a few leading operators at dimension 7.
	\item We assume that all DM states are complete gauge singlets.  Operators where this assumption is relaxed are given in \appref{app:DMwithEW}.
\end{itemize}
It is natural to expect $2 \to 2$ processes to dominate over $2\to 3$, since the latter are phase-space suppressed and often come from higher-dimensional operators.  However, we will find that in certain regions of parameter space $2\to 3$ operators can be more important.  For $2\to 2$ non-relativistic processes with DM mass $m$ mediated by a dimension $d$ operator $\op_d^i$ with Wilson coefficient $c_d^i = 1$, we can estimate the cross section as 
\begin{equation}
	\langle \sigma v \rangle \sim \frac{1}{8\pi} \, \frac{1}{4m^2} \, \biggl( \frac{m}{\Lambda} \biggr)^{2(d-4)} \sim \biggl( \frac{\text{100 GeV}}{m} \biggr)^{2} \biggl( \frac{m}{\Lambda} \biggr)^{2(d-4)} \times 10^{-23} \, \text{cm}^3 \text{s}^{-1} \,.
\end{equation}
Taking $\Lambda > 2 m$ for the EFT to be valid, and $\langle \sigma v \rangle \gtrsim 3 \times 10^{-26}$~cm$^2$\,s$^{-1}$ as an estimate of sensitivity, gives $d \lesssim 8.2$; instead taking $\Lambda = 4m$ gives $d \lesssim 6.1$.  This motivates our upper limit on $d$, as higher-dimension operators are unlikely to lead to observable signals.  

\begin{table}
	\centering
	\begin{tabular}{|c|c||c|c|}
		\hline
		SM Field & $SU(3)_c \times U(1)_{em}$ & SM Field & $SU(3)_c \times U(1)_{em}$ \\
		\hline\hline
		$u$ & $3_{2/3}$ & $h$ & $1_0$ \\
		$d$ & $3_{-1/3}$ & $\gamma,Z$ & $1_0$ \\
		$e$ & $1_{-1}$ & $W$ & $1_1$ \\
		$\nu$ & $1_0$ & $g$ & $8_0$ \\
		\hline
	\end{tabular}
	\caption{Definition of SM fermion (left) and boson (right) fields in the EW broken phase.  We suppress flavour indices.}\label{tab:SMdefBro}
\end{table}

We restrict our focus to gauge singlet DM for two reasons.  First, this maximizes the importance of SA in setting limits on the DM parameter space.  If the DM is charged under the electroweak gauge group, annihilations to pairs of gauge bosons will always be allowed.  We can estimate the relative size of these processes to the SA operators we consider as 
\begin{equation}
	\frac{\sigma^{gauge}}{\sigma^{SA}} \sim g^2 \biggl( \frac{\Lambda}{m} \biggr)^{2(d-4)},
\end{equation}
with $g$ the gauge coupling.  For $g = g_2$ and $d \geq 5$, the gauge processes always dominate.  Second, because all SA operators involve three dark sector fields, the number of possible gauge contractions for non-singlet DM becomes very large, such that enumerating them is less enlightening.

We construct operators in both the EW broken and unbroken phases, together with the explicit relations between the two bases.  The broken phase description allows us to make direct contact with phenomenology, and in particular makes our focus on $2\to 2$ processes most manifest.  Constraints placed in this basis can be mapped to any model, including those with light mediators or where the dark sector fields have EW charges, thanks to the non-relativistic nature of DM today.  Additionally, with one exception all possible SA processes can be generated by operators of dimension 6 or less in this phase.    The unbroken phase description is the natural basis when the dark sector only couples to the SM through mediators with masses large compared to the electroweak scale, and is thus more in the spirit of the EFT description.  We construct all unbroken phase operators up to dimension 6, plus some dimension 7 operators for processes that cannot be generated at lower dimension.  We will find that to this order, the relationship between symmetric and broken phase operators is one-to-many.  We thus also give additional dimension 7 and 8 operators in \appref{app:HigherD} that are required to generate all the broken phase operators individually.

In our operator lists, we use Fierz identities, integration by parts and the equations of motion to eliminate redundant terms.  In particular, we exploit the ability to use the lowest-order equations of motion~\cite{Politzer:1980me,KlubergStern:1975hc,GrosseKnetter:1993td,Arzt:1993gz,Simma:1993ky,Wudka:1994ny} to eliminate any operators containing
\begin{equation}
	D^\mu D_\mu \gfld \,, \quad \gamma^\mu D_\mu F \,, \quad D^\mu \V_{\mu\nu} \,, \quad D^\mu \V_\mu \,.
\end{equation}
with $D_\mu$ the covariant derivative, $\gfld$ any field, $F$ any fermion and $\V$ any vector or field strength tensor.  Additionally, we use the Bianchi identity
\begin{equation}
	\epsilon^{\mu\nu\rho\sigma} D_\nu \V_{\rho\sigma} = 0 \,,
\end{equation}
where $\epsilon^{\mu\nu\rho\sigma}$ is the Levi-Civita tensor, to further eliminate redundant terms.

\begin{table}
  \centering
  \begin{tabular}{|c|c||c|c|}
    \hline
    SM Field & $SU(3)_c \times SU(2)_L \times U(1)_Y$ & SM Field & $SU(3)_c \times SU(2)_L \times U(1)_Y$ \\
    \hline\hline
    $Q_L$ & $(3, 2)_{1/6}$ & $H$ & $(1, 2)_{1/2}$ \\
    $\bar{u}_R$ & $(\bar{3}, 1)_{-2/3}$ & $B$ & $(1, 1)_0$ \\
    $\bar{d}_R$ & $(\bar{3}, 1)_{1/3}$ & $W$ & $(1, 3)_0$ \\
    $L_L$ & $(1,2)_{-1/2}$ & $g$ & $(8, 1)_0$ \\
    $\bar{e}_R$ & $(1, 1)_1$ & & \\
    \hline
  \end{tabular}
  \caption{Definition of SM fermion (left) and boson (right) fields in the EW unbroken phase.  We suppress flavour indices.}\label{tab:SMdefUnb}
\end{table}

For clarity, we define our SM fields in the EW broken and symmetric phases in \tabsref{tab:SMdefBro} and \ref{tab:SMdefUnb}, respectively.  Unlike some similar works~\cite{1412.0520} we do not include a right-handed neutrino among the SM fields.  The broken (unbroken) phase fermions are Dirac (Weyl) spinors.  We normalise the Higgs VEV as $\expect{H^\dagger H} = v^2/2$, \emph{i.e.} $v = 246$~GeV, and define
\begin{equation}
	\tilde{H} = i \sigma^2 H^\ast \,, \qquad \tilde{\V}_{\mu\nu} = i \, \epsilon_{\mu\nu\rho\sigma} \V^{\rho\sigma} \,, \quad \text{and} \quad \brtilde{X} = \{X, \tilde{X} \} \,,
\end{equation}
where $\V_{\mu\nu}$ is any field strength tensor and $X$ is any object we might put a tilde over (including $\V_{\mu\nu}$).  We call a state \emph{light} if it is relevant to ID signals; roughly, this means it has a mass no more than twice the mass of the heaviest stable state.  

All our operators involve either three DM fields, or two DM fields and one dark partner field.  When there are multiple DM fields with the same spin, there will be an exchange symmetry, which the Wilson coefficients must respect.  We find it useful to show this explicitly by contracting our operators with dummy tensors of definite symmetry; that is, we write operators in the form
\begin{equation}
	\op_{SA} = s^{ijk} \dms_i \dms_j \dms_k \, H^\dagger H \,,
\end{equation}
where the Lagrangian term is
\begin{equation}
	\lag \supset \frac{c^{ijk}}{6\Lambda} \, \dms_i \dms_j \dms_k \, H^\dagger H \,,
\end{equation}
and the Wilson coefficient $c^{ijk}$ has the same symmetry representation as the dummy tensor $s$.  The factor of one-sixth is a symmetry factor.  For operators with two DM fields, we contract the DM flavour indices with $s^{ij}$, $a^{ij}$, or $n^{ij} = s^{ij} + a^{ij}$, where $s^{ij} = s^{ji}$ is fully symmetric and $a^{ij} = - a^{ji}$ is fully antisymmetric.  For operators with three DM fields, we use the fully (anti)-symmetric tensors $s^{ijk}$ and $a^{ijk}$, as well as two tensors $x^{ijk}$ and $y^{ijk}$ of mixed symmetry,
\begin{align}
  x^{ijk} &= - x^{jik} \,, & x^{ijk} + x^{jki} + x^{kij} &= 0 \,, \\
  y^{ijk} &= y^{jik} \,, & y^{ijk} + y^{jki} + y^{kij} &= 0 \,.
\end{align}
Some relevant properties of these tensors are listed in \appref{app:TensSym}.  Note that when there is only a single DM field, the only operators which survive are those contracted with $s^{ij}$ and $s^{ijk}$.

In our operator lists, we give each term a unique name of the form $\op_{dP}^{\vis x}$, where $d$ is the operator dimension; $P = B$ ($U$) for operators in the EW broken (unbroken) phase; $\vis$ denotes the SM state that appears in the $2\to 2$ SA process; and $x$ are additional labels to further distinguish the operators.  We use the same labels for the Wilson coefficients.  In the EW broken phase, $f$ stands for any (Dirac) SM fermion, and $\V_\mu \in \{Z_\mu , W_\mu\}$ represents a massive vector field.  In the EW broken phase, we use $F$ ($\bar{f}$) to denote the left-handed doublet (right-handed singlet) SM Weyl fermions.  In both phases, $\V_{\mu\nu}$ represents the field strength tensor of any of the SM vector fields; $\dms$ is a (in general complex) DM scalar; and $\dmf$ is a (in general Dirac) DM fermion with Weyl components $(\dmfL , \dmfR)^T$.  When we introduce dark partners in \secref{sec:DP}, we use $\dps$ for scalars and $\dpf = (\dpfL, \dpfR)^T$ for fermions.  Lastly for any fields $\gfld_{1,2}$ we define
\begin{equation}
  \gfld_1 \Dlr{\mu} \gfld_2 = \gfld_1 \, (D_\mu \gfld_2) - (D_\mu \gfld_1) \, \gfld_2 \,.
\end{equation}

\section{Dark Matter Only Models}\label{sec:DMonly}

We first consider the case where the only light states in the dark sector are stable, \emph{i.e.} DM.  We can subdivide this into three classes, depending on whether these states are scalars, fermions or a mixture of both.  Only the first two possibilities exist if we demand a unique DM state.  For a $2 \to 2$ SA process, the only possible SM final states are the $h$, $\gamma$, $Z$ and $\nu$, while any SM final state can appear in a $2\to 3$ process.

A generic feature of (almost) all the operators we write down in the EW symmetric phase is the presence of $2\to 3$ processes.  Recall that we defined the SM content without a right-handed neutrino, so the only single-field gauge singlet in the visible sector is the hypercharge field strength tensor, $B_{\mu\nu}$.  Since (by assumption) the DM fields are all total gauge singlets, operators that can lead to SA to $h$, $Z_\mu$ or $\nu$ necessarily also lead to $2\to 3$ processes.  Though these are phase-space suppressed by $1/(4\pi)^{2}$, processes with two-body final states are suppressed by $v^2/m^2$:
\begin{equation}
	\amp_{2\to 2} \sim \frac{v\,m^{d-5}}{\Lambda^{d-4}} \,, \qquad \amp_{2\to 3} \sim \frac{m^{d-4}}{\Lambda^{d-4}} \,.
\end{equation}
Na\"\i vely, $2\to 2$ processes will dominate for $m \lesssim 4\pi v \approx 3$~TeV.  The actual condition is usually weaker than this, since the final states from three-body processes are typically softer than from two-body processes, making them harder to distinguish from backgrounds.  Exceptions can occur when the 3-body final state involves a more-easily detected SM particle, but this is not a concern for any of the operators we find.  In \secref{sec:ID}, we will find that the limits at such high masses are weak, justifying our focus on $2\to 2$ processes.

A second generic feature of all DM models is the possibility of annihilation through Higgs portals.  For scalar DM, we can always write the renormalisable coupling
\begin{equation}
	- \lag \supset \lambda^{ij} \, \dms_i^\dagger \dms_j \, H^\dagger H \,,
\label{eq:scaHP}\end{equation}
while for fermion DM, we have the dimension-5 operator~\cite{Kim:2006af}
\begin{equation}
	- \lag \supset \frac{1}{\Lambda} \, \bar{\dmf}_i \bigl( c_S^{ij} + c_P^{ij} \gamma^5 \bigr) \dmf_j \, H^\dagger H \,.
\label{eq:ferHP}\end{equation}
These terms are always allowed for $i = j$, and non-diagonal couplings may exist depending on the $\dsym$ representations.  The presence of annihilation channels is not a theoretical obstacle; it may even be necessary to obtain the correct relic density, as discussed in \secref{sec:method}.  However, the Higgs portals are low dimension operators so could plausibly dominate the DM phenomenology.  Demanding that the SA processes are the most important gives bounds on $\lambda$ and $c_P$, which in turn restricts possible UV completions.  ($c_S$ leads to $p$-wave suppressed annihilation, so it is not constrained by these considerations.)  If the UV physics generates the Higgs portal coupling at $n$-loops, then
\begin{equation}
	\lambda, c_P \sim \frac{N_F}{(4\pi)^{2n}} \,,
\label{eq:HPcoeff}\end{equation}
where $N_F$ is roughly the number of diagrams that contribute.  In particular, for one-loop generation, $N_F$ is the number of fields that run in the loop.  It is straightforward to forbid tree-level generation of Higgs portal couplings: the UV completion should contain no gauge- and $\dsym$-singlet scalars, as well as no hypercharge one-half electroweak doublets with the same spin and conjugate $\dsym$ representation as the DM.  Forbidding one-loop generation while allowing tree-level generation of $\op_{SA}$ is a more serious model-building challenge.  As such we shall assume $\lambda, c_P \sim N_F/(4\pi)^{2}$, and infer constraints on the UV scale $\Lambda$.

These limits depend on the SA final state, but we can derive a simple estimate by requiring that the annihilation cross sections are smaller than those for SA.  For scalars, consider an SA operator of dimension $d = 4 + \delta$, where the $2\to 2$ process is also $v/m$-suppressed.  Then SA dominates if
\begin{equation}
	\Lambda \lesssim c_{SA} \, \frac{v}{\lambda} \, \biggl( \frac{m}{\Lambda}\biggr)^{\delta - 1} \,.
\end{equation}
If $\lambda$ is generated at tree-level, this requires $\Lambda < v$, in which case the EFT framework is unlikely to be valid.  Alternatively, if the Higgs portal is one-loop, $N_F \sim 3$, and $c_{SA} \sim 1$ then we get no bounds on renormalisable SA operators and
\begin{equation}
	\Lambda \lesssim \{ 10, 5, 3 \} \text{ TeV} \quad \text{for } d = 5, 6, 7.
\label{eq:scaHPlimV}\end{equation}
A similar analysis for fermions gives no bounds on dimension-5 $\op_{SA}$ and 
\begin{equation}
	\Lambda \lesssim \{ 10, 5 \} \text{ TeV} \quad \text{for } d = 6, 7.
\label{eq:ferHPlimV}\end{equation}
Finally, for operators where $2\to 2$ processes do not have this suppression, we find do not find absolute bounds, but rather we constrain the ratio $m/\Lambda$:
\begin{align}
	\Lambda & \lesssim \frac{m}{\lambda^{1/\delta}} \sim \{ 50m , 7m, 3m \} \text{ for scalar DM and } d = 5, 6, 7 \,, \label{eq:scaHPlimNV} \\
	\Lambda & \lesssim \{ 50m , 7m \} \text{ for fermion DM and } d = 6, 7 \,.\label{eq:ferHPlimNV} 
\end{align}
Renormalisable scalar and dimension-5 fermion $\op_{SA}$ are unconstrained, and tree-level $\lambda$ or $c_P$ will still always dominate.  As noted, these are only approximate bounds; we comment on their applicability to specific SA operators as relevant.

\subsection{Scalar dark matter}\label{sec:scaDMonly}

\begin{table}
	\centering
	\begin{tabular}{|c|c|}
		\hline
		Operator & Definition \\
		\hline
		$\op^{h}_{4B}$ & $s^{ijk} \dms_i \dms_j \dms_k h$ \\
		$\op^{Z}_{5B}$ & $( x^{ikj} + y^{ijk} ) \, \dms_i \dms_j (\partial_\mu \dms_k) Z^\mu$ \\
		$\op^{h\dms}_{6B}$ & $( x^{ikj} + y^{ijk} ) \,  (\partial_\mu \dms_i) (\partial^\mu \dms_j) \dms_k h$ \\
		\hline
	\end{tabular}
	
	\vspace{8pt}
	
	\begin{tabular}{|c|c|c|}
		\hline
		Operator & Definition & EW Broken Phase \\
		\hline
		$\op^{H}_{5U}$ & $s^{ijk} \dms_i \dms_j \dms_k \, H^\dagger H$ & $\frac{v}{\Lambda} \, \op^{h}_{4B}$ \\
		$\op^{Z}_{7U}$ & $(x^{ikj} + y^{ijk}) \, \dms_i \dms_j (\partial^\mu \dms_k) \bigl( i H^\dagger \Dlr{\mu} H \bigr)$ & $- \frac{v m_Z}{\Lambda^2} \, \op^{Z}_{5B}$ \\
		$\op^{H}_{7U}$ & $(x^{ikj} + y^{ijk}) \, (\partial_\mu \dms_i) (\partial^\mu \dms_j) \dms_k \, H^\dagger H$ & $\frac{v}{\Lambda} \, \op^{h\dms}_{6B}$ \\
		\hline
	\end{tabular}
	\caption{Operators relevant for dark sectors comprised of scalar DM and no other light fields, in the EW broken phase (top) and symmetric phase (bottom).  We also show the relation between the two phases in the lower table.}\label{tab:ScaDM}
\end{table}

We first consider the case where the dark sector consists of one or more stable scalars and no other light states.  In this case, the possible SM final states are the Higgs, $Z$ and photon.  In the EW broken phase, there are only three operators contributing to $2 \to 2$ processes up to dimension 6, which we show in the upper part of \tabref{tab:ScaDM}.  Only $\op^{h}_{4B}$ is non-vanishing when the DM state is unique.  These generate the SA processes $\dms_i \dms_j \to \dms_k^\dagger Z$ and $\dms_i \dms_j \to \dms_k^\dagger h$.  The leading contribution to $\dms_i \dms_j \to \dms_k^\dagger \gamma$ occurs at dimension 7 (and vanishes unless the DM fields are all distinct).  However, as shown in \appref{app:pwave}, all effective operator contributions to this process lead to $p$-wave annihilation.  They therefore do not lead to observable signals today and so we ignore them.  In contrast, all the terms in \tabref{tab:ScaDM} lead to $s$-wave annihilation.

In the EW unbroken phase for gauge singlet DM, there is only a single operator up to dimension 6, which corresponds to $\op_{4B}^h$ after EWSB.  The other two broken phase terms derive from dimension 7 operators, which we also include in the lower part of \tabref{tab:ScaDM}.  We see that the operators in the two phases are in a one-to-one relation in this case.

In constructing the operators in \tabref{tab:ScaDM}, we are implicitly assuming that they are the most important couplings between the visible and dark sectors.  It is necessary to critically examine that assumption, and the implications for the parameter space and possible UV completions.  We have already discussed the existence of $2\to 3$ processes in the EW unbroken phase.  In this case, the final states for both types of processes are similar (Higgses and electroweak gauge bosons) so that the simple bound above, $m \lesssim 3$~TeV, is probably a reasonable approximation for when two-body final states dominate.

Next, we discuss the Higgs portal coupling of \modeqref{eq:scaHP}.  The constraints of \modeqref{eq:scaHPlimV} apply well to the operators in this section, as both annihilation and SA involve similar final states.  Tree-level generation of the quartic will always dominate, while if it is one-loop then SA will dominate for $\Lambda \lesssim \order{10}$~TeV (3~TeV) for $\op_{5U}^H$ ($\op_{7U}^{Z,H}$).  All the operators of \tabref{tab:ScaDM} will generate the quartic coupling at two-loops, which gives a lower bound
\begin{equation}
	\lambda \gtrsim \frac{c_{5U}^{H}/c_{7U}^{Z,H}}{(4\pi)^4} \,,
\end{equation}
independent of the UV completion.  If this value is realised, the upper limit on $\Lambda$ increases by a factor of $(4\pi)^2$.

Lastly we observe that if we can write down any of the operators in \tabref{tab:ScaDM}, then in addition to the Higgs portal term of \modeqref{eq:scaHP} we can also include the renormalisable term
\begin{equation}
  - \lag \supset \frac{1}{6} \, \rho^{ijk} \dms_i \dms_j \dms_k + h.c.
\label{eq:phi3}\end{equation}
The (fully symmetric) cubic term must be a singlet under the dark sector symmetry $\dsym$ for the terms in \tabref{tab:ScaDM} to be allowed, while it is a SM singlet since we restrict ourselves to total gauge singlet DM.  It can only be generated in the UV at tree-level by mixing between the DM and a heavy dark sector field.  If this is forbidden, we expect
\begin{equation}
	\rho \sim \frac{N_F \Lambda}{(4\pi)^2} \,.
\end{equation}
In the low-energy theory, $\op_{5U}^{H}$ generates $\rho$ at one loop, while the dimension 7 operators generate it at four loops.  These contributions give the approximate lower bounds
\begin{equation}
	\rho \gtrsim c_{5U}^{H} \, \frac{\Lambda}{(4\pi)^2} \quad \text{or} \quad  c_{7U}^{Z,H} \, \frac{\Lambda}{(4\pi)^8} \,.
\end{equation}

The cubic and Higgs portal couplings together lead to the SA process $\dms_i\dms_j\to \dms_k^\dagger h$.  This will have the same non-relativistic cross section as $\op^{H}_{5U}$ if
\begin{multline}
  \frac{(c^{H}_{5U})^{ijk}}{\Lambda} = \sum_l \biggl[ \frac{\rho^{ijl}\lambda^{lk}}{(m_i + m_j)^2 - m_l^2} - \frac{\rho^{ikl} \lambda^{lj} (m_i + m_j)}{(m_i + m_j) (m_i m_j + m_l^2) - m_j m_k^2 - m_i m_h^2} \\
  - \frac{\rho^{jkl} \lambda^{li} (m_i + m_j)}{(m_i + m_j) (m_i m_j + m_l^2) - m_i m_k^2 - m_j m_h^2} \biggr] \to - \frac{\rho \lambda}{m^2} \,,
\end{multline}
where in the last step we assumed a single DM species with $m^2 \gg m_h^2$.  This allows us to interpret the limits we derive on $\op^{H}_{5U}$ directly as limits on $\rho \lambda/m^2$ in a renormalizable theory.  It also lets us identify in which regions of parameter space the higher-dimensional operators serve as the dominant couplings between the two sectors:
\begin{equation}
	\Lambda \lesssim c_{5U}^{H} \, \frac{m^2}{\rho \lambda} \qquad \text{or} \qquad \Lambda^3 \lesssim c_{7U}^{H,Z} \, \frac{m^4}{\rho \lambda} \,.
\end{equation}
These inequalities are stronger than the constraints from annihilation if $m^2 < v \rho$.  In contrast, if both $\rho$ and $\lambda$ are generated at one loop then we find the weaker conditions $\Lambda \lesssim 16\pi^2 m/N_F$ ($\Lambda \lesssim 4\pi m/\sqrt{N_F}$) for $\op_{5U}^{H}$ ($\op_{7U}^{H,Z}$).

\subsection{Fermion dark matter}\label{sec:fdmonly}

\begin{table}
	\centering
	\begin{tabular}{|c|c|}
		\hline
		Operator & Definition \\
		\hline
		$\op_{6B}^{\nu L}$ & $\bigl( s^{ijk} + y^{ijk} + x^{ikj} \bigr) \, (\bar{\dmf}_i^c P_L \dmf_j) \, (\bar{\nu} P_R \dmf_k)$ \\
		$\op_{6B}^{\nu R}$ & $\bigl( y^{ijk} + x^{ikj} \bigr) \, (\bar{\dmf}_i^c P_R \dmf_j) \, (\bar{\nu} P_R \dmf_k)$ \\
		\hline
	\end{tabular}
	
	\vspace{8pt}
	
	\begin{tabular}{|c|c|c|}
		\hline
		Operator & Definition & EW Broken Phase \\
		\hline
		$\op_{7U}^{LL}$ & $\bigl( s^{ijk} + y^{ijk} + x^{ikj} \bigr) \, (\dmfL_i \dmfL_j) \, \bigl( (L^\dagger \tilde{H}) \dmfR_k \bigr)$ & $\frac{v}{\sqrt{2}\Lambda} \, \op^{\nu L}_{6B}$ \\
		$\op_{7U}^{LR}$ & $\bigl( y^{ijk} + x^{ikj} \bigr) \, (\bar{\xi}_i^\dagger \bar{\xi}_j^\dagger) \, \bigl( (L^\dagger \tilde{H}) \dmfR_k \bigr)$ & $\frac{v}{\sqrt{2}\Lambda} \, \op^{\nu R}_{6B}$ \\
		\hline
	\end{tabular}
	\caption{Operators relevant for dark sectors comprised of fermion DM and no other light fields, in the EW broken phase (top) and unbroken phase (bottom).}\label{tab:FerDM}
\end{table}

We next consider a dark sector composed of one or mor stable fermions $\dmf_i$ and no other light states.  There is only a single possible $2\to 2$ SA process, $\dmf_i\dmf_j \to \dmf_k^\dagger \nu$.  This is generated by the two dimension-6 operators shown in the upper part of \tabref{tab:FerDM} (six after accounting for neutrino generations).  We have used Fierz identities to reduce these to the minimal set.  In particular, spinor lines involving Lorentz indices all either vanish identically or can be written in terms of the objects in \tabref{tab:FerDM}.  Both operators lead to phenomenologically-relevant $s$-wave annihilation.  Note that, when the DM is unique, only the single operator $\op_{6B}^{\nu L}$ survives.

In the unbroken phase, there are no operators up to dimension 6.  The leading terms are at dimension 7, generated from the broken phase operators through the replacement $\bar{\nu}_L \to L^\dagger \tilde{H}$.  We list these terms in the lower part of \tabref{tab:FerDM}; recall that the four component DM spinor $\dmf$ has Weyl components $(\dmfL, \dmfR)^T$.  As for scalar DM, the unbroken phase operators also lead to $2\to3$ SA processes.  The inequality  $m \lesssim 4\pi v$ likely underestimates the region where $2\to2$ processes dominate, as they have a clear feature (a monochromatic neutrino) that is lacking for three body final states.

These models are simpler than pure scalar DM, in that Lorentz invariance forbids a $\dmf^3$ term so there is no necessary additional source of SA.  We still have the Higgs portal bounds from \modeqref{eq:ferHPlimV}, $\Lambda \lesssim 5$~TeV.  As for the comparison with $2\to3$ processes, this condition is probably conservative: SA leads to a feature, monochromatic neutrinos at $3m_\dmf/4$, while annihilation bounds will derive from a broad excess in $\gamma$ rays at $\order{10}$~GeV.  Finally, we note that the operators of \tabref{tab:FerDM} only generate the Higgs portal coupling at two loops, so there is no inconsistency with this coupling being small in the low energy theory.

\subsection{Scalar and Fermion dark matter}\label{sec:SFonly}

\begin{table}
	\centering
	\begin{tabular}{|c|c||c|c|}
		\hline
		Operator & Definition & Operator & Definition\\
		\hline
		$\op_{5B}^{\nu}$ & $s^{ij} \, \dms_i \dms_j \, \bar{\nu} P_R \dmf$ & $\op_{6B}^{\gamma}$ & $a^{ij} \, \bar{\dmf}^c_i \sigma^{\mu\nu} \dmf_j \, \dms \, \gamma_{\mu\nu}$ \\
		$\op_{6B}^{\nu}$ & $a^{ij} \dms_i (\partial_\mu \dms_j) \, \bar{\nu} \gamma^\mu P_L \dmf$ & $\tilde{\op}_{6B}^{\gamma}$ & $a^{ij} \, \bar{\dmf}^c_i \sigma^{\mu\nu} \dmf_j \, \dms \, \tilde{\gamma}_{\mu\nu}$ \\
		\cline{1-2}
		$\op_{5B}^{hS}$ & $s^{ij} \, \bar{\dmf}_i^c \dmf_j \, \dms \, h$ & $\op_{6B}^{Z}$ & $a^{ij} \, \bar{\dmf}^c_i \sigma^{\mu\nu} \dmf_j \, \dms \, Z_{\mu\nu}$ \\
		$\op_{5B}^{hP}$ & $s^{ij} \, \bar{\dmf}_i^c \gamma^5 \dmf_j \, \dms \, h$ & $\tilde{\op}_{6B}^{Z}$ & $a^{ij} \, \bar{\dmf}^c_i \sigma^{\mu\nu} \dmf_j \, \dms \, \tilde{Z}_{\mu\nu}$ \\
		$\op_{5B}^{Z V}$ & $a^{ij} \, \bar{\dmf}_i^c \gamma^\mu \dmf_j \, \dms \, Z_\mu$ & $\op_{6B}^{ZsS}$ & $s^{ij} \bigl( \dms \dlr{\mu} (\bar{\dmf}_i^c \dmf_j) \bigr) Z^\mu$ \\
		$\op_{5B}^{Z A}$ & $s^{ij} \, \bar{\dmf}_i^c \gamma^\mu \gamma^5 \dmf_j \, \dms \, Z_\mu$ & $\op_{6B}^{ZsP}$ & $s^{ij} \bigl( \dms \dlr{\mu} (\bar{\dmf}_i^c \gamma^5 \dmf_j) \bigr) Z^\mu$ \\
		\cline{1-2}
		$\op_{6B}^{hV}$ & $a^{ij} \, \bar{\dmf}_i^c \gamma^\mu \dmf_j \,  \bigl( \dms \dlr{\mu} h \bigr)$ & $\op_{6B}^{ZaS}$ & $a^{ij} ( \bar{\dmf}_i^c \partial_\mu \dmf_j ) \, \dms \, Z^\mu$ \\
		$\op_{6B}^{hA}$ & $s^{ij} \, \bar{\dmf}_i^c \gamma^\mu \gamma^5 \dmf_j \,  \bigl( \dms \dlr{\mu} h \bigr)$ & $\op_{6B}^{ZaP}$ & $a^{ij} ( \bar{\dmf}_i^c \gamma^5 \partial_\mu \dmf_j ) \, \dms \, Z^\mu$ \\
		\hline
	\end{tabular}
	\caption{Operators relevant for dark sectors comprised of scalar and fermion DM, and no other light fields, in the EW broken phase.  The operators $\op_{5B}^{\nu}$ and $\op_{6B}^{\nu}$ involve two DM scalars and one DM fermion; all others involve two fermions and one scalar.  All operators lead to $s$-wave cross sections except for $\dmf\dmf$-initiated processes from $\op^{h\dmf S}_5$ and $\op^{Z\dmf S}_6$.}\label{tab:MixDMBro}
\end{table}

Finally we consider the case where there are both scalar and fermion DM fields.  The DM is necessarily multicomponent, and all neutral SM particles are possible final states.  In the EW broken phase up to dimension 6 we find a total of 16 operators (20 summing over neutrino generations), as listed in \tabref{tab:MixDMBro}.  Two (three-fold degenerate) operators involve two DM scalar fields, and the remainder involve two fermions.  Generically, either only operators with two scalars \emph{or} only operators with two fermions will be allowed by the dark symmetry $\dsym$.  Each operator leads to SA processes where the initial state particles have the same \emph{and} have different spin.  The latter cross sections always have non-zero $s$-wave piece.  The same is true for all processes with $\dms\dms$ initial states, but two of the operators lead to $p$-wave $\dmf\dmf$-initiated SA.  Lastly, seven operators survive in the minimal case with a single scalar and a single fermion field.

In the EW unbroken phase, there are five operators at dimension 6 and an additional five at dimension 7, as listed in \tabref{tab:MixDMUnb}.  Eight of these are in one-to-one correspondance with the operators in the left column of \tabref{tab:MixDMBro}.  The two remaining operators, $\op_{6U}^B$ and $\tilde{\op}_{6U}^B$, each generate two of the remaining broken phase terms in a fixed ratio through the replacement
\begin{equation}
	B_{\mu\nu} \to \cos \theta_W \, \gamma_{\mu\nu} - \sin\theta_W \, Z_{\mu\nu} \,,
\end{equation}
where $\theta_W$ is the Weinberg angle.  If we want to generate these broken phase operators separately or with arbitrary coefficients, we would need to go to dimension 8 in the unbroken phase.  This is also the dimension at which the operators $\op_{6B}^{Z(s/a)(S/P)}$ are generated for total gauge singlet DM.  These four terms are not the leading contributions to any SA processes, so we defer all these dimension 8 terms to \appref{app:HigherD}.

\begin{table}
	\centering
	\begin{tabular}{|c|c|c|}
		\hline
		Operator & Definition & EW Broken Phase \\
		\hline
		\hline
		$\op_{6U}^{LH^\dagger}$ & $s^{ij} \, \dms_i \dms_j \, \bigl( (L^\dagger \tilde{H}) \dmfR \bigr)$ & $\frac{v}{\sqrt{2}\Lambda} \, \op_{5B}^\nu$ \\
		$\op_{7U}^{L}$ & $a^{ij} \dms_i (\partial_\mu \dms_j) \, \bigl( (L^\dagger \tilde{H}) \bar{\sigma}^\mu \dmfL \bigr)$ & $\frac{v}{\sqrt{2}\Lambda} \, \op_{6B}^\nu$ \\
		\hline
		\hline
		$\op_{6U}^{HS}$ & $s^{ij} \, \bar{\dmf}_i^c \dmf_j \, \dms \, H^\dagger H$ & $\frac{v}{\Lambda} \, \op_{5B}^{hS}$ \\
		$\op_{6U}^{HP}$ & $s^{ij} \, \bar{\dmf}_i^c\gamma^5  \dmf_j \, \dms \, H^\dagger H$ & $\frac{v}{\Lambda} \, \op_{5B}^{hS}$ \\
		$\brtilde{\op}_{6U}^{B}$ & $a^{ij} \, \bar{\dmf}_i^c \sigma^{\mu\nu} \dmf_j \, \dms \, \brtilde{B}_{\mu\nu}$ & $c_W \, \brtilde{\op}_{6B}^{\gamma} - s_W \, \brtilde{\op}_{6B}^{Z}$ \\
		\hline
		$\op_{7U}^{ZV}$ & $a^{ij} \, \bar{\dmf}_i^c \gamma^\mu \dmf_j \, \dms \, \bigl( i H^\dagger \Dlr{\mu} H \bigr)$ & $- \frac{v m_Z}{\Lambda^2} \, \op_{5B}^{ZV}$ \\
		$\op_{7U}^{ZA}$ & $s^{ij} \, \bar{\dmf}_i^c \gamma^\mu \gamma^5 \dmf_j \, \dms \, \bigl( i H^\dagger \Dlr{\mu} H \bigr)$ & $- \frac{v m_Z}{\Lambda^2} \, \op_{5B}^{ZA}$ \\
		$\op_{7U}^{HV}$ & $a^{ij} \, \bar{\dmf}_i^c \gamma^\mu \dmf_j \, \bigl( \dms \dlr{\mu} (H^\dagger H) \bigr)$ & $\frac{v}{\Lambda} \, \op_{6B}^{hV}$ \\
		$\op_{7U}^{HA}$ & $s^{ij} \, \bar{\dmf}_i^c \gamma^\mu \gamma^5 \dmf_j \, \bigl( \dms \dlr{\mu} (H^\dagger H) \bigr)$ & $\frac{v}{\Lambda} \, \op_{6B}^{hV}$ \\
		\hline
	\end{tabular}
	\caption{Operators relevant for dark sectors comprised of scalar and fermion DM, and no other light fields, in the EW unbroken phase.  The upper (lower) section lists operators with two DM scalars and one fermion (one scalar and two fermions).  $c_W$ ($s_W$) is the cosine (sine) of the Weinberg angle.  Recall that $\brtilde{\op} = \{ \op, \tilde{\op}\}$ so that the fifth line represents two different operators.}\label{tab:MixDMUnb}
\end{table}

The majority of the terms in \tabref{tab:MixDMUnb} lead to $2\to 3$ processes, with similar conclusions to those in the pure scalar and pure fermion scenarios.  However, the operators $\op_{6U}^B$ and $\tilde{\op}_{6U}^B$ \emph{only} generate $2\to 2$ processes and so bypass this concern.  With a suitable assignment of $\dsym$ charges, it is possible to write all the operators in this section while forbidding those of \secsref{sec:scaDMonly} and \ref{sec:fdmonly}, or the scalar cubic of \modeqref{eq:phi3}.  However, we can not forbid DM annihilation through the scalar and fermion Higgs portals.  A direct comparison of annihilation and SA signals is complicated by the multicomponent nature of the DM; for example, if the DM today is mostly-fermion then scalar annihilation will be suppressed compared to scalar-fermion SA.  The unavoidable constraints are between processes with identical initial states.

For $\dms\dms$-initiated processes, the SA channels involve neutrino final states.  SA can plausibly set stronger limits even if its cross section is relatively small due to the signal being easily distinguishable from the backgrounds.  If we still demand the SA cross section to be larger, the bounds from \modeqref{eq:scaHPlimV} are $\Lambda \lesssim 5$~TeV (3~TeV) for $\op_{6U}^{LH^\dagger}$ ($\op_{7U}^{L}$).  Additionally, we note that these operators generate the Higgs portal coupling at two loops so there is no problem with $\lambda$ being small.

For $\dmf\dmf$-initiated processes, we have two classes of SA channels.  $\op_{6U}^B$ and $\tilde{\op}_{6U}^{B}$ lead to monochromatic photon final states and are not $v/m$-suppressed.  The simple limits of \modeqref{eq:ferHPlimNV}, $\Lambda \lesssim 50m$, should be conservative.  The other operators involve $h$ and $Z$ final states, where we expect \modeqref{eq:ferHPlimV} to be a reasonable estimate of when SA dominates: $\Lambda \lesssim 10$ (5)~TeV for the dimension 6 (7) operators.  These latter terms also generate the fermion Higgs portal at two loops, so there is no problem with $c^P$ being small.

Finally, we note that for the operators $\op_{6U}^{LH^\dagger}$ and $\op_{7U}^{L}$, it is possible to arrange the dark sector charges under $\dsym$ such that these operators are the leading contributions to SA.  However, for all the other terms in \tabref{tab:MixDMUnb}, we can also write the renormalisable coupling 
\begin{equation}
  - \lag \supset \dms \, \bar{\dmf}^c_i (y_S^{ij} + y_P^{ij} \gamma^5) \dmf_j \,.
\label{eq:ffsterm}\end{equation}
Together with the scalar Higgs portal, these will lead to the SA processes $\dmf\dmf \to \dms^\dagger h$ and $\dmf\dms \to \bar{\dmf}h$. The former is $p$-wave suppressed when $y_P = 0$, but the latter always has a non-zero $s$-wave piece.  For non-relativistic processes, the cross-sections are the same as for the operators $\op_{6U}^{HS/P}$ if we make the relation
\begin{equation}
  c_{6U}^{HS,P} \, \frac{m^2}{\Lambda^2} \approx y_{S,P} \lambda\,.
\end{equation}
As in the pure scalar case, we can interpret limits on the effective operators as limits on $y_{S,P} \lambda$ (and vice versa).  We can also consider when our effective operators are the dominant coupling.  If we assume $\lambda \sim N_F/(4\pi)^2$ to suppress annihilation as discussed above, then $y_{S,P} \sim \order{1}$ is allowed provided that $\Lambda \lesssim 4\pi m/\sqrt{N_F}$.  This is stronger than \modeqref{eq:ferHPlimNV} for $\brtilde{\op}_{6U}^B$, but relatively mild for the $h$ and $Z$ operators.

\section{Dark Partner Models}\label{sec:DP}

In the previous section we considered SA effective operators for $2\to 2$ processes where all external states are either DM or SM.  In doing so, we restricted ourselves to only four possible visible-sector final states: $h$, $\gamma$, $Z$ and $\nu$.  Further, if the DM particle is unique then only two possibilities remain up to dimension 7: $\dms\dms \to \dms^\dagger h$ from $\op_{5U}^{H}$ and $\dmf\dmf \to \bar{\dmf}\nu$ from $\op_{7U}^{\nu L}$.  This would imply a relatively sparse space of SADM phenomenolgy.

However, it is possible that there could exist states charged under both the dark symmetry $\dsym$ and the SM gauge group.  If heavy, these states will serve as mediators between the two sectors, but if light they open new possibilities for SA.  We continue to restrict our focus to $2\to 2$ processes with two DM particles in the initial state, which can potentially produce cosmic ray signals today.  All such channels will take the form $\dm_i\dm_j \to \dkp^\dagger\vis$, where $\dm$ represents scalar and/or fermion DM, and $\dkp$ is charged under $\dsym$ and has the same SM charges as the visible state $\vis$.  Including these dark partners allows all SM particles to appear as SA final states.  Additionally, we will find that the majority of dark partner processes can be generated at the same dimension in the broken and unbroken phases, avoiding the $v/\Lambda$ suppression associated with most DM-only SA.

There is one obstacle to the inclusion of dark partners.  By definition, these states are unstable (and generally must be so to avoid constraints on charged and coloured relics).  They must also be light enough that non-relativistic DM can produce them.  This gives us the bound
\begin{equation}
  m_{\dkp} < m_i + m_j - m_\vis < m_i + m_j \,,
\end{equation}
where $m_{i,j}$ are the DM masses.  When this constraint is satisfied, it means that the dark partner can not decay to any final state involving the two DM particles $\dm_i \, \dm_j$.  In particular, in our minimal theory where the operator $\op_{SA} = c_{SA} \, \dm_i \, \dm_j \, \dkp \, \vis^\dagger/\Lambda^\delta$ is the only non-gauge coupling between the two sectors, $\dkp$ will be stable.  We must expand the connection between two sectors, which in general will modify the dark sector phenomenology.

We add a single term to our theory of the form
\begin{equation}
	\lag \supset \op_{dec} \equiv \frac{c_{dec}}{\Lambda^{\delta_{dec}}} \, \dkp \, \dm_k^\dagger \, \op_{SM}^\dagger \,,
\label{eq:dkpdecay}\end{equation}
where $\op_{SM}$ is an operator built from SM fields with the same quantum numbers as $V$.  This allows the dark partner to decay as $\dkp \to \dm_k + SM$.  The minimal case $\op_{SM} = V$ may or may not be allowed by Lorentz invariance, depending on the spins of $X_k$ and $\dkp$.  In particular, note that if all DM states are scalars (fermions), then $\op_{SM} = V$ is always allowed (forbidden), regardless of the spin of $\dkp$.  $\op_{dec}$ does not break $\dsym$ if $\dkp$ and $\dm_k$ transform identically; so when there are both fermion and scalar DM, $\op_{SM} = V$ can be forbidden or allowed depending on the dark sector representations.

The decay width for $\dkp$ mediated by \modeqref{eq:dkpdecay} is
\begin{equation}
	\Gamma \sim \frac{c_{dec}^2 m_\dkp}{8\pi} \, \biggl( \frac{1}{4\pi} \biggr)^{2(n-1)} \, \biggl( \frac{m_\dkp}{\Lambda} \biggr)^{2\delta_{dec}} \,,
\end{equation}
where $n$ is the number of SM fields in $\op_{SM}$.  There are cosmological lower bounds on $c_{dec}$, depending on its decay modes.  In particular, decays with a lifetime $\tau \gtrsim 0.05$\,s will occur after the onset of big-bang nucleosynthesis, and can spoil the successful predictions of the primordial light element abundance~\cite{Kawasaki:2004qu,Jedamzik:2006xz}.  Demanding that $\dkp$ decay before this leads to a lower bound
\begin{equation}
	c_{dec} \gtrsim 10^{-11} \, (4\pi)^{n-1} \, \biggl( \frac{\Lambda}{m_\Psi} \biggr)^{\delta_{dec}} \,.
\label{eq:cdeclower}\end{equation}
This bound is extremely weak unless either $n$ or $\delta_{dec}$ are large.

This cosmological upper bound on the dark partner lifetime means that, for the purposes of indirect searches using cosmic rays, we can treat the decays as prompt.  The full SA process is $\dm_i \dm_j \to \dkp^\dagger V \to \dm_k^\dagger \op_{SM}^\dagger V$; in particular, if $\op_{SM} = V$ then we pair-produce it.  This is similar to the $2\to 3$ processes we discussed and neglected in \secref{sec:DMonly}.  However, since this is a $2\to 2$ process followed by a decay, we do not have a phase space suppression.  The visible states are also typically produced with larger energies, making them more distinguishable from the backgrounds.  When $\op_{SM} = V$, either visible particle can be more energetic depending on the spectrum; the $V$ produced from SA (decay) will be more energetic when $m_\dkp \approx m$ ($m_\dkp \approx 2m$).  When $\op_{SM} \neq V$, such that $\dkp$ decays to three or more particles, the SA-produced $V$ will tend to be the most energetic state.

As well as the $\dkp$ decay, \modeqref{eq:dkpdecay} will mediate additional annihilation and SA processes, as well as contributing to collider and direct detection signals.  These can potentially be important in the dark sector phenomenology; in particular, when $\op_{SM} = V$ then $\op_{dec}$ is a lower-dimension operator than $\op_{SA}$, which suggests it might be more relevant.  For simplicity, we will demand that processes mediated by $\op_{SA}$ dominate, and derive the resultant upper bounds on $c_{dec}$.  We first list generic bounds, then note specific operator-dependent constraints when we list the SA operators.

First, $\op_{dec}$ directly leads to co-annihilation $\dm_k \dkp^\dagger \to SM$.  This can be important for determining the relic density when $m_\dkp \lesssim 1.05 \, m$, but is irrelevant for ID signals today.  Even when the dark partner is sufficiently light, we can estimate that co-annihilation will be negligible if 
\begin{equation}
	c_{dec} \lesssim c_{SA} \, \biggl( \frac{m}{\Lambda} \biggr)^{\delta - \delta_{dec}} \max \bigl\{ (4\pi)^{n-2}, 1 \bigr\} \, .
\end{equation}
The strongest constraints come for $n = 1$ or 2, and $\delta - \delta_{dec} = 1$; then  $c_{dec} \lesssim c_{SA} \, m/\Lambda$.  For the regions of parameter space where we can set limits on SA (see \secref{sec:ID}), this is no stronger than $c_{dec} \lesssim \order{0.1}$.

\begin{figure}
	\centering
	\begin{tikzpicture}[node distance=0.35cm and 2cm]
		\coordinate (v1);
		\coordinate[above = of v1, label=$c_{dec}/\Lambda^{D_{dec}}$] (va);
		\coordinate[below = of v1] (v3);
		\coordinate[below = of v3, label=left:$\dkp$] (v4);
		\coordinate[below = of v4] (v5);
		\coordinate[below = of v5] (v2);
		\coordinate[below = of v2, label=below:$c_{dec}/\Lambda^{D_{dec}}$] (vb);
		\coordinate[above left = of v1, label=above left:$\dm_k$] (i1);
		\coordinate[below left = of v2, label=below left:$\dm_k^\dagger$] (i2);
		\coordinate[above right = of v1, label=above right:{$V$}] (o1);
		\coordinate[below right = of v2, label=below right:{$V^\dagger$}] (o2);
		\draw[fermionnoarrow] (i1) -- (v1) -- (v2) -- (i2);
		\draw[photon] (o2) -- (v2);
		\draw[photon] (v1) -- (o1);
		\draw[fill = white] (v1) circle (0.25);
		\fill[pattern = north west lines] (v1) circle (0.25);
		\draw[fill = white] (v2) circle (0.25);
		\fill[pattern = north west lines] (v2) circle (0.25);
	\end{tikzpicture}\qquad
	\begin{tikzpicture}[node distance=0.35cm and 0.5cm]
		\coordinate (v1);
		\coordinate[right = of v1] (v125);
		\coordinate[right = of v125, label=above:$\dkp$] (v15);
		\coordinate[right = of v15] (v175);
		\coordinate[right = of v175] (v2);
		\coordinate[below = of v1] (va);
		\coordinate[below = of va, label=left:$V$] (vb);
		\coordinate[below = of vb] (vc);
		\coordinate[below = of vc] (v3);
		\coordinate[right = of v3] (v325);
		\coordinate[right = of v325] (v35);
		\coordinate[right = of v35] (v375);
		\coordinate[right = of v375] (v4);
		\coordinate[below = of v3, label=below:$\phantom{c_{dec}/\Lambda^{D_{dec}}}$] (vp);
		\coordinate[left = of v1] (ia);
		\coordinate[left = of ia] (ib);
		\coordinate[above left = of ib, label=above left:$\dm_k$] (i1);
		\coordinate[left = of v3] (ic);
		\coordinate[left = of ic] (id);
		\coordinate[below left = of id, label=below left:$N$] (i2);
		\coordinate[right = of v2] (oa);
		\coordinate[right = of oa] (ob);
		\coordinate[above right = of ob, label=above right:$\dm_k$] (o1);
		\coordinate[right = of v4] (oc);
		\coordinate[right = of oc] (od);
		\coordinate[below right = of od, label=below right:$N$] (o2);
		\draw[fermionnoarrow] (i1) -- (v1) -- (v2) -- (o1);
		\draw[fermionnoarrow] (i2) -- (v3) -- (v4) -- (o2);
		\draw[photon] (v1) -- (v3);
		\draw[photon] (v2) -- (v4);
		\draw[fill = white] (v1) circle (0.25);
		\fill[pattern = north west lines] (v1) circle (0.25);
		\draw[fill = white] (v2) circle (0.25);
		\fill[pattern = north west lines] (v2) circle (0.25);
		\draw[fill = white] (v3) circle (0.25);
		\fill[pattern = north east lines] (v3) circle (0.25);
		\draw[fill = white] (v4) circle (0.25);
		\fill[pattern = north east lines] (v4) circle (0.25);
	\end{tikzpicture}
	\caption{DM annihilation (left) and inelastic scattering (right) created by the dark partner decay operator $\op_{dec}$.}\label{fig:AnnfromDec}
\end{figure}
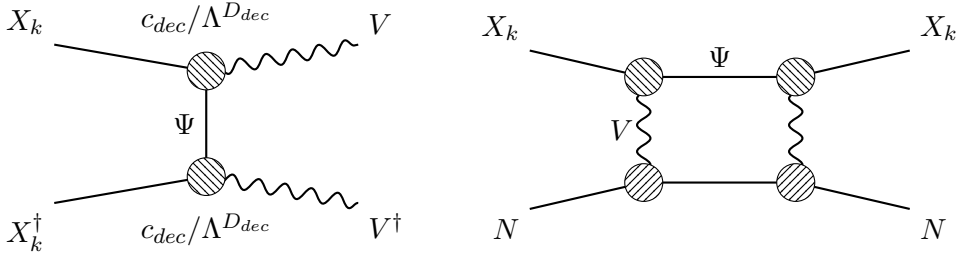

$\op_{dec}$ also contributes to ID signals today via the annihilation $\dm_k \dm_k^\dagger \to \op_{SM} \op_{SM}^\dagger$ through $t$-channel $\dkp$-exchange, see \figref{fig:AnnfromDec}.  Like the co-annihilation bounds, these apply for any $\op_{SM}$.  The cross section for this process will be smaller than that for $\dm_i\dm_j \to \dkp^\dagger\vis$ if
\begin{equation}
	c_{dec}^2 \lesssim (4\pi)^{2(n-1)} \, c_{SA} \, \frac{m_\dkp^2}{m^2} \, \biggl( \frac{\Lambda}{m} \biggr)^{2 \delta_{dec} - \delta} \,.
\label{eq:annfromSA}\end{equation}
Since we expect $\Lambda$, $m_\dkp > m$ this typically allows $c_{dec} > c_{SA}$.  The strongest bound comes for $n = 1$ when $\op_{SA}$ is a scalar quartic coupling and $\op_{dec}$ a scalar cubic.  In that case, we require the modest suppression $c_{dec} \lesssim \sqrt{c_{SA}} \, m_{\dkp}/\Lambda$, or equivalently that the cubic coupling be of comparable size to the dark partner mass.

A potentially strong constraint on $c_{dec}$ can come from direct detection searches, though this is dependent on the decay mode.  In particular, when $\op_{SM} = V$ then $\op_{dec}$ mediates elastic scattering at one-loop with one heavy mediator as shown in \figref{fig:AnnfromDec}, compared to two loops or one loop with two heavy mediators from $\op_{SA}$.  The cross sections and whether the scattering is spin-dependent or independent will depend on the mediator $V$, and so are operator-dependent.  We make a na\"\i ve dimensional estimate for the per-nucleon scattering cross section,
\begin{equation}
	\sigma_n \sim \frac{c_{dec}^4 g_{SM}^4}{(4\pi)^5 m^2} \, \biggl( \frac{\mu_n}{\Lambda} \biggr)^{2} \sim c_{dec}^4 g_{SM}^4 \times 10^{-43} \text{cm}^2 \, \biggl( \frac{100~\text{GeV}}{m} \biggr)^2 \, \biggl( \frac{1~\text{TeV}}{\Lambda} \biggr)^{2} \,,
\end{equation}
where $g_{SM}$ is the $V$-nucleon coupling and $\mu_n$ the reduced nucleon-DM mass.  Comparing to the LUX~2016 bound~\cite{1608.07648} of $\sigma_n \lesssim 10^{-46}$~cm$^2$ for $m = 100$~GeV, we see that even for $g_{SM} \sim 1$ we expect a bound of at most $c_{dec} \lesssim \order{0.1}$.

A further constraint on $c_{dec}$ comes from cubic pure-DM terms.  As noted, $\dkp$ and $\dm_k$ must have the same charges under $\dsym$.  Since we are focusing on gauge-singlet DM, $\dm_i \, \dm_j \, \dm_k$ is a total gauge and $\dsym$ singlet, and will represent a renormalisable source of SA when allowed by Lorentz invariance.  When all DM fields are scalars, we cannot have dark partner SA without also generating the cubic term of \modeqref{eq:phi3}.  Whether the couplings of \modeqref{eq:ffsterm} are allowed for fermion-scalar theories will depend on the form of $\op_{dec}$; however, they are allowed for all the decay operators we give in \secref{sec:MixDMDP}.   These terms can derive from the UV completion, but $\op_{SA}$ and $\op_{dec}$ together generate these cubics at one-loop (scalar) or two-loops (scalar and fermions).  In \secref{sec:DMonly} we found that the bounds on such one-loop terms are weaker than those from Higgs portal annihilation; this implies the weak constraint $c_{dec} \lesssim \order{1}$.

A related observation is that if  $\dm_i \, \dm_j \, \dm_k$ is a total gauge and $\dsym$ singlet, then the operators of \secref{sec:DMonly} are also allowed.  The dark partner operators can dominate through being lower dimension and/or producing charged or coloured states.  However, a simple UV completion that suppresses all the DM-only terms is when the heavy mediator $M$ connecting the two sectors couples to the SM as $M \dkp V^\dagger$.  The dark partner operators can then be generated at tree level, while the DM-only terms are loop-level.

Dark partners can also be relevant for collider phenomenology.  The SA operators we are interested can only lead to triple production of DM states; including an ISR jet or photon leads to a phase-space suppressed four-body final state.  Additionally, because of the large centre of mass energies accessible at \emph{e.g.} the LHC, it is quite possible that the EFT will break down, making it difficult to derive robust limits.  In contrast, dark partners can be pair-produced through (renormalisable) gauge couplings.  This leads to several possible signals, including jets plus missing energy, leptons plus missing energy, or charged tracks depending on the $\dkp$ decay modes and lifetime.  A full study of the resultant limits is beyond the scope of this work.  However, assuming collider-prompt decays, we might expect $m_\dkp \gtrsim 1$--2~TeV (200--500~GeV) for coloured (uncoloured) dark partners, based on simplified SUSY searches~\cite{1403.5294,1405.7570,ATLAS:2016kts,CMS-err,ATLAS-CONF-2016-096,ATLAS:2016tsc}.  Since $m > \frac{1}{2} \, m_\dkp$, this also amounts to lower bounds on the DM mass.

Finally, we briefly review a couple of issues that these models have in common with the pure DM models of \secref{sec:DMonly}.  We noted above that it is usually possible to construct EW symmetric phase operators that only lead to $2\to2$ processes (not including dark partner decays).  However, in enumerating all operators up to dimension 6, we will necessarily list some terms that also lead to $2\to3$ processes.  As before, the two-body final states will be dominant for at least $m \lesssim 4\pi v \approx 3$~TeV.  Similarly, we always have the Higgs portal operators of \modeqsref{eq:scaHP} and~\eqref{eq:ferHP}, with expected coefficient given by \modeqref{eq:HPcoeff}.  The requirement that annihilation mediated through these terms be subdominant to SA remain those of \modeqsref{eq:scaHPlimV}--\eqref{eq:ferHPlimNV}.  These conditions also ensure that SA through the renormalisable cubic terms of \modeqsref{eq:phi3} and~\eqref{eq:ffsterm} is subdominant, provided they are either generated at one-loop, or are tree level and
\begin{equation}
	\Lambda \lesssim 
	\begin{cases}
		(4\pi)^{2/(\delta+1)} m & \text{ (scalars)} \\
		(4\pi)^{2/\delta} m & \text{ (fermions).}
	\end{cases}
\end{equation}

\subsection{Scalar dark matter}\label{sec:SDMDP}

\begin{table}[p]
	\centering
	\begin{tabular}{|c|c|c|}
		\hline
		Operator & Definition & $\op_{dec}$ \\
		\hline
		\hline
		$\op_{4B}^{h\dps}$ & $ s^{ij} \, \dms_i \dms_j \, \dps h$ & $\dms_k^\dagger \, \dps h$ \\
		\hline
		$\op_{5B}^{\V\dps}$ & $s^{ij} \, \dms_i \dms_j (D_\mu \dps) \, \V^\mu$ & $\bigl( \dms_k^\dagger \Dlr{\mu} \dps \bigr) \V^\mu$ \\
		$\op_{5B}^{\V\dms}$ & $a^{ij} \, \dms_i (\partial_\mu \dms_j) \, \dps \, \V^\mu$ & $\bigl( \dms_k^\dagger \Dlr{\mu} \dps \bigr) \V^\mu$ \\
		\hline
		$\op_{6B}^{h\dps\partial}$ & $s^{ij} \, (\partial_\mu \dms_i) (\partial^\mu \dms_j) \, \dps \, h$ & $\dms_k^\dagger \, \dps h$ \\
		$\op_{6B}^{h\partial\dps}$ & $a^{ij} \, \dms_i \, (\partial^\mu \dms_j) \, \bigl( h \dlr{\mu} \dps)$ & $\dms_k^\dagger \, (\partial_\mu \dps) (\partial^\mu h)$ \\
		$\op_{6B}^{\partial h\dps}$ & $s^{ij} \, \dms_i \dms_j \, (\partial_\mu \dps) (\partial^\mu h)$ & $\dms_k^\dagger \, (\partial_\mu \dps) (\partial^\mu h)$ \\
		\hline
		\hline
		$\op_{5B}^{fL\dpf}$ & $s^{ij} \, \dms_i \dms_j \, \bar{f} P_R \dpf$ & $\dms_k^\dagger \, \bar{f} P_R \dpf$ \\
		$\op_{5B}^{fR\dpf}$ & $s^{ij} \, \dms_i \dms_j \, \bar{f} P_L \dpf$ & $\dms_k^\dagger \, \bar{f} P_L \dpf$ \\
		\hline
		$\op_{6B}^{fL\dpf}$ & $a^{ij} \dms_i (\partial_\mu \dms_j) \, \bar{f} \gamma^\mu P_L \dpf$ & $\dms_k^\dagger \, \bar{f} P_R \dpf$ \\
		$\op_{6B}^{fR\dpf}$ & $a^{ij} \dms_i (\partial_\mu \dms_j) \, \bar{f} \gamma^\mu P_R \dpf$ & $\dms_k^\dagger \, \bar{f} P_L \dpf$ \\
		\hline
	\end{tabular}
	\caption{Operators relevant for dark sectors comprised of scalar DM $\dms$ and additional light unstable dark partner scalars $\dps$ (top) and/or fermions $\dpf$ (bottom), in the EW broken phase.  $\V \in \{W, Z\}$ is a massive vector field, and if $\V = W$ then $\dps$ is electrically charged.  In the operators with fermion dark partners, $\dpf$ has the same gauge charges as $f$ where $f$ is any SM fermion.  We also list possible choices for the operator $\op_{dec}$ that mediates dark partner decays; see the text for more details.}\label{tab:SdmBro}
\end{table}

\begin{table}[p]
	\centering
	\begin{tabular}{|c|c|c|c|c|}
		\hline
		Operator & Definition & $\dps$/$\dpf$ & Broken Phase & $\op_{dec}$ \\
		\hline
		\hline
		$\op_{4U}^H$ & $s^{ij} \, \dms_i \dms_j \, (H^\dagger \dps)$ & (1, 2, $\frac{1}{2}$) & $\frac{1}{\sqrt{2}} \, \op_{4B}^{h\dps}$ & $\dms_k^\dagger \, (H^\dagger \dps)$ \\
		\hline
		$\op_{5U}^{|H|^2_1}$ & $s^{ij} \, \dms_i \dms_j \, \dps \, H^\dagger H$ & (1, 1, 0) & $\frac{v}{\Lambda} \, \op_{4B}^{h\dps}$ & $\dms_k^\dagger \, \dps \, H^\dagger H$ \\
		$\op_{5U}^{|H|^2_3}$ & $s^{ij} \, \dms_i \dms_j \, \dps^a \, H^\dagger \sigma^a H$ & (1, 3, 0) & $- \frac{v}{\Lambda} \, \op_{4B}^{h\dps}$ & $\dms_k^\dagger \, \dps^a \, H^\dagger \sigma^a H$ \\
		$\op_{5U}^{H^2}$ & $s^{ij} \, \dms_i \dms_j \, \dps^a \, H^\dagger \sigma^a \tilde{H}$ & (1, 3, 1) & $\frac{v\sqrt{2}}{\Lambda} \, \op_{4B}^{h\dps}$ & $\dms_k^\dagger \, \dps^a \, H^\dagger \sigma^a \tilde{H}$ \\
		\hline
		$\op_{6U}^{Hd}$ & $s^{ij} \, \dms_i \dms_j \, (H^\dagger \dps) (H^\dagger H)$ & (1, 2, $\frac{1}{2}$) & $\frac{3 v^2}{\sqrt{8}\Lambda^2} \, \op_{4B}^{h\dps}$ & $\dms_k^\dagger \, (H^\dagger \dps) (H^\dagger H)$ \\
		$\op_{6U}^{Hq}$ & $s^{ij} \, \dms_i \dms_j \, \dps^{IJK} H^\dagger_I H^\dagger_J \tilde{H}^\dagger_K$ & (1, 4, $\frac{1}{2}$) & $\frac{\sqrt{3} v^2}{\sqrt{8} \Lambda^2} \, \op_{4B}^{h\dps}$ & $\dms_k^\dagger \, \dps^{IJK} H^\dagger_I H^\dagger_J \tilde{H}^\dagger_K$ \\
		$\op_{6U}^{H^3}$ & $s^{ij} \, \dms_i \dms_j \, \dps^{IJK} H^\dagger_I H^\dagger_J H^\dagger_K$ & (1, 4, $\frac{3}{2}$) & $\frac{3 v^2}{\sqrt{8}\Lambda^2} \, \op_{4B}^{h\dps}$ & $\dms_k^\dagger \, \dps^{IJK} H^\dagger_I H^\dagger_J H^\dagger_K$ \\
		$\op_{6U}^{H \partial^2}$ & $s^{ij} \, (\partial_\mu \dms_i) (\partial^\mu \dms_j) (H^\dagger \dps)$ & (1, 2, $\frac{1}{2}$) & $\frac{1}{\sqrt{2}} \, \op_{6B}^{h\dps\partial}$ & $\dms_k^\dagger \, (H^\dagger \dps)$ \\
		$\op_{6U}^{H \partial D}$ & $a^{ij} \, \dms_i (\partial_\mu \dms_j) \, \bigl( H^\dagger \Dlr{\mu} \dps \bigr)$ & (1, 2, $\frac{1}{2}$) & See text & $\dms_k^\dagger \, (D^\mu H)^\dagger (D_\mu \dps)$ \\
		$\op_{6U}^{H D^2}$ & $s^{ij} \, \dms_i \dms_j \, (D^\mu H)^\dagger (D_\mu \dps)$ & (1, 2, $\frac{1}{2}$) & See text & $\dms_k^\dagger \, (D^\mu H)^\dagger (D_\mu \dps)$ \\
		\hline
		\hline
		$\overset{\phantom{\scriptscriptstyle{*}}}{\op}{}_{5U}^{\bar{f}\dpf}$ & $s^{ij} \, \dms_i \dms_j \, \bar{f} \dpfL$ & ($\bar{R}_{\bar{f}}$, 1, $- Y_{\bar{f}}$) & $\op_{5B}^{fR\dpf}$ & $\dms_k^\dagger \, \bar{f} \dpfL$ \\
		$\op_{5U}^{F\dpf}$ & $s^{ij} \, \dms_i \dms_j \, F^\dagger \dpfR$ & ($R_F$, 2, $Y_F$) & $\op_{5B}^{F_uL\dpf} + \op_{5B}^{F_dL\dpf}$ & $\dms_k^\dagger \, F^\dagger \dpfR$ \\
		\hline
		$\overset{\phantom{\scriptscriptstyle{*}}}{\op}{}_{6U}^{\bar{f}H\dpf}$ & $s^{ij} \, \dms_i \dms_j \, \bar{f} (\tilde{H}^\dagger \dpfL) $ & ($\bar{R}_{\bar{f}}$, 2, $- Y_{\bar{f}} - \frac{1}{2}$) & $\frac{v}{\sqrt{2}\Lambda} \, \op_{5B}^{fR\dpf}$ & $\dms_k^\dagger \, \bar{f} (\tilde{H}^\dagger \dpfL) $ \\
		$\op_{6U}^{\bar{f}H^\dagger\dpf}$ & $s^{ij} \, \dms_i \dms_j \, \bar{f} ( H^\dagger \dpfL)$ & ($\bar{R}_{\bar{f}}$, 2, $- Y_{\bar{f}} + \frac{1}{2}$) & $\frac{v}{\sqrt{2}\Lambda} \, \op_{5B}^{fR\dpf}$ & $\dms_k^\dagger \, \bar{f} ( H^\dagger \dpfL)$ \\
		$\op_{6U}^{FH\dpf_1}$ & $s^{ij} \, \dms_i \dms_j \, (F^\dagger H) \dpfR$ & ($R_F$, 1, $Y_F - \frac{1}{2}$) & $\frac{v}{\sqrt{2}\Lambda} \, \op_{5B}^{F_dL\dpf}$ & $\dms_k^\dagger \, (F^\dagger H) \dpfR$ \\
		$\op_{6U}^{FH^\dagger\dpf_1}$ & $s^{ij} \, \dms_i \dms_j \, (F^\dagger \tilde{H}) \dpfR$ & ($R_F$, 1, $Y_F + \frac{1}{2}$) & $\frac{v}{\sqrt{2}\Lambda} \, \op_{5B}^{F_uL\dpf}$ & $\dms_k^\dagger \, (F^\dagger \tilde{H}) \dpfR$ \\
		$\op_{6U}^{FH\dpf_3}$ & $s^{ij} \, \dms_i \dms_j \, (F^\dagger \sigma^a H) \bar{\upsilon}^{a\dagger}$ & ($R_F$, 3, $Y_F - \frac{1}{2}$) & See text & $\dms_k^\dagger \, (F^\dagger \sigma^a H) \bar{\upsilon}^{a\dagger}$ \\
		$\op_{6U}^{FH^\dagger\dpf_3}$ & $s^{ij} \, \dms_i \dms_j \, (F^\dagger \sigma^a \tilde{H}) \bar{\upsilon}^{a\dagger}$ & ($R_F$, 3, $Y_F + \frac{1}{2}$) & See text & $\dms_k^\dagger \, (F^\dagger \sigma^a \tilde{H}) \bar{\upsilon}^{a\dagger}$ \\
		$\op_{6U}^{\bar{f}\partial}$ & $a^{ij} \dms_i (\partial_\mu \dms_j) \, \bar{f} \sigma^\mu \dpfR$ & ($\bar{R}_{\bar{f}}$, 1, $- Y_{\bar{f}}$) & $\op_{6B}^{fR\dpf}$ & $\dms_k^\dagger \, \bar{f} \dpfL$ \\
		$\op_{6U}^{F\partial}$ & $a^{ij} \dms_i (\partial_\mu \dms_j) \, F^\dagger \bar{\dpfL}^\mu \dmfL$ & ($R_F$, 2, $Y_F$) & $\op_{6B}^{F_uL\dpf} + \op_{6B}^{F_dL\dpf}$ & $\dms_k^\dagger \, F^\dagger \dpfR$ \\
		\hline
	\end{tabular}
	\caption{Operators relevant for dark sectors comprised of scalar DM $\dms$ and additional light unstable dark partner scalars $\dps$ (top) and/or fermions $\dpf = (\dpfL, \dpfR)^T$ (bottom), in the EW symmetric phase.  $\bar{f}$ and $F = (F_u, F_d)$ are $SU(2)$-singlet and doublet fields respectively, with $SU(3)_C$ representations $R_{\bar{f}/F}$ and hypercharge $Y_{\bar{f}/F}$.  We also list possibilities for the dark partner decay operator $\op_{dec}$.
	}\label{tab:SdmUnb}
\end{table}

When scalar dark matter is accompanied by light dark partners, then summing over flavour there are 50 operators in the EW broken phase up to dimension 6, as shown in \tabref{tab:SdmBro}.  Specifically, there are four operators involving Higgses; two each involving $W$ and $Z$; and two classes of operators each for left- and right-handed SM fermions, with a multiplicity of 12 ($\{u, d, e, \nu\}$ times three generations) and 9 (no $\nu_R$) respectively.  All operators lead to $s$-wave $\dms\dms$-initiated processes.  When the DM state is unique, three of the Higgs and exactly half of the $W$, $Z$ and fermion operators survive, for a total of 26 terms.  

The operators involving $h$ and $Z$ are closely related to those found already in \secref{sec:scaDMonly}, except that we have replaced one DM field with an unstable scalar.  Likewise, the couplings to $\nu$ are similar to terms already found in \secref{sec:SFonly}.  The phenomenology of these terms differ due to the dark partner decays as discussed above.  More dramatic are the existence of couplings to all SM fields other than the photon and gluon, substantially expanding the range of possible SA processes.

In the EW unbroken phase, we find ten (72) operators with scalar (fermion) dark partners at dimension 6 and below.  We list these in the upper and lower parts respectively of \tabref{tab:SdmUnb}.  We also give their leading expressions after EWSB, except for the following four operators where the relation is too long to fit in the table:
\begin{align}
	\op_{6U}^{H\partial D} & = \frac{1}{\sqrt{2}} \, \op_{6B}^{h\partial\dps} - \frac{2 i m_W}{\Lambda} \, \op_{5B}^{W\dms} + \frac{i \sqrt{2} m_Z}{\Lambda} \, \op_{5B}^{Z\dms} \,, \\
	\op_{6U}^{HD^2} & = \frac{1}{\sqrt{2}} \, \op_{6B}^{\partial h\dps} + \frac{i m_W}{\Lambda} \, \op_{5B}^{W\dps} - \frac{i m_Z}{\sqrt{2}\Lambda} \, \op_{5B}^{Z\dps} \,, \\
	\op_{6U}^{FH\dpf_3} & = \frac{v}{\Lambda} \, \op_{5B}^{F_uL\dpf} - \frac{v}{\sqrt{2}\Lambda} \, \op_{5B}^{F_dL\dpf} \,, \\
	\op_{6U}^{FH^\dagger\dpf_3} & = \frac{v}{\sqrt{2}\Lambda} \, \op_{5B}^{F_uL\dpf} + \frac{v}{\Lambda} \, \op_{5B}^{F_dL\dpf} \,.
\end{align}
All terms in \tabref{tab:SdmBro} can be produced from the unbroken phase operators we list.  However, the broken phase dimension-5 $W/Z$ operators are only generated together with each other and dimension-6 Higgs operators.  Similarly, dimension-6 operators coupling to left-handed fermions come in isospin pairs ($u_L$ and $d_L$, or $e_L$ and $\nu_L$).  To generate all these terms uniquely, we need to go to dimension 7; we defer these operators till \appref{app:HigherD}.  When the DM is unique, 9 (57) of the operators with a scalar (fermion) dark partner survive.

In contrast to \secref{sec:scaDMonly}, almost all the EW broken phase operators can be derived from unbroken phase operators of the same dimension, with the only exceptions being the $W/Z$ operators.  As such $2\to 3$ processes are less of a concern.  However, this statement is dependent on the dark partner quantum numbers.  For example, the lowest dimension SA operator for a scalar triplet $\dps$ with zero hypercharge leads to $v/\Lambda$-suppressed $2\to 2$ and phase-space suppressed $2\to 3$ processes.  These cases lead to the familiar conclusion that $2\to 2$ SA will dominate for $m \lesssim 4\pi v \sim 3$~TeV.

As discussed above, in the minimal theory where the SA operators are the only non-gauge couplings between the two sectors, the top partners $\dps$ and $\dpf$ are stable.  As such, we have included in \tabsref{tab:SdmBro} and~\ref{tab:SdmUnb} possible operators $\op_{dec}$ that can mediate dark partner decay.  These choices are not unique, but are the minimal operators which, as far as possible, retain the Lorentz and gauge contractions of the dark partner and SM fields.  An alternative choice would be to construct the lowest-dimension operator containing $\dms_k^\dagger \dps$ or $\dms_k^\dagger\dpf$ which is consistent with SM gauge symmetries.  This latter approach can give lower dimension $\op_{dec}$, but it can also lead to the SA and decay processes producing different SM states.  Without knowledge of the UV completion, we restrict ourselves to dark partner--SM couplings that already exist in our models.

For the majority of SA operators, we construct $\op_{dec}$ through the simple replacement $\dms_i \dms_j \to \dms_k^\dagger$.  However, this approach fails in the presence of derivatives.  For the two $\dms$-derivative operators $\op_{6B}^{h\partial\dps}$ and $\op_{6U}^{H\partial^2}$, the replacement $(\partial_\mu \dms_i) (\partial^\mu \dms_j) \to \partial^2 \dms_k^\dagger$ has the same number of derivatives and does not modify the SM field Lorentz structure.  We then use the equations of motion to reduce $\op_{dec}$ to a dimension-3 operator.  Another problem arises from single $\dms$-derivative operators, where the natural replacement is $a^{ij} \, \dms_i \partial_\mu \dms_j\to\partial_\mu \dms_k^\dagger$.  For the dimension-6 fermionic operators $\{\op_{6B}^{fL/R\dpf}, \op_{6U}^{\bar{f}/F\partial}\}$, this leads to expressions that can be reduced using the equations of motion.  For the dimension-5 scalar operators $\op_{5B}^{\V\dps}$, we rewrite $\op_{dec}$ using $D_\mu \V^\mu = 0$.  Lastly, for the dimension-6 scalar operators $\{\op_{6B}^{h\partial\dps}, \op_{6U}^{H\partial D}\}$, replacing the DM fields produces a total derivative.  We choose instead the same $\op_{dec}$ as for $\{\op_{6B}^{\partial h \dps}, \op_{6U}^{HD^2}\}$, which feature the same fields and number of derivatives.

An additional consequence of $\op_{dec}$ for any SA operator featuring $H$ (and no derivatives) is that EWSB will induce mass mixing between $\dms$ and $\dps$:
\begin{equation}
	\op_{dec} \sim \frac{c_{dec}}{\Lambda^{\delta_{dec}}} \, \dms_k^\dagger \, \dps H^{\delta_{dec} + 2} \supset c_{dec} \, \biggl( \frac{v}{\Lambda} \biggr)^{\delta_{dec}} \, v^2 \, \dms_k^\dagger \dps^0 \,,
\end{equation}
where $\dps^0$ is the neutral component of the $\dps$ multiplet.  The most important phenomenological effect of this mixing is to create a DM--$Z$ coupling when $\dps$ has non-zero hypercharge.  The resultant elastic $\dms N \to \dms N$ scattering leads to upper bounds the mixing angle, $\sin\theta \lesssim \order{0.01}$~\cite{1609.06555}; and thus also the Wilson coefficient, 
\begin{equation}
	c_{dec} \lesssim 0.01 \, \biggl( \frac{\Lambda}{v} \biggr)^{\delta_{dec}} \frac{m_\dps^2 - m^2}{v^2} \sim 0.003 \, \biggl( \frac{m_\dps - m}{m} \biggr) \biggl( \frac{m}{100~\text{GeV}} \biggr)^2 \biggl( \frac{\Lambda}{v} \biggr)^{\delta_{dec}} \,.
\end{equation}
Since $m_\dps - m < m$, this is the strongest bound on $c_{dec}$ we have found, but still easily compatible with the lower bound of \modeqref{eq:cdeclower}.  The mass mixing will also modify the physical masses of the dark sector states; this results in a lower bound on the mass splitting,
\begin{gather}
	m_\dps - m \gtrsim \frac{c_{dec} v^2}{m} \, \biggl( \frac{v}{\Lambda} \biggr)^{\delta_{dec}} , \\
	c_{dec} \lesssim \frac{m (m_\dps - m)}{v^2} \biggl( \frac{\Lambda}{v} \biggr)^{\delta_{dec}} \sim 0.1 \, \biggl( \frac{m_\dps - m}{m} \biggr) \biggl(\frac{m}{100~\text{GeV}} \biggr)^2 \biggl( \frac{\Lambda}{v} \biggr)^{\delta_{dec}} .
\end{gather}
This bound is only important for $m_\dps - m \ll m$ and/or $\delta_{dec} = -1$

\subsection{Fermion dark matter}\label{sec:FDMDP}

\begin{table}
	\centering
	\begin{tabular}{|c|c|c|}
		\hline
		Operator & Definition & $\op_{dec}$ \\
		\hline
		\hline
		$\op_{5B}^{h\dps S}$ & $s^{ij} \, \bar{\dmf}_i^c \dmf_j \, \dps \, h$ & $\dps \, \bar{\dmf}_k \nu$ \\
		$\op_{5B}^{h\dps P}$ & $s^{ij} \, \bar{\dmf}_i^c \gamma^5 \dmf_j \, \dps \, h$ & $\dps \, \bar{\dmf}_k \nu$ \\
		$\op_{5B}^{\V\dps V}$ & $a^{ij} \, \bar{\dmf}_i^c \gamma^\mu \dmf_j \, \dps \, \V_\mu$ & $\dps \, \bar{\dmf}_k l$ \\
		$\op_{5B}^{\V\dps A}$ & $s^{ij} \, \bar{\dmf}_i^c \gamma^\mu \gamma^5 \dmf_j \, \dps \, \V_\mu $ & $\dps \, \bar{\dmf}_k l$ \\
		\hline
		$\brtilde{\op}{}_{6B}^{\V\dps}$ & $a^{ij} \, \bar{\dmf}^c_i \sigma^{\mu\nu} \dmf_j \, \dps \, \brtilde{\V}_{\mu\nu}$ & See text \\
		$\op_{6B}^{h\dps V}$ & $a^{ij} \, \bar{\dmf}_i^c \gamma^\mu \dmf_j \bigl( h \dlr{\mu} \dps \bigr)$ & $\dps \, \bar{\dmf}_k \nu$ \\
		$\op_{6B}^{h\dps A}$ & $s^{ij} \, \bar{\dmf}_i^c \gamma^\mu \gamma^5 \dmf_j \bigl( h \dlr{\mu} \dps \bigr)$ & $\dps \, \bar{\dmf}_k \nu$ \\
		$\op_{6B}^{\V\dps sS}$ & $s^{ij} \bigl( \dps \dlr{\mu} (\bar{\dmf}_i^c \dmf_j) \bigr) \V^\mu$ & $\bigl( \dps \dlr{\mu} (\bar{\dmf}_k \nu) \bigr) \V^\mu$ \\
		$\op_{6B}^{\V\dps sP}$ & $s^{ij} \bigl( \dps \dlr{\mu} (\bar{\dmf}_i^c \gamma^5 \dmf_j) \bigr) \V^\mu$ & $\bigl( \dps \dlr{\mu} (\bar{\dmf}_k \nu) \bigr) \V^\mu$ \\
		$\op_{6B}^{\V\dps aS}$ & $a^{ij} \, (\bar{\dmf}_i^c \partial_\mu \dmf_j) \, \dps \, \V^\mu$ & $\dps \, \V^\mu \bigl(\bar{\dmf}_k \dlr{\mu} \nu \bigr)$ \\
		$\op_{6B}^{\V\dps aP}$ & $a^{ij} \, (\bar{\dmf}_i^c \gamma^5 \partial_\mu \dmf_j) \, \dps \, \V^\mu$ & $\dps \, \V^\mu \bigl(\bar{\dmf}_k \dlr{\mu} \nu \bigr)$ \\
		\hline
		\hline
		$\op_{6B}^{f\dpf LR}$ & $s^{ij} \, (\bar{\dmf}_i^c P_L \dmf_j) \, (\bar{f} P_R \dpf)$ & $(\bar{\dmf}_k P_L f_2) \, (\bar{f}_1 P_R \dpf)$ \\
		$\op_{6B}^{f\dpf RR}$ & $s^{ij} \, (\bar{\dmf}_i^c P_R \dmf_j) \, (\bar{f} P_R \dpf)$ & $(\bar{\dmf}_k P_R f_2) \, (\bar{f}_1 P_R \dpf)$ \\
		$\op_{6B}^{f\dpf RL}$ & $s^{ij} \, (\bar{\dmf}_i^c P_R \dmf_j) \, (\bar{f} P_L \dpf)$ & $(\bar{\dmf}_k P_R f_2) \, (\bar{f}_1 P_L \dpf)$ \\
		$\op_{6B}^{f\dpf LL}$ & $s^{ij} \, (\bar{\dmf}_i^c P_L \dmf_j) \, (\bar{f} P_L \dpf)$ & $(\bar{\dmf}_k P_L f_2) \, (\bar{f}_1 P_L \dpf)$ \\
		$\op_{6B}^{f\dmf LR}$ & $n^{ij} \, (\bar{\dmf}_i^c P_L \dpf) \, (\bar{f} P_R \dmf_j)$ & $(\bar{\dmf}_k P_L \dpf) \, (\bar{f}_1 P_R f_2)$ \\
		$\op_{6B}^{f\dmf RR}$ & $n^{ij} \, (\bar{\dmf}_i^c P_R \dpf) \, (\bar{f} P_R \dmf_j)$ & $(\bar{\dmf}_k P_R \dpf) \, (\bar{f}_1 P_R f_2)$ \\
		$\op_{6B}^{f\dmf RL}$ & $n^{ij} \, (\bar{\dmf}_i^c P_R \dpf) \, (\bar{f} P_L \dmf_j)$ & $(\bar{\dmf}_k P_R \dpf) \, (\bar{f}_1 P_L f_2)$ \\
		$\op_{6B}^{f\dmf LL}$ & $n^{ij} \, (\bar{\dmf}_i^c P_L \dpf) \, (\bar{f} P_L \dmf_j)$ & $(\bar{\dmf}_k P_L \dpf) \, (\bar{f}_1 P_L f_2)$ \\
		\hline
	\end{tabular}
	\caption{Operators relevant for dark sectors comprised of fermion DM $\dmf$ and additional light unstable dark partner scalars $\dps$ (top) and/or fermions $\dpf$ (bottom), in the EW broken phase.  We also list possible (non-unique) choices for the operator $\op_{dec}$ that mediates $\dpf$ decays, where $l = e, \nu$ according to the charge of $\dps$.  See the text for details and definitions of $f_{1,2}$.}\label{tab:FdmBro}
\end{table}

\begin{table}
	\centering
	\begin{tabular}{|c|c|c|c|c|}
		\hline
		Operator & Definition & $\dps$/$\dpf$ & Broken Phase & $\op_{dec}$ \\
		\hline
		\hline
		$\op_{5U}^{H\dps S}$ & $s^{ij} \, \bar{\dmf}_i^c \dmf_j \, H^\dagger \dps$ & (1, 2, $\frac{1}{2}$) & $\frac{1}{\sqrt{2}} \, \op_{5B}^{h\dps S}$ & $H^\dagger \dps \, \bigl(\dmfRbar_k (\tilde{H}^\dagger \! L)\bigr)$ \\
		$\op_{5U}^{H\dps P}$ & $s^{ij} \, \bar{\dmf}_i^c \gamma^5 \dmf_j \, H^\dagger \dps$ & (1, 2, $\frac{1}{2}$) & $\frac{1}{\sqrt{2}} \, \op_{5B}^{h\dps P}$ & $H^\dagger \dps \, \bigl(\dmfRbar_k (\tilde{H}^\dagger \! L)\bigr)$ \\
		\hline
		$\op_{6U}^{|H|^2_1\dps S}$ & $s^{ij} \, \bar{\dmf}_i^c \dmf_j \, \dps \, H^\dagger H$ & (1, 1, 0) & $\frac{v}{\Lambda} \, \op_{5B}^{h\dps S}$ & $\dps \, H^\dagger H \, \bigl(\dmfRbar_k (\tilde{H}^\dagger \! L)\bigr)$ \\
		$\op_{6U}^{|H|^2_3\dps S}$ & $s^{ij} \, \bar{\dmf}_i^c \dmf_j \, \dps^a H^\dagger \sigma^a H$ & (1, 3, 0) & $\frac{v}{\Lambda} \, \op_{5B}^{h\dps S}$ & $\dps^a \, H^\dagger \sigma^a H \, \bigl(\dmfRbar_k (\tilde{H}^\dagger \! L)\bigr)$ \\
		$\op_{6U}^{H^2\dps S}$ & $s^{ij} \, \bar{\dmf}_i^c \dmf_j \, \dps^a \tilde{H}^\dagger \sigma^a H$ & (1, 3, 1) & $\frac{v\sqrt{2}}{\Lambda} \, \op_{5B}^{h\dps S}$ & $\dps^a \, \tilde{H}^\dagger \sigma^a H \, \bigl(\dmfRbar_k (\tilde{H}^\dagger \! L)\bigr)$ \\
		$\op_{6U}^{|H|^2_1\dps P}$ & $s^{ij} \, \bar{\dmf}_i^c\gamma^5  \dmf_j \, \dps \, H^\dagger H$ & (1, 1, 0) & $\frac{v}{\Lambda} \, \op_{5B}^{h\dps P}$ & $\dps \, H^\dagger H \, \bigl(\dmfRbar_k (\tilde{H}^\dagger \! L)\bigr)$ \\
		$\op_{6U}^{|H|^2_3\dps P}$ & $s^{ij} \, \bar{\dmf}_i^c \gamma^5 \dmf_j \, \dps^a H^\dagger \sigma^a H$ & (1, 3, 0) & $\frac{v}{\Lambda} \, \op_{5B}^{h\dps P}$ & $\dps^a \, H^\dagger \sigma^a H \, \bigl(\dmfRbar_k (\tilde{H}^\dagger \! L)\bigr)$ \\
		$\op_{6U}^{H^2\dps P}$ & $s^{ij} \, \bar{\dmf}_i^c \gamma^5 \dmf_j \, \dps^a \tilde{H}^\dagger \sigma^a H$ & (1, 3, 1) & $\frac{v\sqrt{2}}{\Lambda} \, \op_{5B}^{h\dps P}$ & $\dps^a \, \tilde{H}^\dagger \sigma^a H \, \bigl(\dmfRbar_k (\tilde{H}^\dagger \! L)\bigr)$ \\
		$\op_{6U}^{h\dps V}$ & $a^{ij} \bar{\dmf}_i^c \gamma^\mu \dmf_j \bigl( H^\dagger \Dlr{\mu} \dps \bigr)$ & (1, 2, $\frac{1}{2}$) & See Text & $\bigl( H^\dagger \Dlr{\mu} \dps \bigr) \bigl(\dmfL^\dagger_k \bar{\sigma}^\mu (\tilde{H}^\dagger \! L)\bigr)$ \\
		$\op_{6U}^{h\dps A}$ & $s^{ij} \bar{\dmf}_i^c \gamma^\mu \gamma^5 \dmf_j \bigl( H^\dagger \Dlr{\mu} \dps \bigr)$ & (1, 2, $\frac{1}{2}$) & See Text & $\bigl( H^\dagger \Dlr{\mu} \dps \bigr) \bigl(\dmfL^\dagger_k \bar{\sigma}^\mu (\tilde{H}^\dagger \! L)\bigr)$ \\
		$\brtilde{\op}_{6U}^{G\dps}$ & $a^{ij} \, \bar{\dmf}_i^c \sigma^{\mu\nu} \dmf_j \, \dps^A \, \brtilde{G}^A_{\mu\nu}$ & (8, 1, 0) & $\brtilde{\op}_{6B}^{G\dps}$ & See Text \\
		$\brtilde{\op}_{6U}^{W\dps}$ & $a^{ij} \, \bar{\dmf}_i^c \sigma^{\mu\nu} \dmf_j \, \dps^a \, \brtilde{W}^a_{\mu\nu}$ & (1, 3, 0) & See Text & $\dps^a \, \brtilde{W}^a_{\mu\nu}\bigl( \dmfRbar_k \sigma^{\mu\nu} (\tilde{H}^\dagger \! L) \bigr)$ \\
		$\brtilde{\op}_{6U}^{B\dps}$ & $a^{ij} \, \bar{\dmf}_i^c \sigma^{\mu\nu} \dmf_j \, \dps \, \brtilde{B}_{\mu\nu}$ & (1, 1, 0) & $c_W \, \brtilde{\op}_{6B}^{\gamma\dps} - s_W \, \brtilde{\op}_{6B}^{Z\dps}$ & $\dps^a \, \brtilde{B}^a_{\mu\nu} \bigl( \dmfRbar_k \sigma^{\mu\nu} (\tilde{H}^\dagger \! L) \bigr)$ \\
		\hline
		\hline
		$\overset{\phantom{\scriptscriptstyle{*}}}{\op}{}_{6U}^{\bar{f}\dpf L}$ & $s^{ij} \, (\dmfL_i \dmfL_j) \, (\bar{f} \dpfL)$ & ($\bar{R}_{\bar{f}}$, 1, $-Y_{\bar{f}}$) & $\op_{6B}^{f\dpf LL}$ & $(\dmfRbar_k \mathcal{F}_2) \, (\mathcal{F}_1 \dpfL)$ \\
		$\overset{\phantom{\scriptscriptstyle{*}}}{\op}{}_{6U}^{\bar{f}\dpf R}$ & $s^{ij}\, (\dmfR_i \dmfR_j) \, (\bar{f} \dpfL)$ & ($\bar{R}_{\bar{f}}$, 1, $-Y_{\bar{f}}$) & $\op_{6B}^{f\dpf RL}$ & $(\dmfL^\dagger_k \mathcal{F}_2^\dagger) \, (\mathcal{F}_1 \dpfL)$ \\
		$\overset{\phantom{\scriptscriptstyle{*}}}{\op}{}_{6U}^{F\dpf L}$ & $s^{ij}\, (\dmfL_i \dmfL_j) \, (F^\dagger \dpfR)$ & ($R_F$, 2, $Y_F$) & $\op_{6B}^{F_u\dpf LR} + \op_{6B}^{F_d\dpf LR}$ & $(\dmfRbar_k \mathcal{F}_2) \, (\mathcal{F}_1^\dagger \dpfR)$ \\
		$\overset{\phantom{\scriptscriptstyle{*}}}{\op}{}_{6U}^{F\dpf R}$ & $s^{ij} \, (\dmfR_i \dmfR_j) \, (F^\dagger \dpfR)$ & ($R_F$, 2, $Y_F$) & $\op_{6B}^{F_u\dpf RR} + \op_{6B}^{F_d\dpf RR}$ & $(\dmfL^\dagger_k \mathcal{F}_2^\dagger) \, (\mathcal{F}_1^\dagger \dpfR)$ \\
		$\overset{\phantom{\scriptscriptstyle{*}}}{\op}{}_{6U}^{\bar{f}\dmf L}$ & $n^{ij} \, (\dmfL_i \dpfL) \, (\bar{f} \dmfL_j)$ & ($\bar{R}_{\bar{f}}$, 1, $-Y_{\bar{f}}$) & $\op_{6B}^{f\dmf LL}$ & $(\dmfRbar_k \dpfL) \, (\mathcal{F}_1 \mathcal{F}_2)$ \\
		$\overset{\phantom{\scriptscriptstyle{*}}}{\op}{}_{6U}^{\bar{f}\dmf R}$ & $n^{ij} \, (\dmfR_i \dpfR) \, (\bar{f} \dmfL_j)$ & ($\bar{R}_{\bar{f}}$, 1, $-Y_{\bar{f}}$) & $\op_{6B}^{f\dmf RL}$ & $(\dmfL_k^\dagger \dpfR) \, (\mathcal{F}_1 \mathcal{F}_2)$ \\
		$\overset{\phantom{\scriptscriptstyle{*}}}{\op}{}_{6U}^{F\dmf L}$ & $n^{ij} \, (\dmfL_i \dpfL) \, (F^\dagger \bar{\xi}_j^\dagger)$ & ($R_F$, 2, $Y_F$) & $\op_{6B}^{F_u\dmf LR} + \op_{6B}^{F_d\dmf LR}$ & $(\dmfRbar_k \dpfL) \, (\mathcal{F}_1^\dagger \mathcal{F}_2^\dagger)$ \\
		$\overset{\phantom{\scriptscriptstyle{*}}}{\op}{}_{6U}^{F\dmf R}$ & $n^{ij} \, (\bar{\xi}_i^\dagger \dpfR) \, (F^\dagger \bar{\xi}_j^\dagger)$ & ($R_F$, 2, $Y_F$) & $\op_{6B}^{F_u\dmf RR} + \op_{6B}^{F_d\dmf RR}$ & $(\dmfL_k^\dagger \dpfR) \, (\mathcal{F}_1^\dagger \mathcal{F}_2^\dagger)$ \\
		\hline
	\end{tabular}
	\caption{Operators relevant for dark sectors comprised of fermion DM $\dmf = (\dmfL, \dmfR)^T$ and additional light unstable dark partner scalars $\dps$ (top) and $\dpf = (\dpfL, \dpfR)^T$ (bottom), in the EW unbroken phase.  $\bar{f}$ and $F = (F_u, F_d)$ are $SU(2)$-singlet and doublet fields respectively.  We also list possible choices for the  dark partner decay operators $\op_{dec}$; see the text for details and definitions of $\mathcal{F}_{1,2}$.}\label{tab:FdmUnb}
\end{table}

When fermion dark matter is accompanied by scalar (fermion) dark partners, we have a total of 22 (84) operators up to dimension 6 in the EW broken phase.  Specifically, for scalar dark partners we have four operators involving the Higgs; 6 each for $Z_\mu$ and $W_\mu$; and two for any field strength tensor.  For fermion dark partners we have eight classes of operators: four with left-handed SM fields that have flavour and generation multiplicity 12, and four with right-handed fields with multiplicity 9 (no $\nu_R$).  When the DM is unique, three of the Higgs, three of the $W$/$Z$, and all of the fermion operators survive.

The operators involving $\nu$ are closely related to those already discussed in \secref{sec:fdmonly}; we have simply replaced one DM field with an unstable neutral fermion.  Similarly, the couplings to scalar dark partners and $h$, $Z$ and $\gamma$ all have equivalents in the two-fermion operators from \secref{sec:SFonly}, where the scalar DM has been replaced by an unstable field.  Our phenomenology is again heavily expanded by the possibility of charged and coloured SM final states, as well as by dark partner decay.

In the EW unbroken phase, we find 13 (60) operators with scalar (fermion) dark partners at dimension 6 and below.  We list these in the upper and lower sections respectively of \tabref{tab:FdmUnb}.  We also give their connection to the broken phase operators, except in three cases where there is insufficient space:
\begin{align}
	\op_{6U}^{h\dps V} & = \frac{1}{\sqrt{2}} \, \op_{6B}^{h\dps V} + \frac{i m_W}{\Lambda} \, \op_{5B}^{W\dps V} - \frac{i m_Z}{\sqrt{2}\Lambda} \, \op_{5B}^{Z\dps V} \,, \\
	\op_{6U}^{h\dps A} & = \frac{1}{\sqrt{2}} \, \op_{6B}^{h\dps A} + \frac{i m_W}{\Lambda} \, \op_{5B}^{W\dps A} - \frac{i m_Z}{\sqrt{2}\Lambda} \, \op_{5B}^{Z\dps A} \,, \\
	\brtilde{\op}_{6U}^{W\dps} & = \brtilde{\op}_{6B}^{W\dps} + \brtilde{\op}_{6B}^{W^\dagger\dps}  + s_W \, \brtilde{\op}_{6B}^{\gamma\dps} + c_W \, \brtilde{\op}_{6B}^{Z\dps} \,.
\end{align}
In the last line, $\brtilde{\op}_{6B}^{W^\dagger\dps}$ is given by replacing $W_{\mu\nu} \to W^\dagger_{\mu\nu}$ in $\brtilde{\op}_{6B}^{W\dps}$.  All but four of the terms in \tabref{tab:FdmUnb} persist when the DM is unique.  We can generate most terms in the broken phase using these operators; however the dimension-6 $W/Z$ operators $\op_{6B}^{\V\dps(s/a)(S/P)}$ first appear at dimension 7 in the unbroken phase.  Since these operators are not the leading contributions to any SA process, we defer them to \appref{app:HigherD}.  We also defer the terms needed to generate individual couplings to left-handed fermions (which appear at dimension 7) and to electroweak gauge bosons (which appear at dimensions 7 and 8).  As in the scalar case, most operators can be generated at the same dimension in the broken and unbroken phases so $2\to 3$ processes are less of a concern.  

We must supplement our theories with operators that allow dark partner decay.  This is a more sever problem when there are no scalar DM particles: if the SA operator has the form $\dmf_i \dmf_j \, \dkp \op_{SM}$, then an operator of the form $\dmf^\dagger_k \dkp \op_{SM}$ will not be a Lorentz scalar.  This leads to an essential ambiguity in the dark partner decays, at least in the absence of a concrete UV model.  One possible resolution is to simply construct the lowest dimension operator consistent with gauge symmetries and $\dsym$ that will allow the dark partner to decay.  However, we will persist in constructing decay operators that resemble the associated SA terms as far as possible, in an attempt to remain agnostic about the UV completion.

In the EW broken phase, we can always construct a decay operator by the replacement $\dmf_i \dmf_j \to \dmf_k^\dagger \nu$.  However, this is not always the minimal or most logical approach.  For SA processes where the SM final state is $h$ or $Z_\mu$, we can construct lower-dimension operators without the bosonic field, \emph{i.e.} that lead to the two-body decay $\dps \to \dmf_k \nu$.  In particular, we always expect $h$ to appear in the full theory as $v+h$ so that the two-body decay is allowed.  For $Z_\mu$, the unbroken phase operators we discuss below lead to two-body decays for the dimension-5 operators, but not for the dimension-6 operators.  Similarly, when the SM final state is $W_\mu$, the unbroken phase can motivate the two-body decay $\dps \to \dmf_k e$ over the three-body decay $\dps \to W_\mu \dmf_k \nu$.  This covers all possibilities for scalar dark partners except the operators $\brtilde{\op}_{6B}^{\V\dps}$, where we did not list $\op_{dec}$ in \tabref{tab:FdmBro} for reasons of space.  This set of 8 operators includes four possible SM final states: $\gamma_{\mu\nu}$, $W_{\mu\nu}$, $Z_{\mu\nu}$, and $G^A_{\mu\nu}$.  For the first three, a three-body decay including a neutrino is the most natural choice:
\begin{equation}
	\op_{dec} = \dps \, \brtilde{\V}_{\mu\nu} \, \bar{\dmf}_k \sigma^{\mu\nu} \nu \quad \text{where } \V = \gamma, W, Z \text{ as appropriate.}
\end{equation}
For the last case, a decay to gluino plus neutrino is allowed but requires assuming that the UV completion has couplings to both coloured and colourless states.  However, there is no other three-body decay consistent with all symmetries.  The minimal decay to coloured states is $\dps \to \dmf u d d$, mediated by the dimension-7 operator
\begin{equation}
	\op_{dec} = \epsilon^{\alpha\beta\gamma} \, \dps^A \, \bigl(\bar{\dmf}_k (t^A d^c)_\alpha\bigr) \, \bigl(\bar{u}_\beta d^c_\gamma\bigr) \,.
\label{eq:decGlupartner}\end{equation}
Here $t^A$ is an $SU(3)$ generator, $\alpha, \beta, \gamma$ are $SU(3)$ fundamental indices, and $\epsilon^{\alpha\beta\gamma}$ is the fully antisymmetric tensor.  If we want to include the decay to $G_{\mu\nu}^A$, we must go to dimension 9.

Broken phase operators with fermion dark partners are generally simpler.  The only case where a two-body decay is possible is for the SA process $\dmf_i \dmf_j \to \bar{\dpf} \nu$; there, $\dpf \to \dmf_k h$ is allowed.  However, to be consistent with all operators in this class we consider only three-body decays to two fermions, as follows:
\begin{align}
	\text{SA Process: } & \dmf_i \dmf_j \to \bar{\dpf} \nu \,, & \text{Decay: } & \dpf \to \dmf_k \nu \bar{\nu} \,, \\
	\text{SA Process: } & \dmf_i \dmf_j \to \bar{\dpf} e \,, & \text{Decay: } & \dpf \to \dmf_k e \bar{\nu} \,, \\
	\text{SA Process: } & \dmf_i \dmf_j \to \bar{\dpf} u \,, & \text{Decay: } & \dpf \to \dmf_k \bar{d} \bar{d} \,, \\
	\text{SA Process: } & \dmf_i \dmf_j \to \bar{\dpf} d \,, & \text{Decay: } & \dpf \to \dmf_k \bar{u} \bar{d} \,.
\end{align}
The associated decay operators are listed \tabref{tab:FdmBro} with undefined fields $f_{1,2}$.  When the SA fermion $f = e, \nu$, then $f_1 = f$ and $f_2 = \nu$ or $\nu^c$, according to the projection operator that acts on it (recall that $\nu^c$ is a right-handed fermion).  When instead $f = u$ ($d$), then $f_1 = d^c$ ($u^c$) and $f_2 = d$ ($d$).

In the EW unbroken phase, similar results hold.  In \tabref{tab:FdmUnb}, for all but one scalar dark partner operators we make the replacement $\dmf_i \dmf_j \to \bar{\dmf}_k H L$.  Note that this can always lead to two-body decays by replacing all Higgses by their VEVs, motivating our choices in the broken phase.  The exception is $\brtilde{\op}_{6U}^{G\dps}$; as in the broken phase, there is no particularly good choice for the dark partner decay, but we can write down the analogue of \modeqref{eq:decGlupartner}:
\begin{equation}
	\op_{dec} = \epsilon^{\alpha\beta\gamma} \, \dps^A \, \bigl(\dmfRbar_k (t^A \bar{d})_\alpha\bigr) \, \bigl(\bar{u}_\beta \bar{d}_\gamma\bigr) \,.
\end{equation}
For fermion dark partners, we essentially reverse-engineer the broken phase results.  We give decay operators in \tabref{tab:FdmUnb} in terms of objects $\mathcal{F}_{1,2}$.  For leptonic processes, these are simply defined: $\mathcal{F}_1 = \bar{e}/L$ is the same field that appears in the SA operator, and $\mathcal{F}_2 = \tilde{H}^\dagger L$.  For hadronic processes, $\mathcal{F}_1$ is determined solely by the visible sector final state: operators of the form $\op_{6U}^{\bar{u}\cdots}$ ($\op_{6U}^{\bar{d}\cdots}$, $\op_{6U}^{Q\cdots}$) have $\mathcal{F}_1 = H^\dagger Q$ ($\tilde{H}^\dagger Q$, $\bar{u} H + \bar{d} \tilde{H}$).  $\mathcal{F}_2$ takes one of two values: $\mathcal{F}_2 = \bar{d}$ ($H^\dagger Q$) for operators of the form $\op_{6U}^{\cdots \dpf R}$ or $\op_{6U}^{Q\dmf\cdots}$ ($\op_{6U}^{\cdots \dpf L}$ or $\op_{6U}^{\bar{u}/\bar{d}\dmf\cdots}$).

\subsection{Scalar and Fermion dark matter}\label{sec:MixDMDP}

\begin{table}[p]
	\centering
	\begin{tabular}{|c|c|c||c|c|c|}
		\hline
		Operator & Definition & $\op_{dec}$ & Operator & Definition & $\op_{dec}$ \\
		\hline
		$\op_{5B}^{fL\dps}$ & $\dms \, \dps \, \bar{f} P_R \dmf$ & $\dps \, \bar{f} P_R \dmf^c_k$ & $\op_{6B}^{fL\dps}$ & $\bigl( \dms \Dlr{\mu} \dps \bigr) \, \bar{f} \gamma^\mu P_L \dmf$ & $\dps \, \bar{f} P_R \dmf^c_k$ \\
		$\op_{5B}^{fR\dps}$ & $\dms \, \dps \, \bar{f} P_L \dmf$ & $\dps \, \bar{f} P_L \dmf^c_k$ & $\op_{6B}^{fR\dps}$ & $\bigl( \dms \Dlr{\mu} \dps \bigr) \, \bar{f} \gamma^\mu P_R \dmf$ & $\dps \, \bar{f} P_L \dmf^c_k$ \\
		\hline
		\hline
		$\op_{5B}^{h\dpf S}$ & $\bar{\dmf}^c \dpf \, \dms h$ & $\bar{\dmf}_k \dpf \, h$ & $\op_{6B}^{h\dpf V}$ & $\bar{\dmf}^c \gamma^\mu \dpf \, \bigl( \dms \dlr{\mu} h\bigr) $ & $\bar{\dmf}_k \dpf \, h$ \\
		$\op_{5B}^{h\dpf P}$ & $\bar{\dmf}^c \gamma^5 \dpf \, \dms h$ & $\bar{\dmf}_k \gamma^5 \dpf \, h$ & $\op_{6B}^{h\dpf A}$ & $\bar{\dmf}^c \gamma^\mu \gamma^5 \dpf \, \bigl( \dms \dlr{\mu} h\bigr)$ & $\bar{\dmf}_k \gamma^5 \dpf \, h$ \\
		$\op_{5B}^{\V\dpf V}$ & $\bar{\dmf}^c \gamma^\mu \dpf \, \dms \V_\mu$ & $\bar{\dmf}_k \gamma^\mu \dpf \, \V_\mu$ & $\op_{6B}^{\V\dpf sS}$ & $\bigl( \dms \Dlr{\mu} (\bar{\dmf}^c \dpf) \bigr) \V^\mu$ & $\bigl( \bar{\dmf}_k \Dlr{\mu} \dpf \bigr) \, \V^\mu$ \\
		$\op_{5B}^{\V\dpf A}$ & $\bar{\dmf}^c \gamma^\mu \gamma^5 \dpf \, \dms \V_\mu$ & $\bar{\dmf}_k \gamma^\mu \gamma^5 \dpf \, \V_\mu$ & $\op_{6B}^{\V\dpf sP}$ & $\bigl( \dms \Dlr{\mu} (\bar{\dmf}^c \gamma^5 \dpf) \bigr) \V^\mu$ & $\bigl( \bar{\dmf}_k \gamma^5 \Dlr{\mu} \dpf \bigr) \, \V^\mu$ \\
		$\op_{6B}^{\V\dpf}$ & $\bar{\dmf}^c \sigma^{\mu\nu} \dpf \, \dms \V_{\mu\nu}$ & $\bar{\dmf}_k \sigma^{\mu\nu} \dpf \, \V_{\mu\nu}$ & $\op_{6B}^{\V\dpf aS}$ & $\bigl( \bar{\dmf}^c \Dlr{\mu} \dpf \bigr) \, \dms \, \V^\mu$ & $\bigl( \bar{\dmf}_k \Dlr{\mu} \dpf \bigr) \, \V^\mu$ \\
		$\tilde{\op}_{6B}^{\V\dpf}$ & $\bar{\dmf}^c \sigma^{\mu\nu} \dpf \, \dms \tilde{\V}_{\mu\nu}$ & $\bar{\dmf}_k \sigma^{\mu\nu} \dpf \, \tilde{\V}_{\mu\nu}$ & $\op_{6B}^{\V\dpf aP}$ & $\bigl( \bar{\dmf}^c \gamma^5 \Dlr{\mu} \dpf \bigr) \, \dms \, \V^\mu$ & $\bigl( \bar{\dmf}_k \gamma^5 \Dlr{\mu} \dpf \bigr) \, \V^\mu$ \\
		\hline
	\end{tabular}
	\caption{Operators relevant for dark sectors comprised of scalar DM $\dms$, fermion DM $\dmf$, and additional light unstable dark partner scalars $\dps$ (top) and/or fermions $\dpf$ (bottom), in the EW broken phase.}\label{tab:MixdmBro}
\end{table}

\begin{table}[p]
	\centering
	\begin{tabular}{|c|c|c|c|c|}
		\hline
		Operator & Definition & $\dps$ & Broken Phase & $\op_{dec}$ \\
		\hline
		\hline
		$\overset{\phantom{\scriptscriptstyle{*}}}{\op}{}_{5U}^{\bar{f}\dps}$ & $\dms \, \dps \, \bar{f} \dmfL$ & ($\bar{R}_{\bar{f}}$, 1, $-Y_{\bar{f}}$) & $\op_{5B}^{fR\dps}$ & $\dps \, \bar{f} \dmfRbar_k$ \\
		$\overset{\phantom{\scriptscriptstyle{*}}}{\op}{}_{5U}^{F\dps}$ & $\dms \, \dps \, F^\dagger \dmfR$ & ($R_F$, 2, $Y_F$) & $\op_{5B}^{F_uL\dps} + \op_{5B}^{F_dL\dps}$ & $\dps \, F^\dagger \dmfL^\dagger_k$ \\
		\hline
		$\overset{\phantom{\scriptscriptstyle{*}}}{\op}{}_{6U}^{\bar{f}H\dps}$ & $\dms \, (\tilde{H}^\dagger \dps) \, \bar{f} \dmfL$ & ($\bar{R}_{\bar{f}}$, 2, $- \frac{1}{2} - Y_{\bar{f}}$) & $\frac{v}{\sqrt{2}\Lambda} \, \op_{5B}^{fR\dps}$ & $(\tilde{H}^\dagger \dps) \, \bar{f} \dmfRbar_k$ \\
		$\overset{\phantom{\scriptscriptstyle{*}}}{\op}{}_{6U}^{\bar{f}H^\dagger\dps}$ & $\dms \, (H^\dagger \dps) \, \bar{f} \dmfL$ & ($\bar{R}_{\bar{f}}$, 2, $\frac{1}{2} - Y_{\bar{f}}$) & $\frac{v}{\sqrt{2}\Lambda} \, \op_{5B}^{fR\dps}$ & $(H^\dagger \dps) \, \bar{f} \dmfRbar_k$ \\
		$\op_{6U}^{FH^\dagger\dps_1}$ & $\dms \, \dps \, \bigl( (F^\dagger \tilde{H}) \dmfR \bigr)$ & ($R_F$, 1, $Y_F + \frac{1}{2}$) & $\frac{v}{\sqrt{2}\Lambda} \, \op_{5B}^{F_uL\dps}$ & $\dps \, \bigl( (F^\dagger \tilde{H}) \dmfL^\dagger_k \bigr)$ \\
		$\op_{6U}^{FH\dps_1}$ & $\dms \, \dps \, \bigl( (F^\dagger H) \dmfR \bigr)$ & ($R_F$, 1, $Y_F - \frac{1}{2}$) & $\frac{v}{\sqrt{2}\Lambda} \, \op_{5B}^{F_dL\dps}$ & $\dps \, \bigl( (F^\dagger H) \dmfL^\dagger_k \bigr)$ \\
		$\op_{6U}^{FH^\dagger\dps_3}$ & $\dms \, \dps^a \, \bigl( (F^\dagger \sigma^a \tilde{H}) \dmfR \bigr)$ & ($R_F$, 3, $Y_F + \frac{1}{2}$) & $\frac{v}{\sqrt{2}\Lambda} \, \op_{5B}^{F_dL\dps}$ & $\dps^a \, \bigl( (F^\dagger \sigma^a \tilde{H}) \dmfL^\dagger_k \bigr)$ \\
		$\op_{6U}^{FH\dps_3}$ & $\dms \, \dps^a \, \bigl( (F^\dagger \sigma^a H) \dmfR \bigr)$ & ($R_F$, 3, $Y_F - \frac{1}{2}$) & $\frac{v}{\sqrt{2}\Lambda} \, \op_{5B}^{F_dL\dps}$ & $\dps^a \, \bigl( (F^\dagger \sigma^a H) \dmfL^\dagger_k \bigr)$ \\
		$\op_{6U}^{\bar{f}D}$ & $\bigl( \dms \Dlr{\mu} \dps \bigr) \, \bar{f} \sigma^\mu \dmfR$ & ($\bar{R}_{\bar{f}}$, 1, $-Y_{\bar{f}}$) & $\op_{6B}^{fR\dps}$ & $\dps \, \bar{f} \dmfRbar_k$ \\
		$\op_{6U}^{FD}$ & $\bigl( \dms \Dlr{\mu} \dps \bigr) \, F^\dagger \bar{\sigma}^\mu \dmfL$ & ($R_F$, 2, $Y_F$) & $\op_{6B}^{F_uL\dps} + \op_{6B}^{F_dL\dps}$ & $\dps \, F^\dagger \dmfL^\dagger_k$ \\
		\hline
		\hline
		$\op_{5U}^{H\dpf S}$ & $\bigl(\bar{\dmf}^c (H^\dagger \dpf) \bigr) \, \dms$ & (1, 2, $\frac{1}{2}$) & $\frac{1}{\sqrt{2}} \, \op_{5B}^{h\dpf S}$ & $\bar{\dmf}_k (H^\dagger \dpf)$ \\
		$\op_{5U}^{H\dpf P}$ & $\bigl(\bar{\dmf}^c \gamma^5 (H^\dagger \dpf) \bigr) \, \dms$ & (1, 2, $\frac{1}{2}$) & $\frac{1}{\sqrt{2}} \, \op_{5B}^{h\dpf P}$ & $\bar{\dmf}_k \gamma^5 (H^\dagger \dpf)$ \\
		\hline
		$\op_{6U}^{|H|^2_1\dpf S}$ & $\bar{\dmf}^c \dpf \, \dms \, H^\dagger H$ & (1, 1, 0) & $\frac{v}{\Lambda} \, \op_{5B}^{h\dpf S}$ & $\bar{\dmf}_k \dpf \, H^\dagger H$ \\
		$\op_{6U}^{|H|^2_3\dpf S}$ & $\bar{\dmf}^c \dpf^a \, \dms \, H^\dagger \sigma^a H$ & (1, 3, 0) & $\frac{v}{\Lambda} \, \op_{5B}^{h\dpf S}$ & $\bar{\dmf}_k \dpf^a \, H^\dagger \sigma^a H$ \\
		$\op_{6U}^{H^2\dpf S}$ & $\bar{\dmf}^c \dpf^a \, \dms \, H^\dagger \sigma^a \tilde{H}$ & (1, 3, 1) & $\frac{v\sqrt{2}}{\Lambda} \, \op_{5B}^{h\dpf S}$ & $\bar{\dmf}_k \dpf^a \, H^\dagger \sigma^a \tilde{H}$ \\
		$\op_{6U}^{|H|^2_1\dpf P}$ & $\bar{\dmf}^c \gamma^5 \dpf \, \dms \, H^\dagger H$ & (1, 1, 0) & $\frac{v}{\Lambda} \, \op_{5B}^{h\dpf P}$ & $\bar{\dmf}_k \gamma^5 \dpf \, H^\dagger H$ \\
		$\op_{6U}^{|H|^2_3\dpf P}$ & $\bar{\dmf}^c \gamma^5 \dpf^a \, \dms \, H^\dagger \sigma^a H$ & (1, 3, 0) & $\frac{v}{\Lambda} \, \op_{5B}^{h\dpf P}$ & $\bar{\dmf}_k \gamma^5 \dpf^a \, H^\dagger \sigma^a H$ \\
		$\op_{6U}^{H^2\dpf P}$ & $\bar{\dmf}^c \gamma^5 \dpf^a \, \dms \, H^\dagger \sigma^a \tilde{H}$ & (1, 3, 1) & $\frac{v\sqrt{2}}{\Lambda} \, \op_{5B}^{h\dpf P}$ & $\bar{\dmf}_k \gamma^5 \dpf^a \, H^\dagger \sigma^a \tilde{H}$ \\
		$\op_{6U}^{h\dpf V}$ & $\bar{\dmf}^c \gamma^\mu \dpf \, \bigl( H^\dagger \Dlr{\mu} \dms \bigr)$ & (1, 2, $\frac{1}{2}$) & $\frac{1}{\sqrt{2}} \, \op_{6B}^{h\dpf V}$ & $\bar{\dmf}_k (H^\dagger \dpf)$ \\
		$\op_{6U}^{h\dpf A}$ & $\bar{\dmf}^c \gamma^\mu \gamma^5 \dpf \, \bigl( H^\dagger \Dlr{\mu} \dms \bigr)$ & (1, 2, $\frac{1}{2}$) & $\frac{1}{\sqrt{2}} \, \op_{6B}^{h\dpf A}$ & $\bar{\dmf}_k \gamma^5 (H^\dagger \dpf)$ \\		
		$\brtilde{\op}_{6U}^{G\dpf}$ & $\bar{\dmf}^c \sigma^{\mu\nu} \dpf^A \, \dms \, \brtilde{G}^A_{\mu\nu}$ & (8, 1, 0) & $\brtilde{\op}_{6B}^{G\dpf}$ & $\bar{\dmf}_k \sigma^{\mu\nu} \dpf^A \, \brtilde{G}^A_{\mu\nu}$ \\
		$\brtilde{\op}_{6U}^{W\dpf}$ & $\bar{\dmf}^c \sigma^{\mu\nu} \dpf^a \, \dms \, \brtilde{W}^a_{\mu\nu}$ & (1, 3, 0) & See Text & $\bar{\dmf}_k \sigma^{\mu\nu} \dpf^a \, \brtilde{W}^a_{\mu\nu}$ \\
		$\brtilde{\op}_{6U}^{B\dpf}$ & $\bar{\dmf}^c \sigma^{\mu\nu} \dpf \, \dms \, \brtilde{B}_{\mu\nu}$ & (1, 1, 0) & $c_W \, \brtilde{\op}_{6B}^{\gamma\dpf} - s_W \, \brtilde{\op}_{6B}^{Z\dpf}$ & $\bar{\dmf}_k \sigma^{\mu\nu} \dpf \, \brtilde{B}_{\mu\nu}$ \\
		\hline
	\end{tabular}
	\caption{Operators relevant for dark sectors comprised of scalar DM $\dms$, fermion DM $\dmf = (\dmfL, \dmfR)^T$, and additional light unstable dark partner scalars $\dps$ (bottom) and fermions $\dpf = (\dpfL, \dpfR)^T$ (top), in the EW unbroken phase.  $\bar{f}$ and $F = (F_u, F_d)$ are $SU(2)$-singlet and doublet fields respectively, with $SU(3)_C$ representations $R_{\bar{f}/F}$ and hypercharge $Y_{\bar{f}/F}$.}\label{tab:MixdmUnb}
\end{table}

When there are both scalar and fermion DM fields, then in the EW broken phase we have the 66 additional possible operators listed in \tabref{tab:MixdmBro}.  Specifically, we have four operators involving Higgses; 6 each for $W_\mu/Z_\mu$; 2 each for any of the four field strength tensors; and two types of operators each for right- and left-handed fermions, with multiplicity 12 and 9 respectively.  For appropriate $\dsym$ representations, we can prevent the scalar- and fermion-only operators of the previous two subsections, so that these are the dominant contributions to SA processes.  The operators involving neutral SM fields are all related to those of \secref{sec:SFonly}, but again the dark partners open the possibility of charged and coloured final states.

In the EW unbroken phase, we find 72 (16) operators with scalar (fermion) dark partners, which we list in \tabsref{tab:MixdmUnb}.  We also list their connection to the broken phase operators, with one exception where there is insufficient space:
\begin{equation}
	\brtilde{\op}_{6U}^{W\dpf} = \brtilde{\op}_{6B}^{W\dpf} + \brtilde{\op}_{6B}^{W^\dagger\dpf} + s_W \, \brtilde{\op}_{6B}^{\gamma\dpf} + c_W \, \brtilde{\op}_{6B}^{Z\dpf} \,.
\end{equation}
Here, $\brtilde{\op}_{6B}^{W^\dagger\dpf}$ is the same as $\brtilde{\op}_{6B}^{W\dpf}$ under the replacement $W_{\mu\nu} \to W_{\mu\nu}^\dagger$.  We can generate all broken phase terms except for the subleading $\op_{6B}^{\V\dpf(s/a)(S/P)}$; these terms appear at dimension 7, which we defer to \appref{app:HigherD}.  We also defer operators necessary to generate all the left-handed fermion and gauge boson operators individually.

We include in \tabsref{tab:MixdmBro} and~\ref{tab:MixdmUnb} possible decay operators $\op_{dec}$ for the dark partners.  We construct these in the minimal way, namely replacing $\dms \dmf \to \bar{\dmf}_k$, where $\dmf_k$ and $\dmf$ may be the same or distinct fermions.  For operators involving the derivative of $\dms$, we are left with a derivative we can integrate by parts and then reduce with equations of motion.  The only exceptions are the operators $\op_{6B}^{\V\dpf s(S/P)}$, where doing this would lead to operators that vanish on-shell by $D_\mu \V^\mu = 0$.  Instead, we use the decay operators for the similar SA operators $\op_{6B}^{\V\dmf a(S/P)}$.

For SA operators with Higgs final states, the $\op_{dec}$ constructed this way will lead to mass mixing between $\dmf$ and the neutral component of $\dpf$:
\begin{equation}
	\op_{dec} \sim \frac{c_{dec}}{\Lambda^{\delta_{dec}}} \, \bar{\dmf} \dpf \, H^{\delta_{dec} + 1} \supset c_{dec} \biggl( \frac{v}{\Lambda} \biggr)^{\delta_{dec}} v \, \bar{\dmf} \dpf \,.
\end{equation}
As before, the mass mixing is most important when $\dpf$ has non-zero hypercharge, as it allows $Z$-mediated elastic nuclear scattering.  From the mixing angle bound $\sin\theta \lesssim \order{0.01}$~\cite{1609.06555}, we derive 
\begin{equation}
	c_{dec} \lesssim 0.01 \, \biggl( \frac{\Lambda}{v} \biggr)^{\delta_{dec}} \frac{m_\psi - m}{v} \sim 0.004 \biggl( \frac{m_\psi - m}{m} \biggr) \, \biggl( \frac{m}{100~\text{GeV}} \biggr) \, \biggl( \frac{\Lambda}{v} \biggr)^{\delta_{dec}} .
\end{equation}
For small $\delta_{dec}$ this is the strongest upper bound we have on $c_{dec}$, but still easily compatible with the lower bound of \modeqref{eq:cdeclower}.

\section{Constraints from Indirect Detection and Astrophysical Obeservation}\label{sec:ID}
Dark matter indirect detection experiments observe cosmic rays, such as gamma rays, neutrinos, positrons and anti-protons,
from dense environments.   
Excesses in cosmic ray fluxes over the presumed astrophysical backgrounds can be interpreted as
the result of dark matter (semi)-annihilation or decay.
Non-observation of such an excess will instead impose constraints on the parameter space of 
various dark matter models.    
 
Among all the cosmic rays, neutrinos and gamma rays travel through space almost undisturbed 
and the cosmic ray flux can be simply expressed 
\begin{align}
\frac{d\Phi}{d E} = \frac{1}{16\pi}\frac{\langle \sigma v\rangle}{ m_{\rm DM}^2}\frac{d N}{ d E} \; J \; ,
\end{align}
where $\langle \sigma v\rangle$ is the thermally averaged (semi)-annihilation cross section, 
$m_{\rm DM}$ 
is the DM mass, and $\frac{dN}{dE}$ is the differential flux per (semi)-annihilation. 
We use \texttt{PYTHIA 8}~\cite{0710.3820} to generate the differential flux at production.   
For the discussion of DM indirect detection, it suffices to include only the $s$-wave contribution of $\langle \sigma v\rangle$
and neglect the higher order corrections, such as the $p$-wave contribution, which can play 
an important role in the DM relic density calculation.
$J$ represents the astrophysical factor
\begin{align}
J=\int_{\Delta \Omega} \int_{l=0}^\infty dl \; d\Omega \; \rho^2(l),
\end{align} 
where $\rho$ is the DM density along the line of sight, and $\Delta\Omega$ is the solid angle integrated over in the observation.

For charged cosmic rays, the flux can be affected by diffusive reacceleration, and also suffer from energy loss during propagation as well as solar modulation.
\texttt{PYTHIA 8} is again used to generate flux at production, while the effects of propagation to the observer is computed with \texttt{DRAGON}~\cite{0807.4730}.

\begin{figure}
\centering
\includegraphics[width=0.44\textwidth]{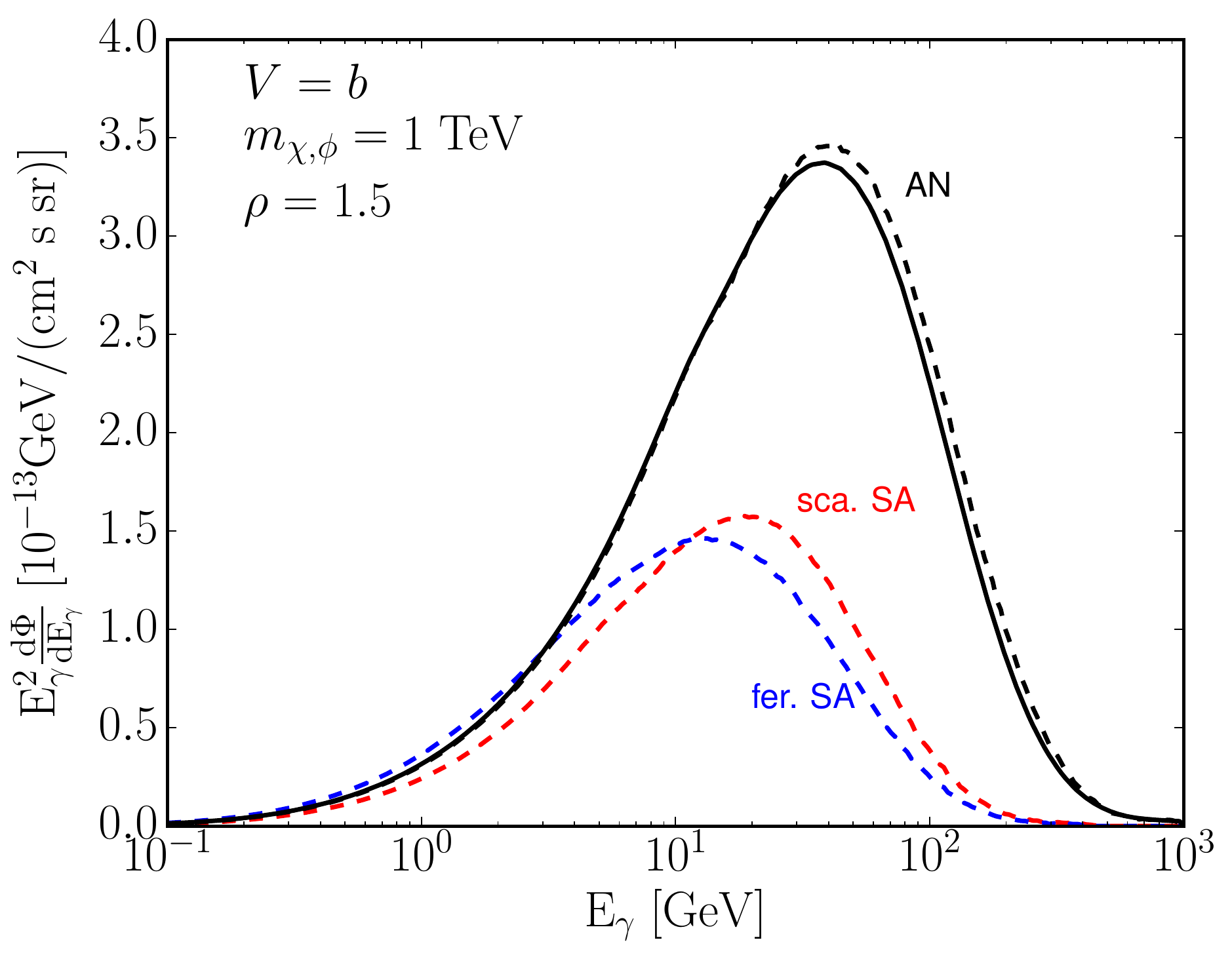}
\hfill
\includegraphics[width=0.51\textwidth]{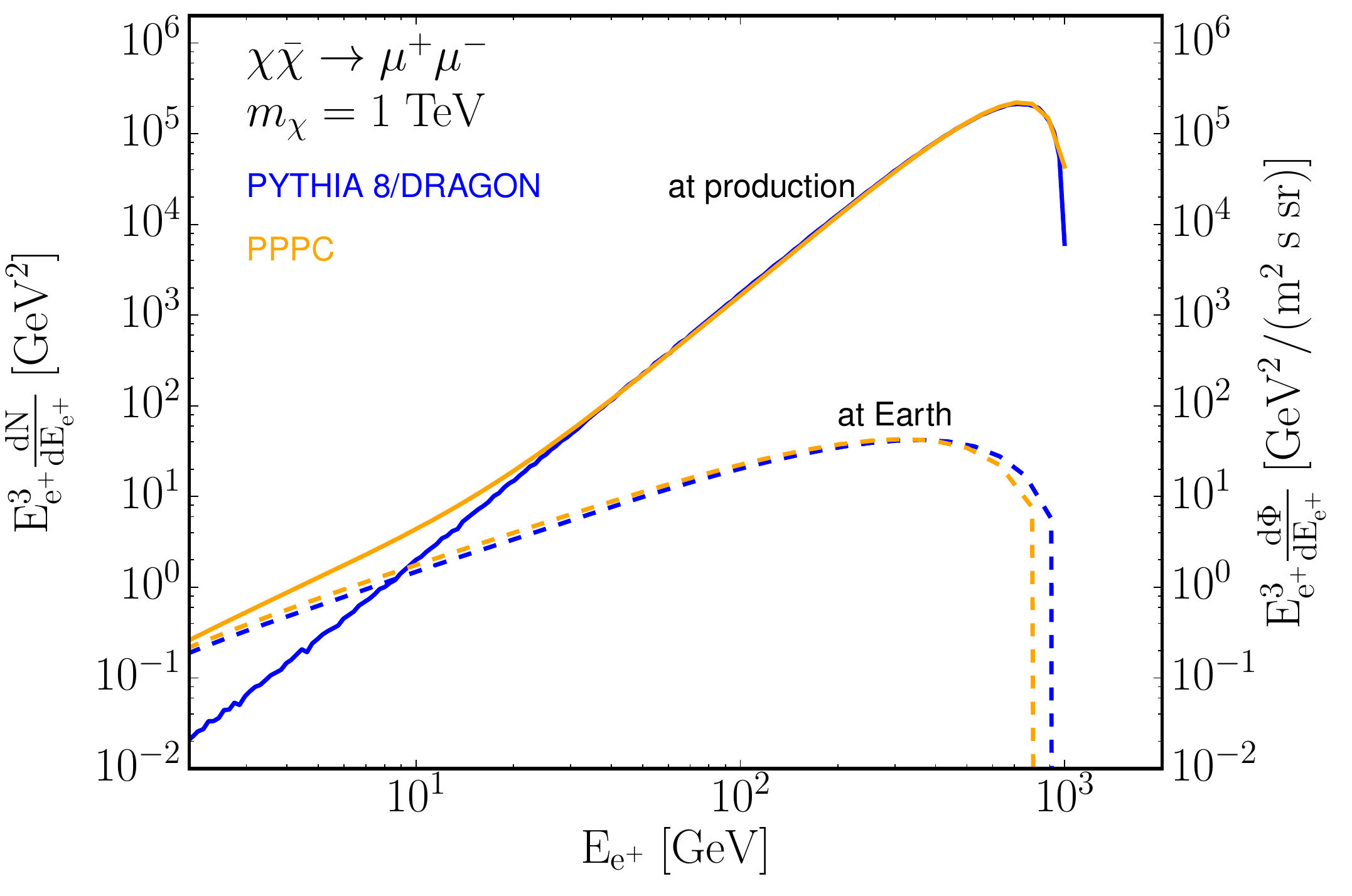}
\caption{Left: Differential gamma flux for dark matter annihilation and semi-annihilation
in black and color curves respectively. 
The solid black curve is the spectrum calculated with \texttt{PPPC} while the dashed curves with \texttt{PYTHIA 8}. 
Right: Differential flux of positrons for dark matter annihilating into a pair of muons per annihilation
at production generated with \texttt{PYTHIA 8} and \texttt{PPPC} 
are shown for in solid blue and orange lines respectively.
The dashed orange line denotes the positron flux at the Earth generated with \texttt{PPPC}, while 
the dashed blue line is the one generated with \texttt{DRAGON} from the spectrum shown in solid blue line. 
Dark matter masses in both panels are set to 1 TeV.
}
\label{fig:validation}
\end{figure}

To validate our simulation of the cosmic ray flux, we show in \figref{fig:validation} fluxes generated with the machinery described 
here and those from \texttt{PPPC 4 DM ID}~\cite{1012.4515}. 
On the left panel of \figref{fig:validation}, the gamma ray fluxes for dark matter annihilating into a pair of $b$ quarks 
produced with \texttt{PYTHIA 8} and \texttt{PPPC} are shown in black solid and dashed curves, while 
the red or blue dashed curves denote the gamma ray fluxes for semi-annihilation operators with a $b$ quark 
and scalar or fermion dark matter.  
In the right panel of \figref{fig:validation}, the positron fluxes per annihilation for $\chi\bar{\chi}\to \mu^+\mu^-$ with
dark matter mass $m_\chi = 1 \; \rm{TeV}$ generated with \texttt{PYTHIA 8} and \texttt{PPPC}
are shown in solid blue and orange lines, which are mostly in good agreement except at low positron energy due to electroweak corrections included in \texttt{PPPC}.
This discrepancy at low energies, however, does not affect very much the final positron fluxes at the Earth after propagation,
which can be seen from the good agreement of the two dashed lines in the right panel of \figref{fig:validation}.   
    
Besides all the indirect detection experiments observing cosmic rays at present, 
the presence of dark matter can also modify the properties of various astrophysical
objects which can in turn impose constraints on dark matter theories.
The most relevant one is the observation of the Cosmic Microwave Background (CMB). 
Charged particles from dark matter annihilation around the recombination epoch inject energy
into thermal plasma and change the CMB anisotropy. 
Accurate measurement of the CMB can then lead to stringent constraints.     

We will discuss the most stringent constraints at present and any substantial improvement of sensitivities in the future from gamma ray, positron, neutrino and CMB observations, and apply them to those operators we constructed in the previous sections that exist for single-component DM.  Following the same procedure, we can in principle derive the limit on any operator detailed in this work.  An important assumption we will make is that the dark matter relic density is saturated.  As discussed in \secref{sec:method}, this may require additional model-dependent self-annihilation.  These extra processes will enhance the cosmic ray flux and lead to stronger bounds.  Thus by considering only the model-independent SA channels, we are taking a conservative approach.

\subsection{Gamma rays}
We start our discussion with the golden channel of indirect detection: cosmic gamma rays. 
The most stringent limits at present derive from the observation of gamma rays from the dwarf spheriodal 
satellite galaxies (dSphs) of the Milky Way at the Fermi Large Area Telescope (LAT)~\cite{1503.02641}.
Dwarf galaxies are dark-matter dominated systems.  Therefore, the background from baryonic matter is much weaker compared with the Galactic center, 
which makes dSphs an ideal probe.    

Fermi-LAT has presented the limits on dark matter annihilation cross-sections from the 
combined 15-dSph analysis for the channels of $e^+e^-$, $\mu^+\mu^-$, $\tau^+\tau^-$, $u\bar{u}$, 
$b\bar{b}$ and $W^+W^-$. 
The 15 dSphs include Bootes I, Canes Venatici II, Carina, Coma Berenices, Draco, Fornax, Hercules, Leo II, Leo IV,
Sculptor, Segue 1, Sextans, Ursa Major II, Ursa Minor and Willman 1, whose $J$-factors are
estimated assuming the dark matter distribution in dSphs follows a Navarro-Frenk-White (NFW) profile~\cite{Navarro:1995iw}.
Note that the $J$-factors are found to be quite insensitive to the choice of dark matter density profile.   
The numerical values and the uncertainties of the $J$-factors are taken directly from Table I of Ref~\cite{1503.02641}.   
Since the SA operators we considered also have SM particles in the final states other than the ones presented by Fermi-LAT,
such as $Z$ and $h$, we make use of the publicly-available likelihood functions provided by Fermi-LAT to derive the limits. 
Incorporating the uncertainty of the $J$-factor for the $i$-th dSph 
with a log-normal distribution with central value $J_{obs,i}$ and width of $\sigma_i$, 
we define the likelihood function for the $i$-th dSph as
\begin{align}
\tilde{L}_i\left(\mu, J_i | \mathcal{D}_i\right) = L_i \left(\mu|\mathcal{D}_i\right)
\frac{1}{ J_i \sqrt{2\pi} \sigma_i \ln(10)} e^{-\left(\log_{10} J_i -\log_{10} J_{obs,i} \right)^2/2\sigma_i^2} \; , 
\end{align}  
where $L_i$ is the product of the binned likelihood function for the $i$-th dSph
provided by Fermi-LAT and $\mu$ is the signal strength of the dark matter (semi)-annihilation.
The total likelihood function is achieved by multiplying all 15 likelihoods together
\begin{align}
L = \Pi_i \; \tilde{L}_i\left( \mu, J_i | \mathcal{D}_i \right)  \; .  
\end{align}
Treating $J_i$ as the nuisance parameter, the profile likelihood method is used to derive the constraints 
on the model parameters.
The $95\%$ confidence level (CL) upper limit on the (semi)-annihilation cross section is found 
where the profiled negative log likelihood function $-2 \log L$ is 2.7 bigger than the minimal value.   
Using the spectrum of dark matter annihilating to a $b\bar{b}$ pair supplied by Fermi-LAT, we compared the constraints achieved with the method here and the ones given by Fermi-LAT and found they agree within 10\%.       

The MAGIC Cherenkov telescope has also studied cosmic gamma rays~\cite{1312.1535}. A combined study has 
been performed for the 158 hours of Segue 1 observations with MAGIC and the 6-year observations of 15 dSphs by Fermi-LAT~\cite{1601.06590}.
Compared with the Fermi-LAT dSphs limit, the combined limits are improved substantially for dark matter mass $\gtrsim 1$ TeV where MAGIC has better sensitivity.  
However, the likelihood functions for MAGIC are not publicly available
and it is unlikely we can recast the combined analysis properly. 
 
We also present the projected limit from the Cherenkov Telescope Array (CTA) which will improve the existing
constraints substantially for dark matter masses above $\mathcal{O}(10^2)$ GeV.
We make use of the {\it morphological analysis}  where a {\it realistic} estimate
including the uncertainties from the Galactic diffuse emission is taken into account~\cite{1408.4131}.
The $J$-factor is calculated with the Einasto profile~\cite{Navarro:2003ew,Graham:2005xx} and takes values
$J_{\rm{ON}}=7.4\times 10^{21} \; {\rm GeV}^2 {\rm cm}^{-5}$ for the signal region
and $J_{\rm OFF}=1.2\times 10^{22} \; {\rm GeV}^2 {\rm cm}^{-5} $ for the background region.
We compare the predicted gamma-ray spectra and the differential sensitivities 
assuming 100 hours observation and 1\% systematic uncertainty
to extract limits in 15 energy bins of size $\delta \log_{10} (E/\rm{GeV}) = 0.173$ 
between $E=25$ GeV and $E=10^4$ GeV.  
The best among the 15 individual limits is used.      
As a cross check, 
we derive limits on the cross section of 
dark matter self-annihilation to a pair of bottom quarks with $J=J_{\rm ON} + J_{\rm OFF}$
with this simple method, 
which agree roughly with those in Ref.~\cite{1408.4131} within $\sim 50-150\%$.
Since the detailed configuration of CTA is not yet fully settled, which may vary the sensitivity by a factor of a few,
this method is enough to give a rough estimate of the projected limits. 
We leave a more accurate study including full analysis of the background to future work.   
         
\subsection{Positrons and Electrons}
Another type of constraint we consider is from the observation of cosmic positrons.
The Alpha Magnetic Spectrometer (AMS-02) has observed an excess of cosmic positrons at energies $\gtrsim 10$ GeV over the standard propagation model~\cite{Aguilar:2014mma, Accardo:2014lma}, 
confirming similar observations by PAMELA~\cite{1308.0133} and HEAT~\cite{DuVernois:2001bb}.
However, the absence of a bump feature in the spectrum leads to, instead of the discovery of dark matter, rather tight limits on dark matter annihilation.

Upper limits on the dark matter annihilation cross section 
have been derived based on either the positron fraction or positron flux~\cite{1306.3983, 1309.2570}.
Annihilation channels considered include  
$e^+e^-$, $\mu^+\mu^-$, $\tau^+\tau^-$, $b\bar{b}$, $W^+W^-$ and $e^+e^-\gamma$.
We choose to recast the limits to cover all relevant channels in this work based 
on positron flux following the method in~\cite{1309.2570},
as the analysis of the positron fraction would involve extra astrophysical uncertainties related to the electron flux.       
Assuming a background positron flux from astrophysical sources such as pulsars
and from spallations of cosmic rays with the interstellar medium,
we parameterize the interstellar positron flux without solar modulation effects as 
\begin{align}
\Phi_{e^+}^{\rm IS} =\Phi_{e^+}^{\rm sec, IS} + \Phi_{e^+}^{\rm source, IS} + \Phi_{e^+}^{\rm DM, IS} \; , 
\end{align} 
where the terms denote secondary positrons, source contributions and a contribution from dark matter annihilations respectively.
The secondary positrons can be expressed in terms of a simple power law
\begin{align}
\Phi_{e^+}^{\rm sec, IS} = C_{e^+} E^{-\gamma_{e^+}}\; ,
\end{align} 
while for the source contribution we use a simple power law with an exponential cut-off
\begin{align}
\Phi_{e^+}^{\rm source, IS} = C_s E^{-\gamma_s} e^{-\frac{E}{E_s}}\; .
\end{align}
With \texttt{PYTHIA 8} and \texttt{DRAGON} the contribution from dark matter (semi)-annihilation for a given dark matter mass and 
annihilation cross section is straightforward to generate.
We have chosen the NFW profile,
a local dark matter density of $\rho_\odot=0.41 \,{\rm GeV}/{\rm cm}^3$,
and two-dimensional propagation in \texttt{DRAGON}. 
Finally the solar modulation effects are parameterized with a single variable $\phi_{e^+}$ under 
the force field approximation and the positron flux at the top of the atmosphere (TOA) is
\begin{align}
\Phi_{e^+}^{\rm TOA} (E)= \frac{E^2}{\left(E+\phi_{e^+}\right)^2} \Phi_{e^+}^{\rm IS}(E+\phi_{e^+}) \; .
\end{align}      
The final positron flux $\Phi_{e^+}^{\rm TOA}$ is a function of six parameters in total: 
$C_{e^+}$, $C_s$, $\gamma_{e^+}$, $\gamma_s$, $E_s$ and $\phi_{e^+}$.

AMS-02 has published measurements of the positron flux covering the energy range 0.5 to 500\,GeV
together with the systematic errors~\cite{Aguilar:2014mma}. 
We will only use data with energy $>2$\,GeV where the secondary positrons
follow a simple power law as described above.
For energies between 2 GeV and 500 GeV,
the best fit for the positron flux $\Phi_{e^+}^{\rm TOA}$ without a dark matter contribution 
can be achieved using a $\chi^2$ test, 
with bounds $C_{e^+},\; C_s \geq 0$, $3.3<\gamma_{e^+} < 3.7$, $\gamma_s\leq \gamma_{e^+}$, $E_s>0$ and 
$0.5 \; \rm{GV} <\phi_{e^+}<1.3 \;\rm{GV}$. 
The total errors are calculated by adding 
the systematic and statistical errors in quadrature $\sigma = \sqrt{\sigma_{stat}^2 + \sigma_{syst}^2}$. 
The $95\%$ CL upper limit on the dark matter annihilation cross section 
is found when the best fit model including dark matter contribution has $\delta \chi^2 = 4$.

\subsection{Neutrinos}\label{subsec:neu}
For operators involving neutrinos, neutrino detectors like Super Kamiokande and IceCube     
would naively be the ideal probe to give the most stringent limits.
If the dark matter masses are above the weak scale, however, 
monochromatic neutrinos in the final state are usually accompanied by gamma rays via electroweak bremsstrahlung.
Cosmic gamma rays can also be used to impose constraints on the model parameter space~\cite{1602.05966}.

The current most stringent limits come from neutrino telescopes IceCube~\cite{1505.07259},
Super Kamio\-kande~\cite{1510.07999} and ANTARES~\cite{1505.04866}.
Both IceCube and ANTARES detect muon neutrinos while Super Kamio\-kande also detects electron neutrinos.    
We will adopt the limits at $90\%$ CL from  ANTARES and the future sensitivities of CTA derived in \cite{1602.05966}.      

Neutrino oscillation plays an important role in setting the limit.
The probability to observe neutrino oscillation $\nu_\alpha \to \nu_\beta$ from the Galactic center is simply
\begin{align}
P(\nu_\alpha \to \nu_\beta) = \sum_i \left|U_{\beta i}\right|^2 \left|U_{\alpha i}\right|^2 \; ,
\end{align}    
as the oscillating part averages to zero due to the extremely long baseline.
Applying the best fit value of the PMNS matrix by NuFit~\cite{1409.5439}, we have
$P(\nu_e\to \nu_\mu)\approx 0.23$,  $P(\nu_\mu\to \nu_\mu)\approx 0.40$,  and $P(\nu_\tau\to \nu_\mu)\approx 0.35$. 
The limit presented by IceCube assumed a relative neutrino flavor ratio of $1:1:1$ at Earth,
which can be translated to the limit on the cross section of dark matter annihilating into neutrino $\nu_i$ 
by a rescaling factor $\frac{1}{3 P(\nu_i \to \nu_\mu)}$.
ANTARES presented the limits for dark matter annihilating into muon neutrinos.
So the limits for annihilations to other neutrino species $\nu_i$ can be easily calculated by multiplying $\frac{P(\nu_i \to \nu_\mu)}{P(\nu_\mu\to \nu_\mu)}$.         

The future sensitivity on operator with neutrinos can be substantially improved by CTA.
The limits at $95\%$ CL can be read directly from Ref.~\cite{1602.05966} for neutrinos of different flavors.

\subsection{Cosmic Microwave Background}
Dark matter (semi)-annihilation injects ionizing particles in the early universe 
which can modify the anisotropies of the cosmic microwave background (CMB).
The effects can be quantified with 
\begin{align}
p_{\rm ann} \equiv f_{\rm eff} \frac{\langle \sigma v\rangle}{ m_{\rm DM}} \; ,
\label{eqn:pann}
\end{align}        
where $f_{\rm eff}$ denotes the efficiency of the energy absorption at the recombination.
$f_{\rm eff}$ is a function of dark matter mass defined as
\begin{align}
f_{\rm eff} \left( m_{\rm DM}\right) = \frac{ \int_0^\infty 
dE E  \left[2 f_{\rm eff}^{e}  \left(\frac{dN}{dE}\right)_{e} + f_{\rm eff}^\gamma \left(\frac{d N}{dE}\right)_\gamma \right]
}{2 m_{\rm DM}} \;
\label{eqn:feff}
\end{align}  
where $f_{\rm eff}^{e,\gamma}$ are the differential efficiency function as a function of electron or photon energies.
Planck~\cite{1502.01589} gave a $95\%$ CL upper limits with TT, TE and lowP data
\begin{align}
p_{\rm ann} < 4.1\times 10^{-28} \; {\rm cm}^3 {\rm s}^{-1} {\rm GeV}^{-1} \; .
\end{align}

In general, we need to derive $f_{\rm eff}$ with the differential flux resulting from dark matter annihilation.
For this work, however, we can make several very efficient simplifications to avoid the tedious calculation and yet 
still lead to satisfactory limits.
First of all, we will apply the constraints from CMB only to operators with light charged leptons, i.e. electrons and muons, 
as they are much less competitive than the ones from cosmic gamma ray observation for operators with quarks and taus.     
Therefore, the contribution from the second term in \modeqref{eqn:feff} is negligible. 
Furthermore, we will choose $\rho$ such that the differential flux of the SA leptons are much stronger
than that from the decaying leptons, so the contribution from the decaying leptons is also negligible.    
Finally, the constraints from CMB are most relevant when the dark matter mass is heavier than a couple of 
hundred GeV where $f_{\rm eff}$ is approximately a constant~\cite{1506.03811}.  
We will use $f_{\rm eff}=0.40$ and $0.15$ for operators with electrons and muons,
which has to be further multiplied by a factor of $1/2$ as only one electron, instead of an electron-positron pair,
is produced in each semi-annihilation.  
 
\subsection{Results}
We now present the limits from indirect detection and astrophysical observations for single-component DM operators.
An important assumption we make is that the relic density is saturated.   
The annihilation cross sections presented in this section 
frequently involve the K{\"a}ll{\'e}n $\lambda$ function defined as
\begin{align}
\lambda(x, y, z)\equiv x^2 + y^2 + z^2 - 2 x y - 2 y z - 2 z x \; .
\end{align}   
Additionally, when considering operators including dark partners we define
\begin{equation}
	\rho \equiv \frac{m_{\dkp}}{m_{\text{DM}}} \,,
\end{equation}
where $m_{\dkp}$ is the dark partner mass.
Additionally, we will also sketch two constraints for each chosen operator: 
the thermal relic cross section, with which the relic density is approximately saturated;
and the perturbativity bound, beyond which the dark matter mass is likely to be heavier than the 
mediator mass in the UV theory invalidating the EFT description.  
 
\subsubsection{D4 operators}
The operators with the lowest mass dimension are $\mathcal{O}_{4B}^{h}$ and $\mathcal{O}_{4B}^{h\omega}$
which have dimensionless couplings. 
The only SM particle involved is the Higgs boson and the most sensitive limit is currently set by 
the dSphs observation of Fermi-LAT.

The $s$-wave annihilation cross section for $\mathcal{O}_{4B}^{h}$ without a dark partner is
\begin{align}
\langle \sigma v\rangle \left(\mathcal{O}_{4B}^h \right) = 
\frac{\left|c_{4B}^h\right|^2}{128\pi} 
\frac{\sqrt{\lambda\left( 4 m_\phi^2, m_\phi^2, m_h^2 \right)}}{m_\phi^4}
\end{align}
and with a scalar dark partner $\omega$ is 
\begin{align}
\langle \sigma v\rangle \left(\mathcal{O}_{4B}^{h\omega} \right) = 
\frac{\left|c_{4B}^{h\omega}\right|^2}{128\pi} 
\frac{\sqrt{\lambda\left(4m_\phi^2, m_\omega^2, m_h^2 \right) }}
{m_\phi^4}  \; ,
\end{align}    
where $c_{4B}^{h, h\omega}$ is the dimensionless coupling for the relevant operator.
For the case with dark partner $\omega$, the annihilation process $\phi\phi\to \omega h$ will
be followed by subsequent decay of $\omega$. The decay of $\omega$ is completely model-dependent
and we will restrict ourselves to the minimal case where $\omega$ decays to $\phi h$ with $100\%$ branching ratio.
To allow this simplest decay scenario, $m_\phi > 2 m_h$ and $m_\omega>m_\phi+m_h$ should be satisfied.

\begin{figure}
\centering
\includegraphics[width=0.48\textwidth]{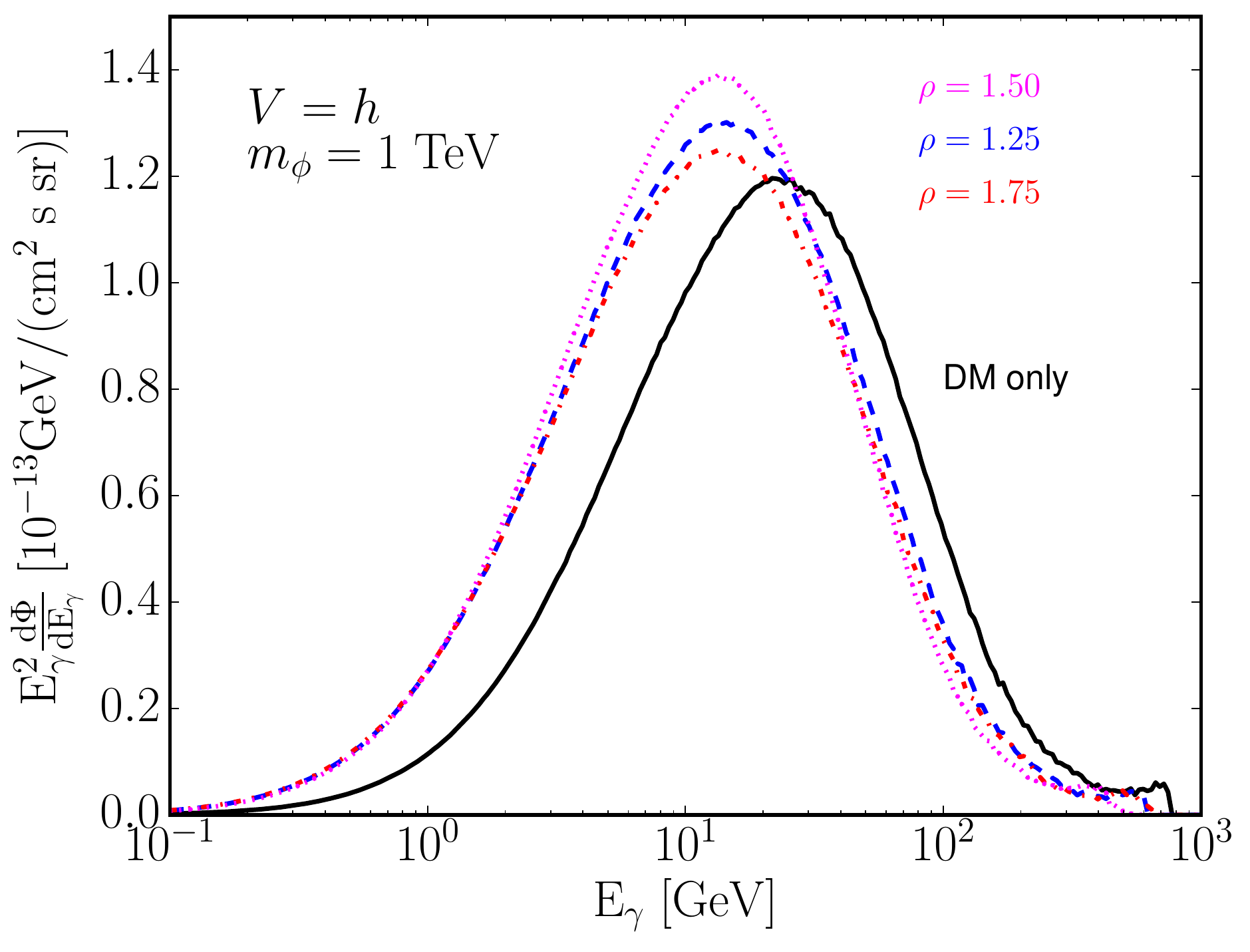}
\hfill
\includegraphics[width=0.48\textwidth]{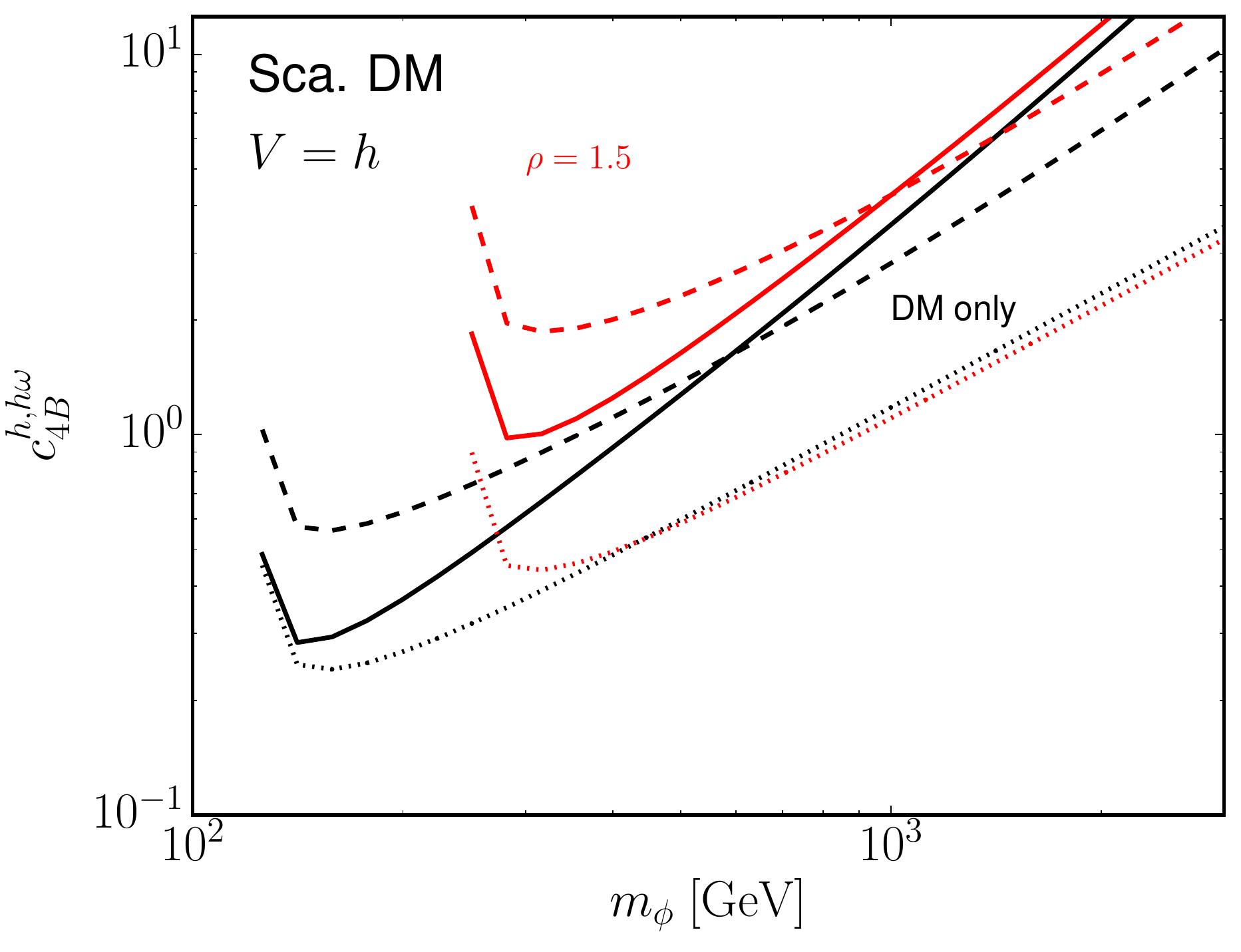}
\caption{Left: cosmic gamma ray spectra for various dimension-4 operators
with the scalar dark matter mass $m_\phi= 1 \; \rm{TeV}$. 
The dark solid line shows the spectrum for SA without a dark partner, while
the blue dashed, magenta dotted and red dot-dashed lines represent SA with different dark partner mass $m_\omega/m_\phi =
1.25,\; 1.50$ and $1.75$ respectively.
We have assumed $\langle \sigma v\rangle = 1.0\times 10^{-25} \rm{cm}^3/\rm{s}$ and $J=1.0\times 10^{18} \rm{GeV}^2/\rm{cm}^5$.
Right: the current constraints from Fermi-LAT dSphs and the future sensitivities from CTA in solid and dashed lines,
where the dark and red lines denote a model without and with a dark partner.  
The dotted lines show approximately the right relic density. }
\label{fig:O4Bh}
\end{figure}

In the left panel of \figref{fig:O4Bh}, gamma ray fluxes for dark matter annihilating into dark partner and a Higgs 
are plotted in blue, magenta and red lines for $\rho=1.25, \; 1.50$ and $1.75$ with
dark matter mass $m_{\phi, \chi} = 1 \; {\rm TeV}$.
Obviously $\rho=1.50$ will yield the strongest bounds with the same dark matter mass.
In \figref{fig:O4Bh} we show the current limits from dSphs and future sensitivities from CTA in 
solid and dashed black lines for DM only models,
and solid and dashed red lines for models with dark partner with $\rho=1.50$.     
The constraints for dark partner models are marginally weaker than for dark matter only models for the same dark matter mass.  This is a consequence of the spectrum in the latter case being slightly harder, as shown in the left panel of \figref{fig:O4Bh}.  Since the SM particle is the Higgs, which has a negligible coupling to the electrons, no constraints from AMS are shown in \figref{fig:O4Bh}.  We also show black and red dotted lines for dark matter only and dark partner models where approximately the correct relic density is  achieved from SA alone.  The Fermi bounds and CTA prospects are always weaker than these relic density cross sections.

\subsubsection{Nonrenormalizable Operators}
We next study the limits on nonrenormalizable operators according to the associated SM annihilation products.

\vspace{0.5cm}
\noindent
\textbf{Higgs}: The relevant operators for scalar DM are
$\mathcal{O}_{6B}^{h\omega \partial, \partial h \omega}$ with annihilation cross sections
\begin{align}
\langle \sigma v\rangle \left( \mathcal{O}_{6B}^{h\omega \partial}\right) &=
\frac{\left|c_{6B}^{h\omega\partial}\right|^2}{128\pi \Lambda^4} 
\sqrt{\lambda\left(4m_\phi^2, m_\omega^2, m_h^2 \right) }  \\
\langle \sigma v\rangle \left( \mathcal{O}_{6B}^{\partial h\omega}\right) &=
\frac{\left|c_{6B}^{\partial h\omega}\right|^2}{512\pi \Lambda^4} 
   \frac{\left( m_h^2 + m_\omega^2 -4 m_\phi^2 \right)^2}{m_\phi^4 } 
\sqrt{\lambda\left(4m_\phi^2, m_\omega^2, m_h^2 \right) }  \; . 
\end{align} 
The decay of the dark partner $\omega$ is model dependent as discussed in \secref{sec:DP}, but conveniently chosen to be $\omega\to \phi^\dagger h$.
So a pair of $\phi$ will annihilate to $\phi^\dagger h h$.  

For fermion DM, the relevant operators are  
$\mathcal{O}_{5B}^{h\omega S(P)}$ and $\mathcal{O}_{6B}^{h\omega A}$, 
among which $\mathcal{O}_{5B}^{h\omega S}$ is $p$-wave suppressed. 
The $s$-wave annihilation cross sections of the other operators are 
\begin{align}
\langle \sigma v\rangle \left( \mathcal{O}_{5B}^{h\omega P}\right) & = 
      \frac{\left|c_{5B}^{h\omega P}\right|^2}{64\pi \Lambda^2}
\frac{\sqrt{\lambda\left(4m_\chi^2, m_\omega^2, m_h^2 \right) } }{m_\chi^2} \\
\langle \sigma v\rangle \left( \mathcal{O}_{6B}^{h\omega A}\right) &=
     \frac{\left|c_{6B}^{h\omega A}\right|^2}{256\pi  \Lambda^4 m_\chi^4}
\sqrt{\lambda\left(4m_\chi^2, m_\omega^2, m_h^2 \right) }\; .
\end{align}
The dark partner $\omega$ is chosen to decay through 
$\omega\to \bar{\chi}  \nu_i$ in this scenario with $100\%$ branching ratio.
 Therefore a pair of 
fermion DM $\chi$ will annihilate to $\bar{\chi} \nu_i h $.

\begin{figure}
\centering
\includegraphics[width=0.48\textwidth]{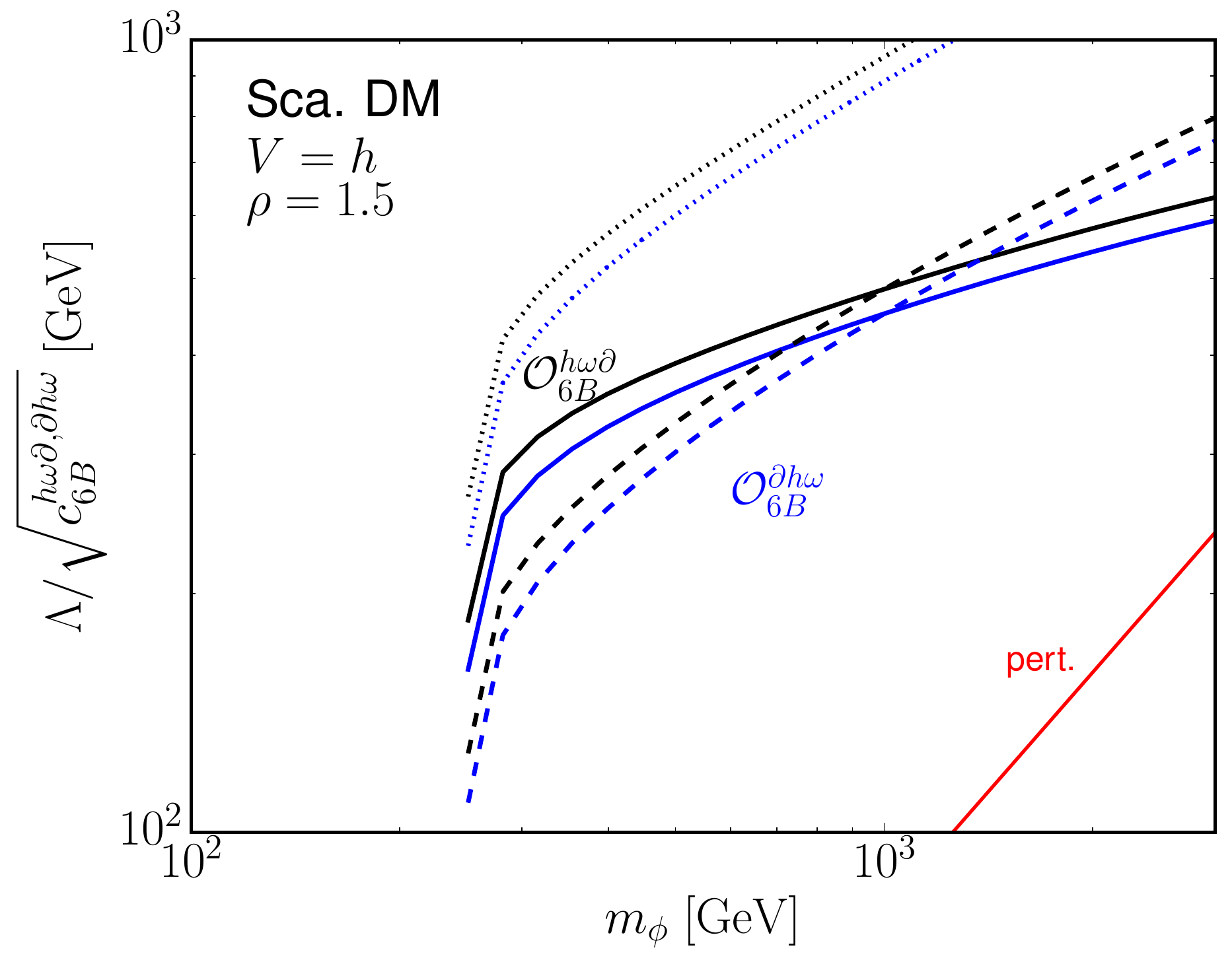}
\hfill
\includegraphics[width=0.48\textwidth]{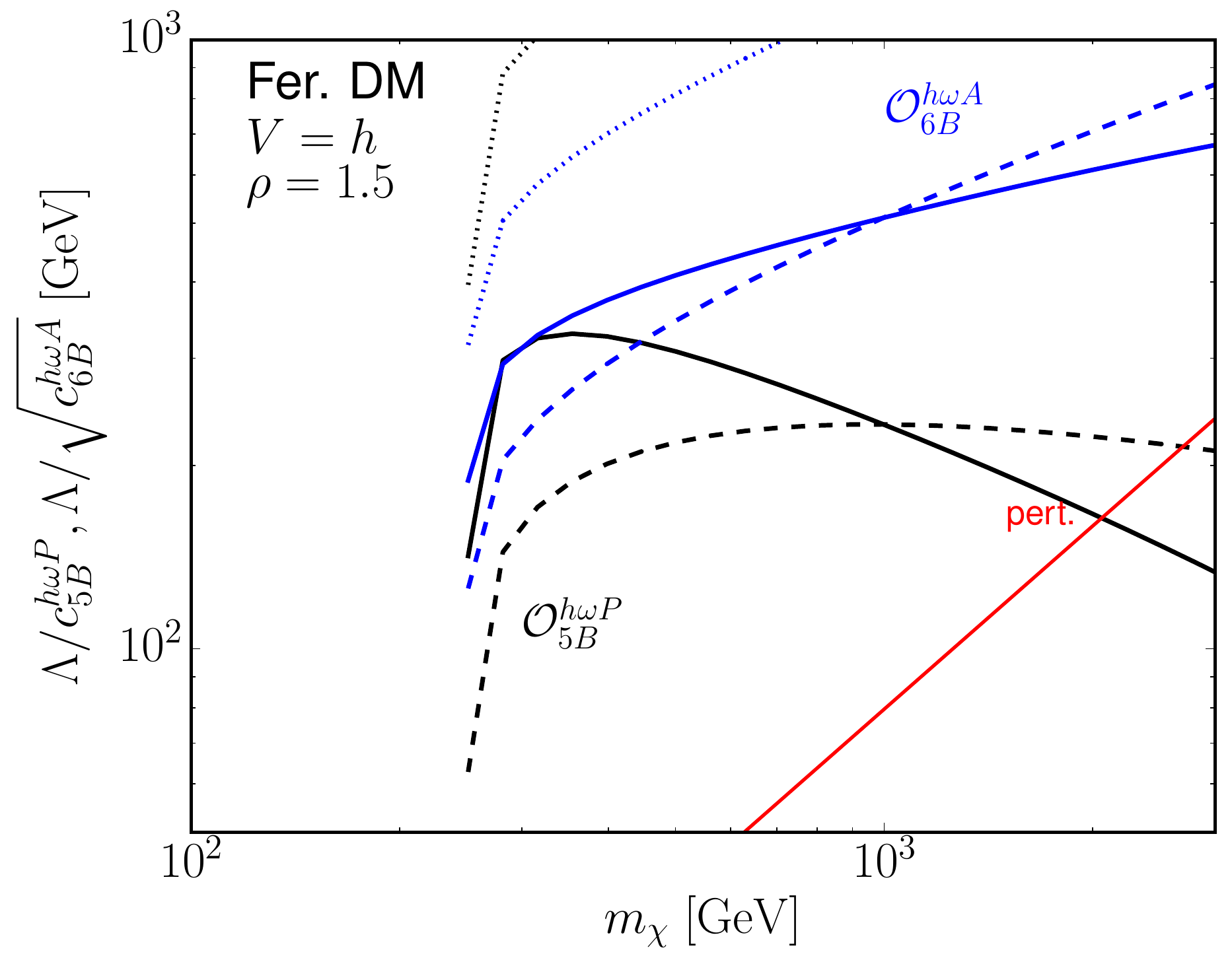}
\caption{Current best limits from Fermi-LAT dSphs observations and future sensitivities of CTA 
on wilson coefficients of SA operators containing SM Higgs and 
are shown as solid and dashed lines for 
$\mathcal{O}_{6B}^{h\omega \partial}$ and $\mathcal{O}_{5B}^{h\omega P}$ in black,
and $\mathcal{O}_{6B}^{\partial h \omega}$ and $\mathcal{O}_{6B}^{h\omega A}$ in blue.
The dotted lines denote the thermal relic cross section in both plots.
The ratio of the dark partner mass and the dark matter mass is set to $\rho=1.5$.   
An estimate of non-perturbativity bound is shown in red solid line.
}
\label{fig:higgs}
\end{figure}

Cosmic gamma ray searches can be applied to both cases, while AMS is only relevant for fermion DM $\chi$. 
The constraints from Fermi-LAT dSphs and future sensitivity from CTA are plotted in \figref{fig:higgs}.
In principle, neutrino telescope experiments can also be utilized to set the limits on the fermion DM operators 
$\mathcal{O}_{5B}^{h\omega P}$ and $\mathcal{O}_{6B}^{h\omega A}$. The limits on the annihilation cross section
 derived from neutrinos, however, are much weaker than the ones from dSphs, and so we do not show them.

\vspace{0.5cm}
\noindent
\textbf{Vector bosons:}
The nonrenomalizable operators with a SM vector boson and scalar DM are
$\mathcal{O}_{5B}^{\mathcal{V}\omega}$
with annihilation cross section,
\begin{align}
\langle \sigma v\rangle \left( \mathcal{O}_{5B}^{\mathcal{V}\omega}\right) = 
\frac{\left|c_{5B}^{\mathcal{V}\omega}\right|^2}{512 \pi \Lambda^2 m_\phi^4 m_\mathcal{V}^2}
   \left[ \lambda\left( 4 m_\phi^2, m_\omega^2, m_\mathcal{V}^2\right)\right]^\frac{3}{2} \; . 
\end{align}
The dark partner $\omega$ will decay to $\mathcal{V} \phi^\dagger$ for scalar DM, leading to a $\mathcal{VV}\phi^\dagger$ final state.

For fermion DM operators, the relevant operators include  $\mathcal{O}_{5B}^{\mathcal{V}\omega A}$
and $\mathcal{O}_{6B}^{\mathcal{V}\omega s S(P)}$, among which $\mathcal{O}_{6B}^{\mathcal{V}\omega s S}$ is $p$-wave suppressed. 
The $s$-wave annihilation cross sections for these operators are 
\begin{align}
\langle \sigma v\rangle(\mathcal{O}_{5B}^{\mathcal{V}\omega A})  & = 
     \frac{\left|c_{5B}^{\mathcal{V}\omega A}\right|^2}{1024\pi \Lambda^2 m_\chi^4 m_{\mathcal{V}}^2}
   \left[ \lambda\left( 4 m_\chi^2, m_\omega^2, m_\mathcal{V}^2\right)\right]^\frac{3}{2} \\ 
\langle \sigma v\rangle \left( \mathcal{O}_{6B}^{\mathcal{V}\omega sP}\right) &=
     \frac{\left|c_{6B}^{\mathcal{V}\omega sP}\right|^2}{64\pi \Lambda^4 m_\chi^2 m_{\mathcal{V}}^2}
   \left[ \lambda\left( 4 m_\chi^2, m_\omega^2, m_\mathcal{V}^2\right)\right]^\frac{3}{2} \; . 
\end{align}
The dark partner $\omega$ in $\mathcal{O}_{5B}^{\mathcal{V}\omega A}$ will decay to $\nu_i \bar{\chi}$ if $\mathcal{V}=Z$ or $l_i \bar{\chi}$ if $\mathcal{V}=W$,
while in $\mathcal{O}_{6B}^{\mathcal{V}\omega sP}$ it will decay to  $\bar{\chi} \nu_i \mathcal{V}$. 
We have chosen the first generation of leptons, i.e. $i=1$, in this analysis. 

\begin{figure}
\centering
\includegraphics[width=0.48\textwidth]{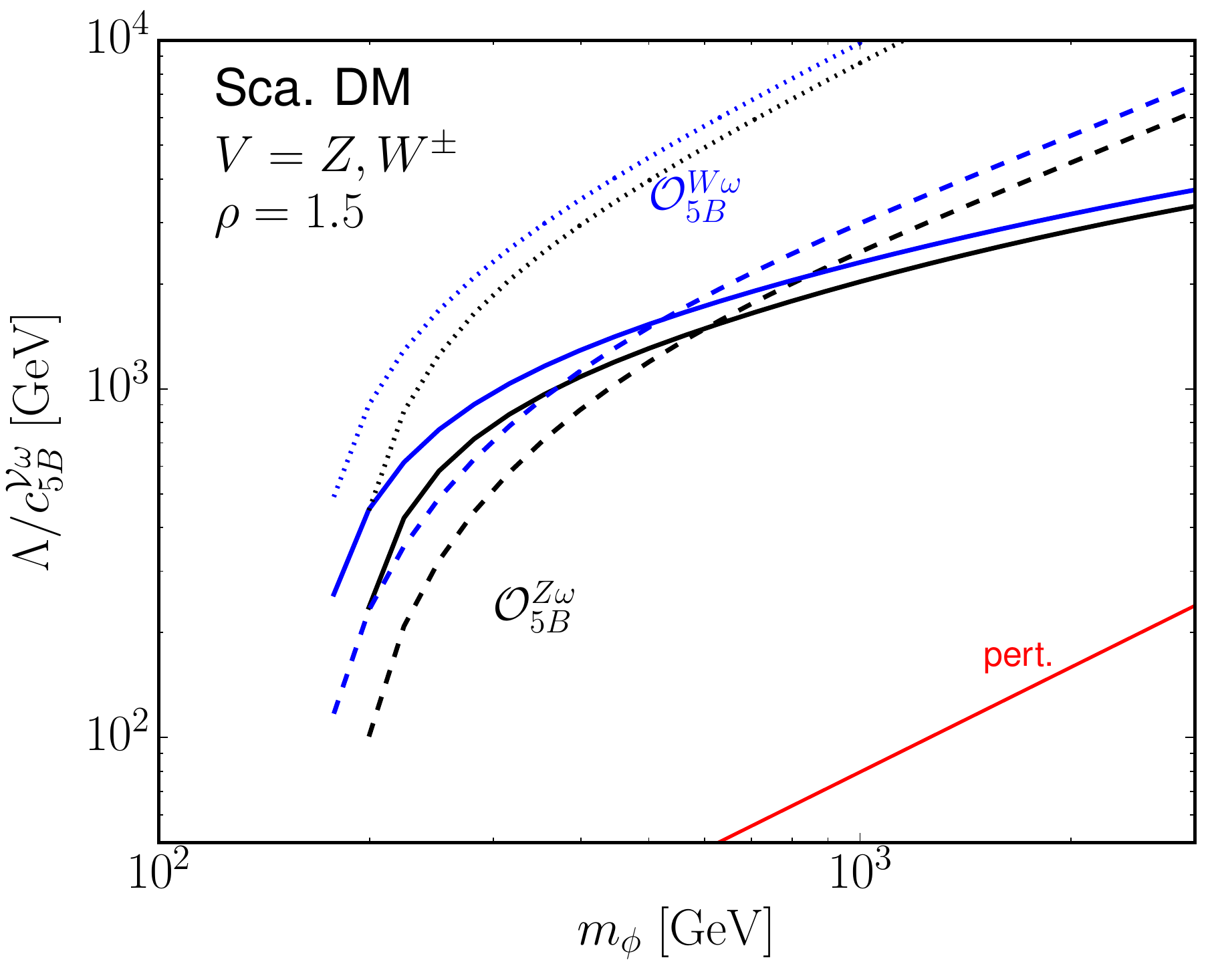}
\hfill
\includegraphics[width=0.48\textwidth]{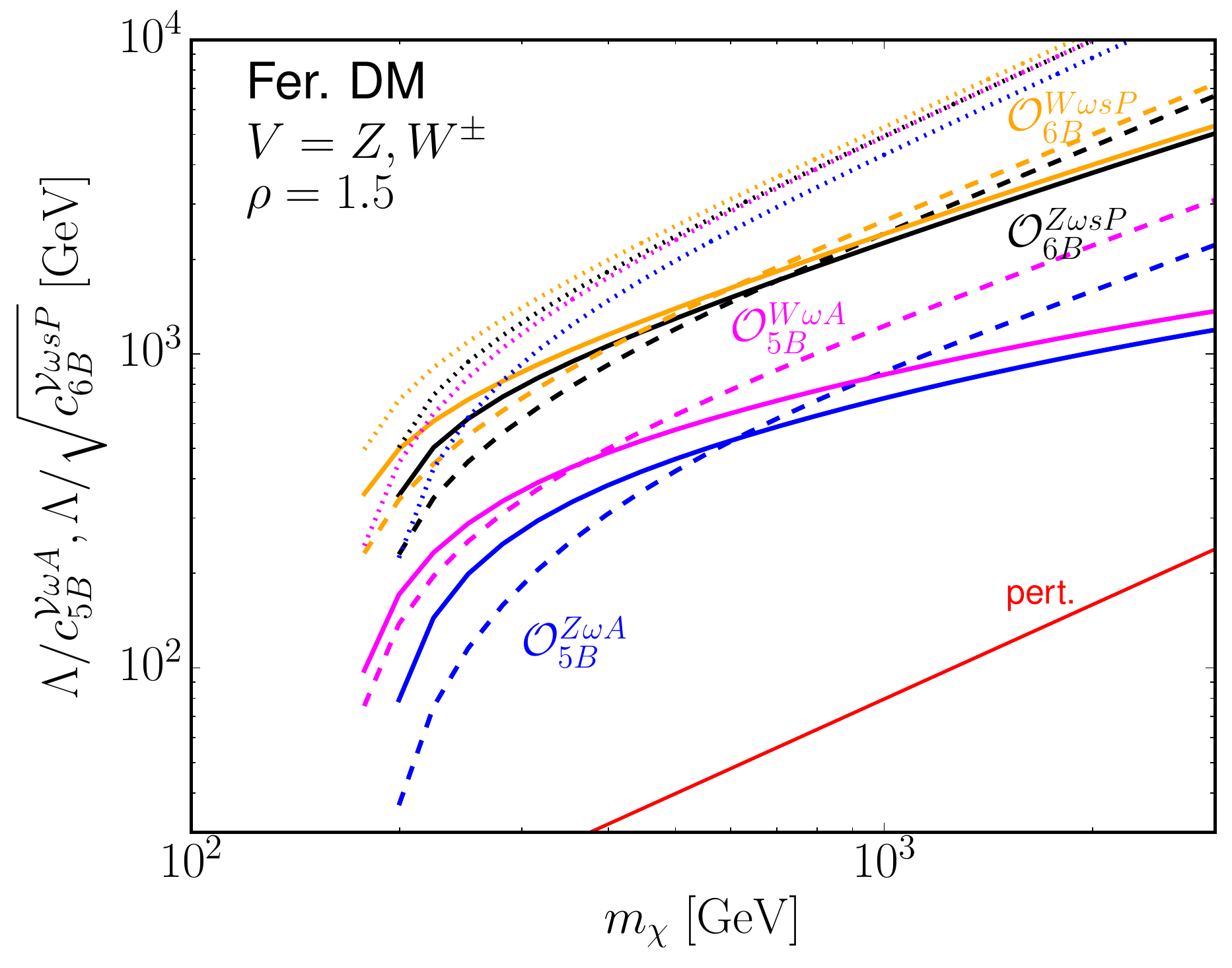}
\caption{Current best limits from Fermi-LAT dSphs observations and future sensitivities of CTA
are shown in solid and dashed lines for operators containing vector bosons.
The left and right panels show the results for scalar and fermion dark matter operators respectively.
The mass ratio between the dark partner and the dark matter is set to 1.5.
The dotted lines denote the thermal relic cross section for different operators,
with the same color codes as the gamma ray limits.   }
\label{fig:vomega}
\end{figure}

We plot the constraints from Fermi-LAT and future sensitivities from CTA for operators with SM vector bosons in 
\figref{fig:vomega}.  We see that the current bounds are much weaker than the thermal relic cross sections, except very close to the kinematic threshold.

\vspace{0.5cm}
\noindent
\textbf{Neutrinos:}
The nonrenomalizable operators with a neutrino for scalar DM are $\mathcal{O}_{5B}^{fL\psi}$ and $\mathcal{O}_{5B}^{fR\psi}$ 
with annihilation cross section,
\begin{align}
\langle \sigma v\rangle \left( \mathcal{O}_{5B}^{fL(R)\psi}\right)  =  
 \frac{\left|c_{5B}^{fL(R)\psi}\right|^2}{128 \pi  \Lambda^2 m_\phi^4}
             \left(4 m_\phi^2 - m_{\psi}^2 - m_f^2 \right)
  \sqrt{\lambda\left( 4 m_\phi^2, m_\psi^2, m_f^2 \right)} \; , 
\label{eqn:sv5Bs}
\end{align} 
where $f=\nu_i$. 
Similarly to other scalar DM scenarios, the dark partner will decay through  $\psi\to \phi^\dagger \nu_i$.

For fermion DM, there is one operator without dark partner,  
$\mathcal{O}_{6B}^{\nu L}$, with $s$-wave annihilation cross section
\begin{align}
\langle \sigma v\rangle \left( \mathcal{O}_{6B}^{\nu L}\right) = \frac{9\left|c_{6B}^{\nu L}\right|^2 m_\chi^2 }{1024 \pi \Lambda^4} \; . 
\end{align}
The operators with dark partners are
$\mathcal{O}_{6B}^{f\psi LL(LR, RL, RR)}$  and $\mathcal{O}_{6B}^{f\chi LL(LR, RL, RR)}$. 
The $s$-wave annihilation cross sections can expressed as
\begin{align}
\label{eqn:sv6BfpsiLL}
\langle \sigma v\rangle \left( \mathcal{O}_{6B}^{f\psi LL}\right) &=
    \frac{\left|c_{6B}^{f\psi LL}\right|^2}{256 \pi \Lambda^4  m_\chi^2}(4m_\chi^2 - m_\psi^2 - m_f^2) 
  \sqrt{\lambda\left( 4 m_\chi^2, m_\psi^2, m_f^2 \right)} \\ 
\label{eqn:sv6BfchiLR}
\langle \sigma v\rangle \left( \mathcal{O}_{6B}^{f\chi LR}\right) &=
    \frac{\left|c_{6B}^{f\chi LR}\right|^2}{4096 \pi \Lambda^4  m_\chi^4}
    \left(4 m_\chi^2(m_\psi^2+m_f^2)-(m_f^2-m_\psi^2)^2\right) 
  \sqrt{\lambda\left( 4 m_\chi^2, m_\psi^2, m_f^2 \right)} \; ,
\end{align}
while the rest are similarly written as
\begin{align}
\label{eqn:rln1}
\langle \sigma v\rangle \left( \mathcal{O}_{6B}^{f\psi LR, f\psi RL, f\psi RR}\right) &=
        \left. \langle \sigma v\rangle \left( \mathcal{O}_{6B}^{f\psi LL}\right)  \right|_{f\psi LL\to f\psi LR, f\psi RL, f\psi RR}  \\ 
\label{eqn:rln2}
\langle \sigma v\rangle \left( \mathcal{O}_{6B}^{f\chi LL, f\chi RR}\right) &=
       \frac{1}{4} \left. \langle \sigma v\rangle \left( \mathcal{O}_{6B}^{f\psi LL}\right)  \right|_{f\psi LL\to f\chi LL, f\chi RR} \\ 
\label{eqn:rln3}
\langle \sigma v\rangle \left( \mathcal{O}_{6B}^{f\chi RL}\right) &=
      \left. \langle \sigma v\rangle \left( \mathcal{O}_{6B}^{f\chi LR}\right)  \right|_{f\chi LR\to f\chi RL}  
\end{align}
with $f=\nu_i$. 
Since the neutrino masses are tiny, we will have for the same Wilson coeffiencts 
\begin{align}
\langle \sigma v\rangle \left( \mathcal{O}_{6B}^{\nu L}\right) :
\langle \sigma v\rangle \left( \mathcal{O}_{6B}^{\nu\psi LL}\right) :
\langle \sigma v\rangle \left( \mathcal{O}_{6B}^{\nu\chi LL}\right) :
\langle \sigma v\rangle \left( \mathcal{O}_{6B}^{\nu\chi LR}\right) =
\frac{9}{4(\rho^2-4)^2}:1:\frac{1}{4}:\frac{\rho^2}{16} \; .
\label{eqn:cxratio}
\end{align} 
$\mathcal{O}_{6B}^{\nu L}$ contains no dark partner,
while the dark partner in the other fermion DM operators will decay
through three-body decay $\psi \to \nu\nu \bar{\chi}$.

\begin{figure}
\centering
\includegraphics[width=0.48\textwidth]{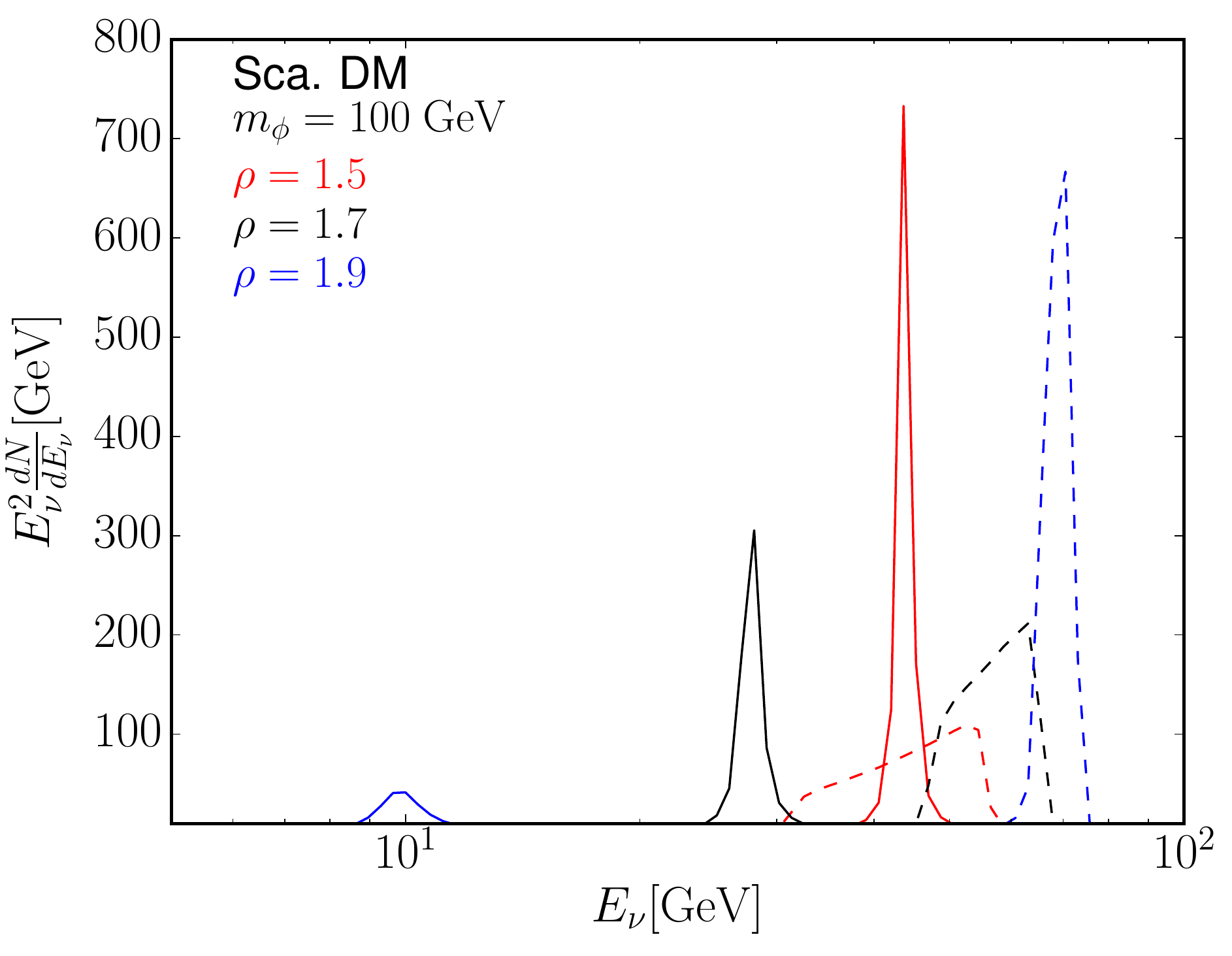}
\hfill
\includegraphics[width=0.48\textwidth]{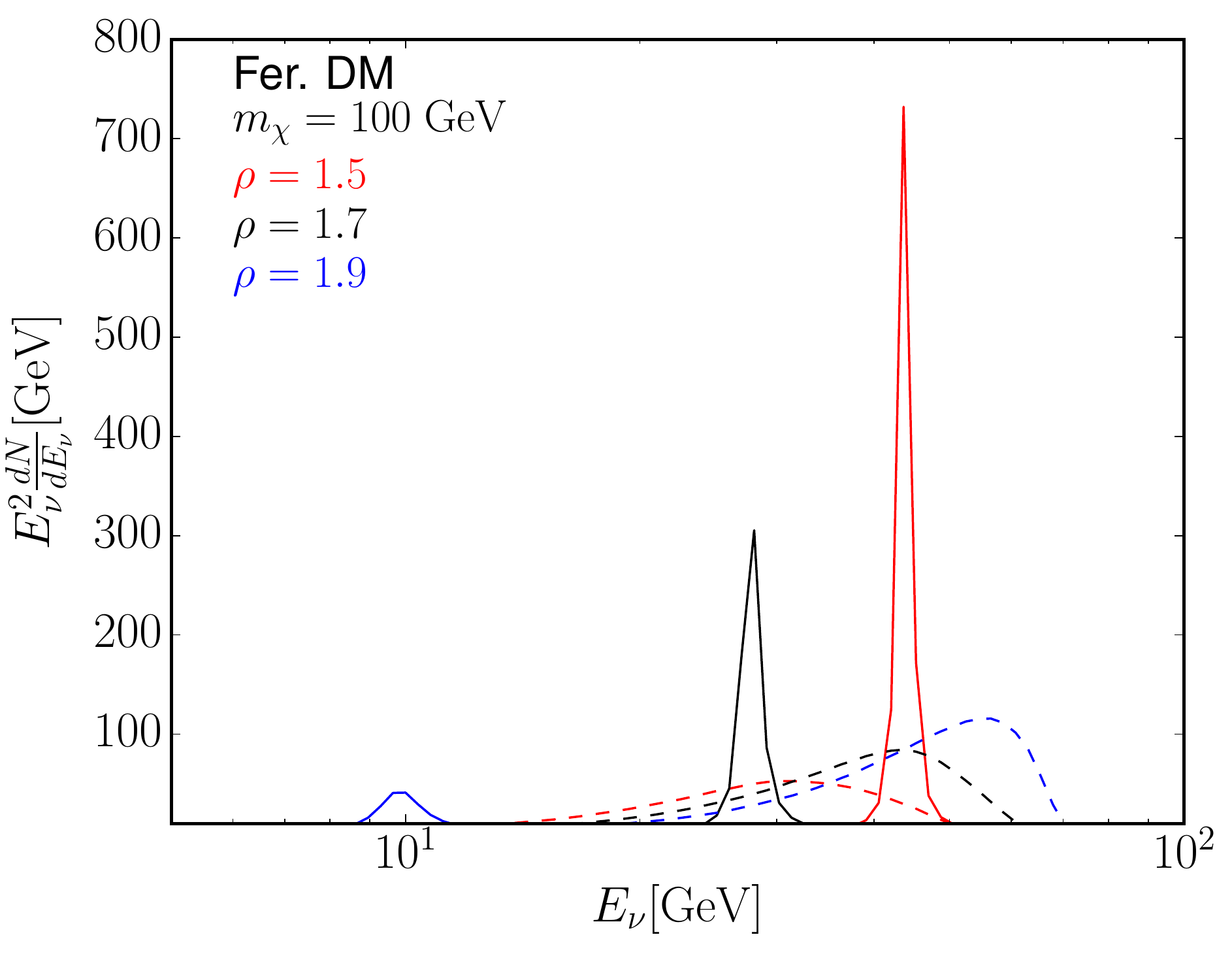}
\caption{Differential spectrum of neutrinos per annihilation from operators with dark partner and neutrino for dark matter mass 
$m_\chi=m_\phi=100\, \rm{GeV}$ in red, black and blue for $\rho=1.5$, 1.7 and 1.9. The solid and dashed lines denote the spectra for 
neutrino from SA and decay of the dark partner respectively.
 }
\label{fig:nuspec}
\end{figure}

As shown in \figref{fig:nuspec}, the neutrinos from the subsequent decays of the dark partners 
can be softer or harder depending on the choice $\rho$.
If $\rho\sim 2.0$, the SA neutrino will be much softer and 
the scenario can be simplified as the decay of the dark partner alone.
If on the contrary $\rho\lesssim 1.5$, we will have mostly hard and dominant neutrino line from 
semi-annihilation.  To demonstrate the derivation of the limits from neutrino telescope experiments 
in a simple manner, we will choose $\rho=1.5$ such that the decaying neutrino can be safely ignored.  
 
\begin{figure}[htp]
\centering
\includegraphics[width=0.48\textwidth]{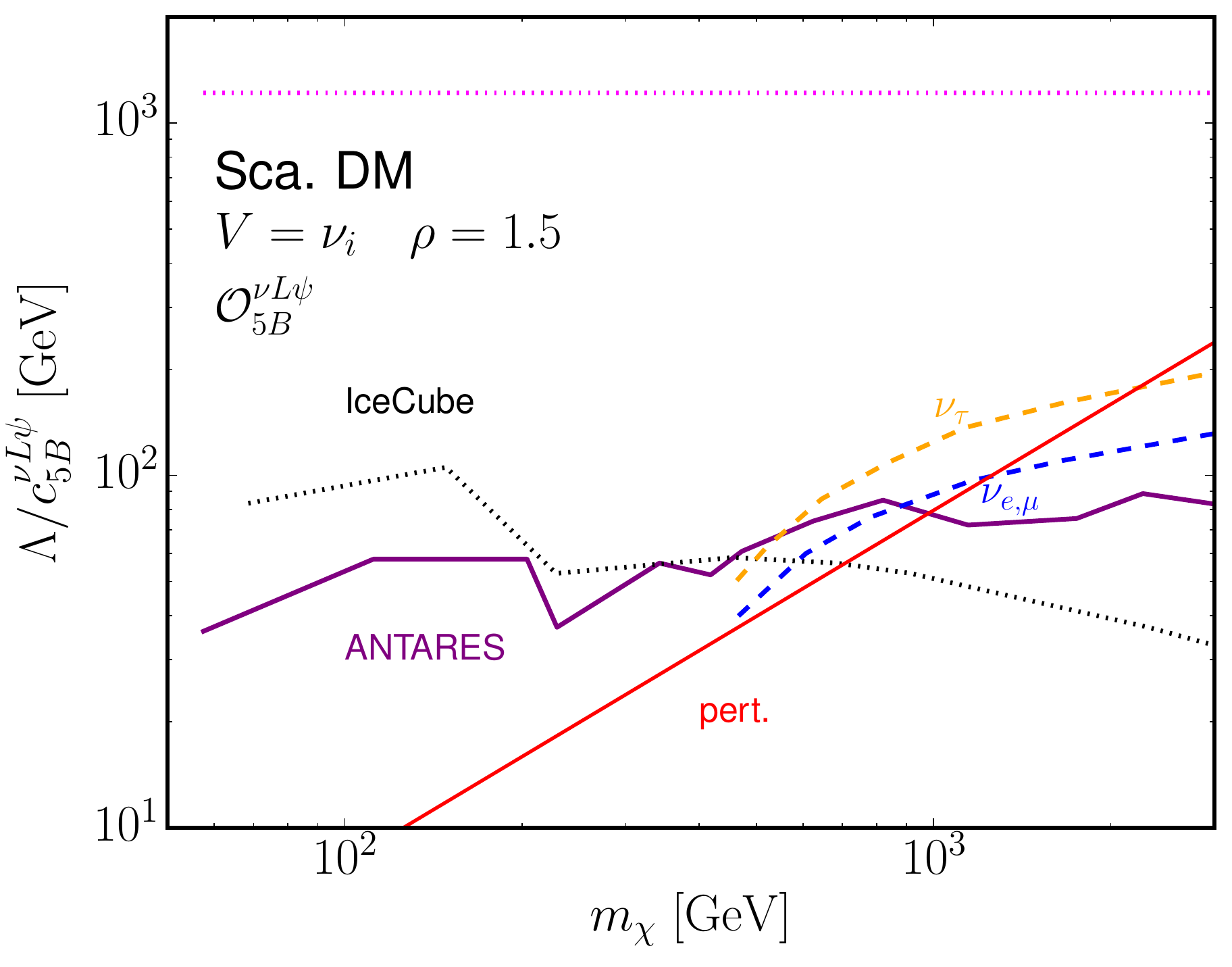}
\hfill
\includegraphics[width=0.48\textwidth]{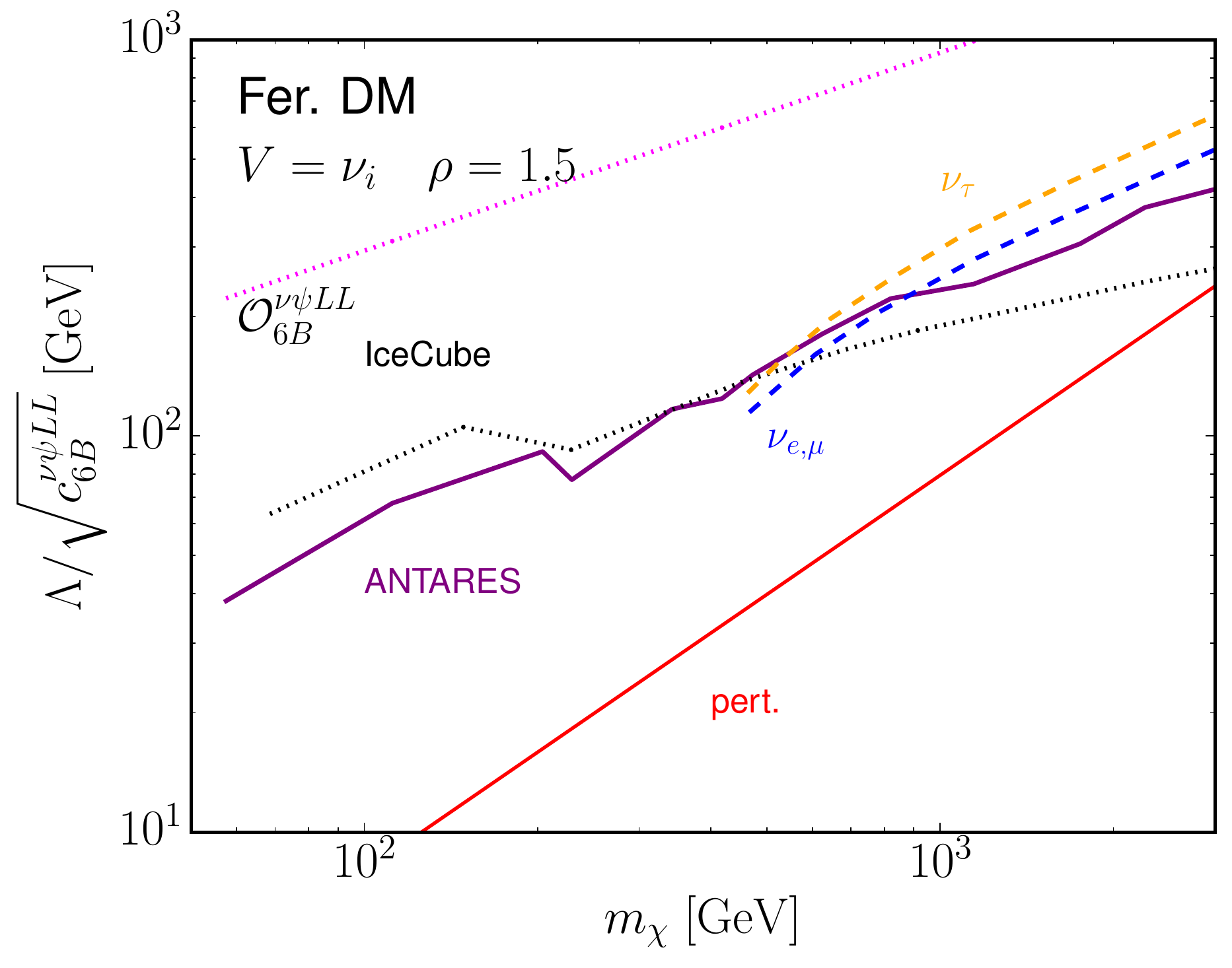}
\caption{Limits and future sensitivities on the Wilson coefficients of operators with neutrino
$\mathcal{O}_{5B}^{\nu L\psi}$ (left) and $\mathcal{O}_{6B}^{\nu\psi LL}$ (right).
The limits for ANTARES and IceCube are shown in purple solid and black dotted lines for
default assumptions of neutrino flavor. The future sensitivities of CTA are shown in 
blue and orange dashed lines for $\nu_{e, \mu}$ and $\nu_\tau$.
Again the ratio of dark partner mass (if any) and the dark matter mass is set to 1.5. 
The thermal relic contour for $\mathcal{O}_{5B}^{\nu L\psi}$
is a horizontal line shown in the left panel 
as the $s$-wave cross section 
is independent of dark matter mass.}
\label{fig:nu}
\end{figure}

In \figref{fig:nu}, we present the current limits from both IceCube and ANTARES observations and future sensitivities from 
CTA for SA operators with neutrinos. 
First of all, we only present results for the scalar DM operator $\mathcal{O}_{5B}^{\nu L\psi}$ and 
the fermion DM operator $\mathcal{O}_{6B}^{\nu \psi LL}$, as the limits can be easily derived for other operators
according to  \modeqref{eqn:cxratio}.
As discussed in \secsref{subsec:neu}, the choice of neutrino flavors will modify the limit 
for IceCube and ANTARES with a factor of $\mathcal{O}(1)$
and it would be too crowded to show the limit for all flavors. 
Thus we choose to demonstrate the limits from IceCube and ANTARES with their own default assumptions,
i.e. $1:1:1$ at Earth and pure $\nu_\mu$ at the Galactic Center.
To translate the limits of IceCube shown as purple lines in \figref{fig:nu} to a chosen flavor, 
we need to divide the limit with a factor of $1.44$,  $1.90$ and $1.77$  for $\nu_{e}$, $\nu_\mu$ and $\nu_\tau$ 
in the left panel and the square root of these factors in the right panel.
Similarly for the ANTARES limits, the factors are $1.32$, $1$ and $1.07$ for $\nu_e$, $\nu_\mu$ and $\nu_\tau$.
We also show the CTA projections for different flavors in dashed lines in \figref{fig:nu}.

\begin{figure}
\centering
\includegraphics[width=0.48\textwidth]{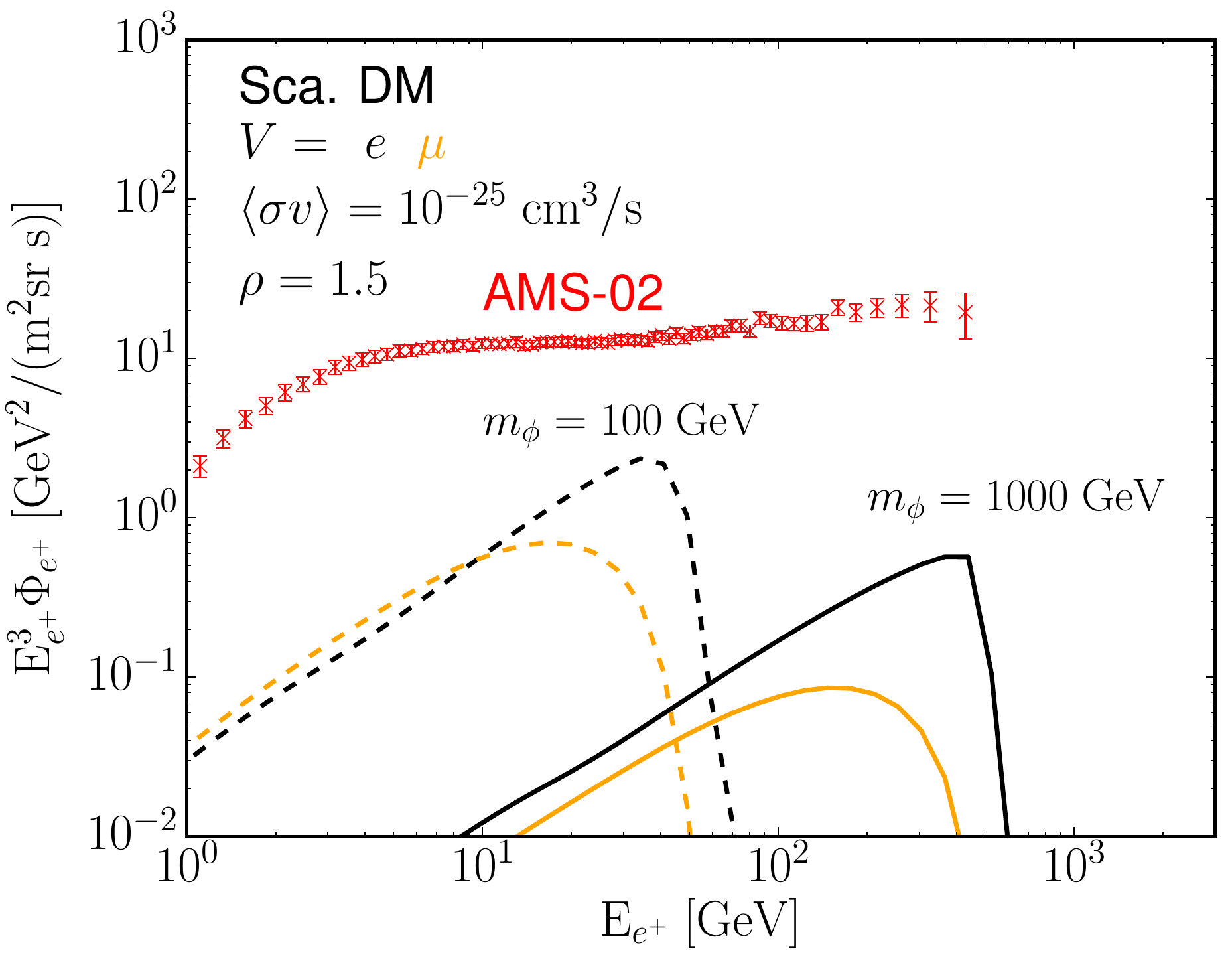}
\hfill
\includegraphics[width=0.48\textwidth]{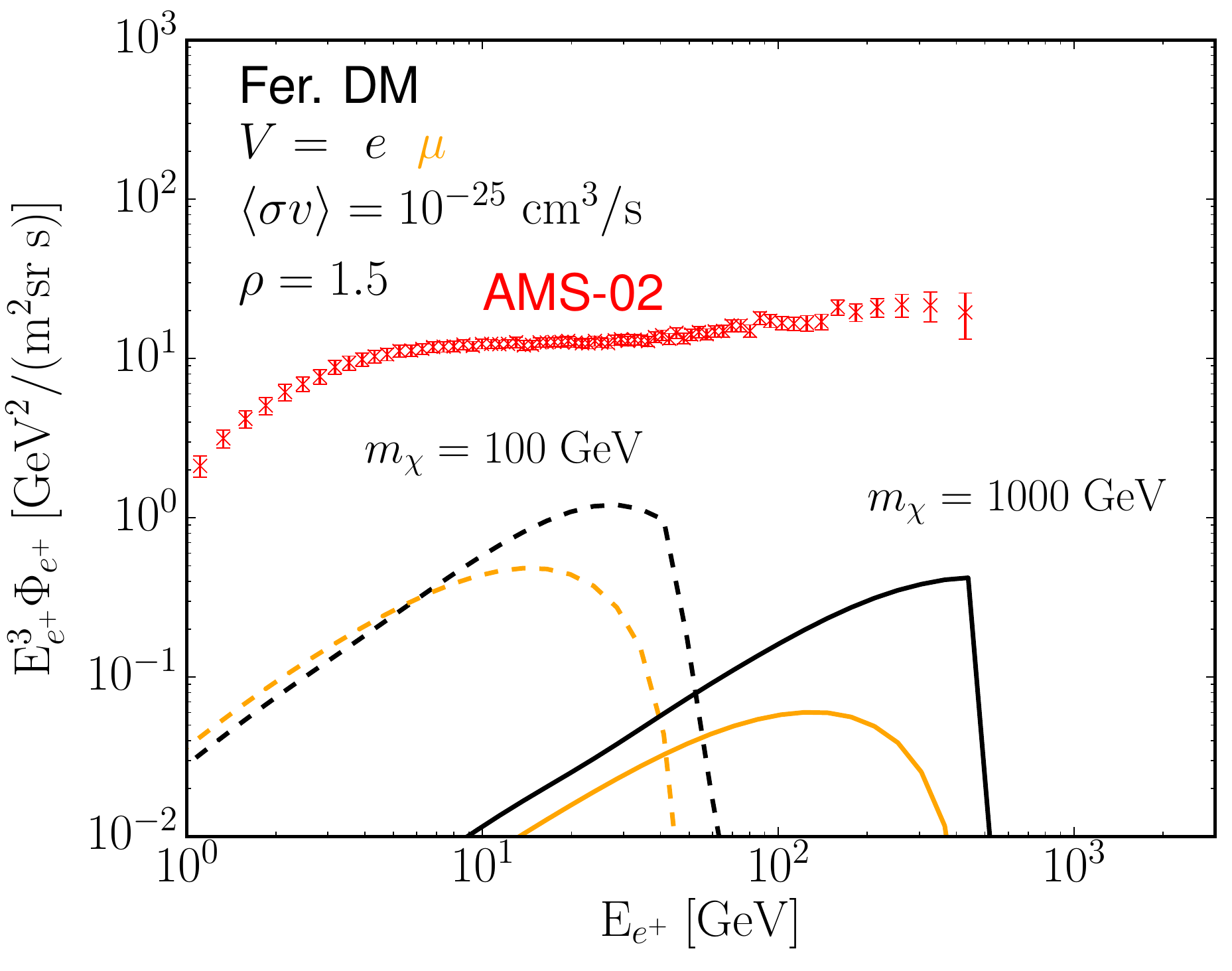}
\caption{Positron flux at the Earth without solar modulation for operators with light charged letpons and
scalar (fermion) dark matter on the left (right) pannel.
The solid (dashed) black and orange lines are fluxes for operator containing
eletron and muon respectively with dark matter mass 1000 GeV (100 GeV).   
The AMS-02 data with errors are shown in red. The annihilation cross section is set to 
$\langle \sigma v\rangle = 10^{-25} {\rm cm}^3/{\rm s}$. }
\label{fig:fluxams}
\end{figure}

\vspace{0.5cm}
\noindent
\textbf{Light charged leptons:}
The light charged leptons discussed here refer to only electrons and muons.
The relevant operators are the same as for neutrinos except $\mathcal{O}_{6B}^{\nu L}$.
The $s$-wave annihilation cross section can also be read directly from \modeqref{eqn:sv5Bs} and
\modeqsref{eqn:sv6BfpsiLL}--\eqref{eqn:rln3}.
The dark partner $\psi$ for scalar DM decays through  $\psi \to \phi^\dagger l_i$, 
while for fermion DM $\psi$ decays through  $\psi \to \bar{\chi} l_i \nu_{l_i}$.
Both the charged leptons and neutrinos from the dark partner decays are much softer than the charged lepton directly from 
the dark matter annihilation, similar to the 
situation with neutrinos shown in \figref{fig:nuspec}.
They can be safely ignored to obtain conservative limits from Planck.
We plot the positron flux $E_{e^+} \Phi_{e^+}$ after propagation at the Earth without solar modulation in 
\figref{fig:fluxams}.

\begin{figure}
\centering
\includegraphics[width=0.48\textwidth]{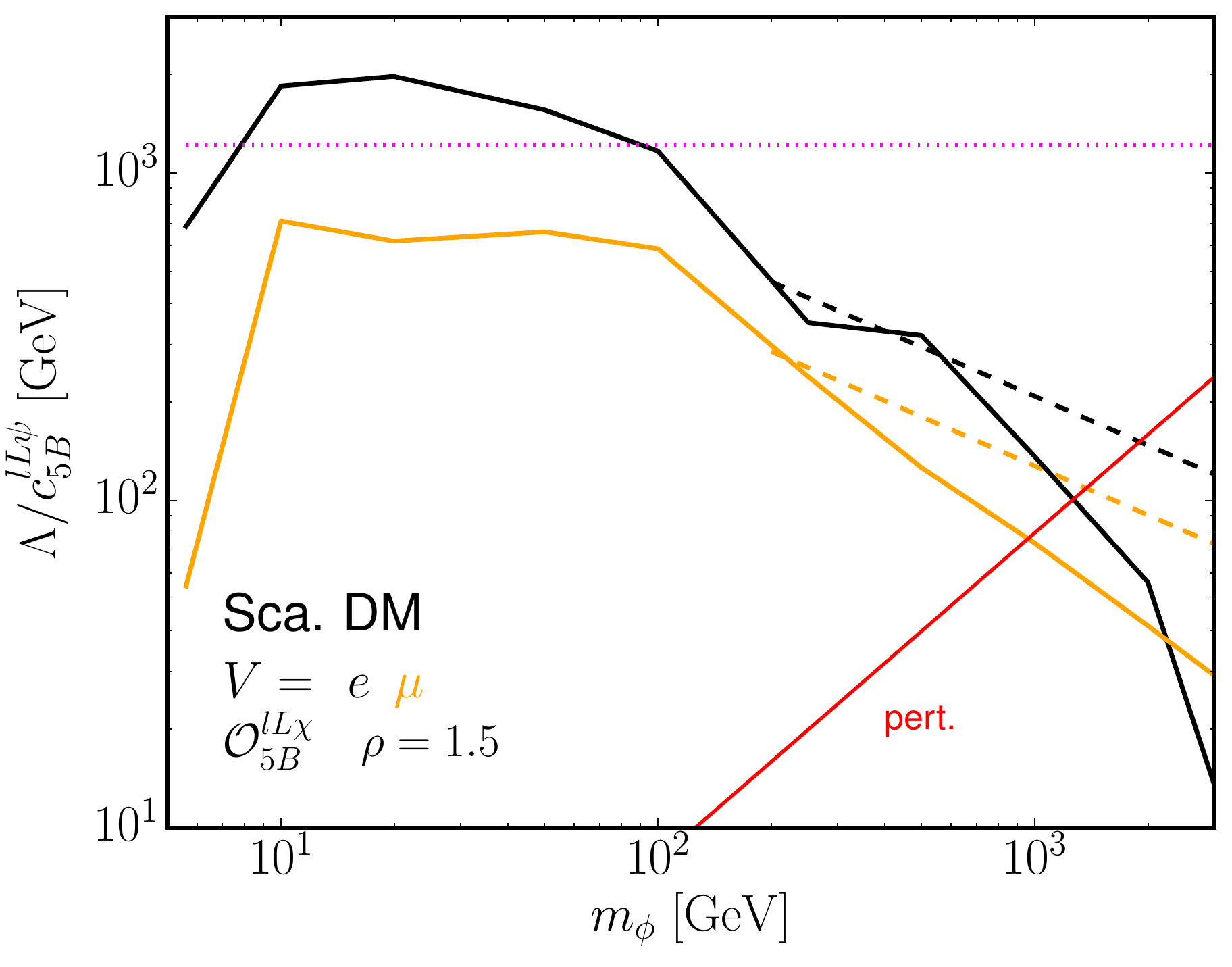}
\hfill
\includegraphics[width=0.49\textwidth]{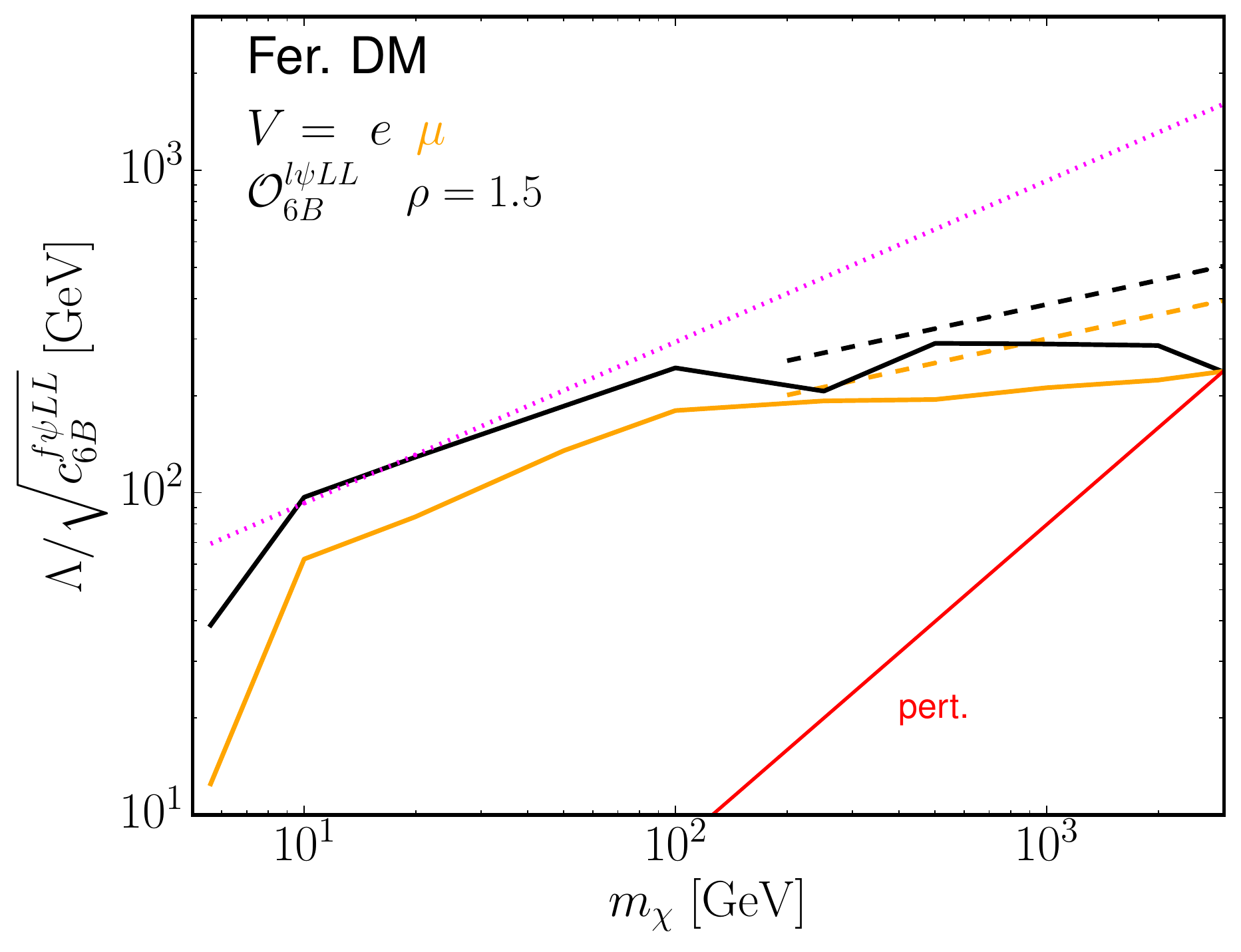}
\caption{Limits from AMS (CMB) are shown in black and orange solid (dashed) lines for
operators with electron and muon respectively for scalar dark matter in the left panel 
and fermion dark matter in the right panel. The ratio of the dark partner mass and the dark matter 
mass is again set to $\rho=1.5$. Dotted magenta lines in both panels denote the thermal relic cross section, 
while the red solid line is the perturbativity limit. }
\label{fig:llimit}
\end{figure}

For operators with all flavors of leptons, Planck imposes strong constraints at higher DM masses. 
For operators with light leptons, i.e. electron and muon, the most stringent constraint for lower DM masses 
come from AMS-02.
We plot the limits for operators with light leptons in \figref{fig:llimit}.  For SA to electrons, we can exclude or nearly exclude the thermal relic cross sections for 10~GeV~$\lesssim m_{\text{DM}} \lesssim 100$~GeV, while the limits on muons are about a factor of 2 weaker.

\vspace{0.5cm}
\noindent
\textbf{Quarks and Tau:}
As tau decays predominantly to hadronic final states, 
its treatment is similar to quarks in indirect detection of dark matter.   
The relevant operators are the same as the ones with charged leptons (with $f$ replaced by quarks or tau) 
and the $s$-wave annihilation cross sections for operators with quarks 
should 
be multiplied by a color factor of $3$.
Similarly the dark partner $\psi$ for scalar DM will decay through $\psi\to \phi^\dagger q$, where $q$ is a SM quark. 
For fermion DM, the dark partner $\psi$ will decay as $\psi \to \bar{\chi} d_i d_i$ 
($\psi \to \bar{\chi} u_i d_i$) if the SA operator contains $u_i$ or $d_i$.      

\begin{figure}[htp]
\centering
\includegraphics[width=0.48\textwidth]{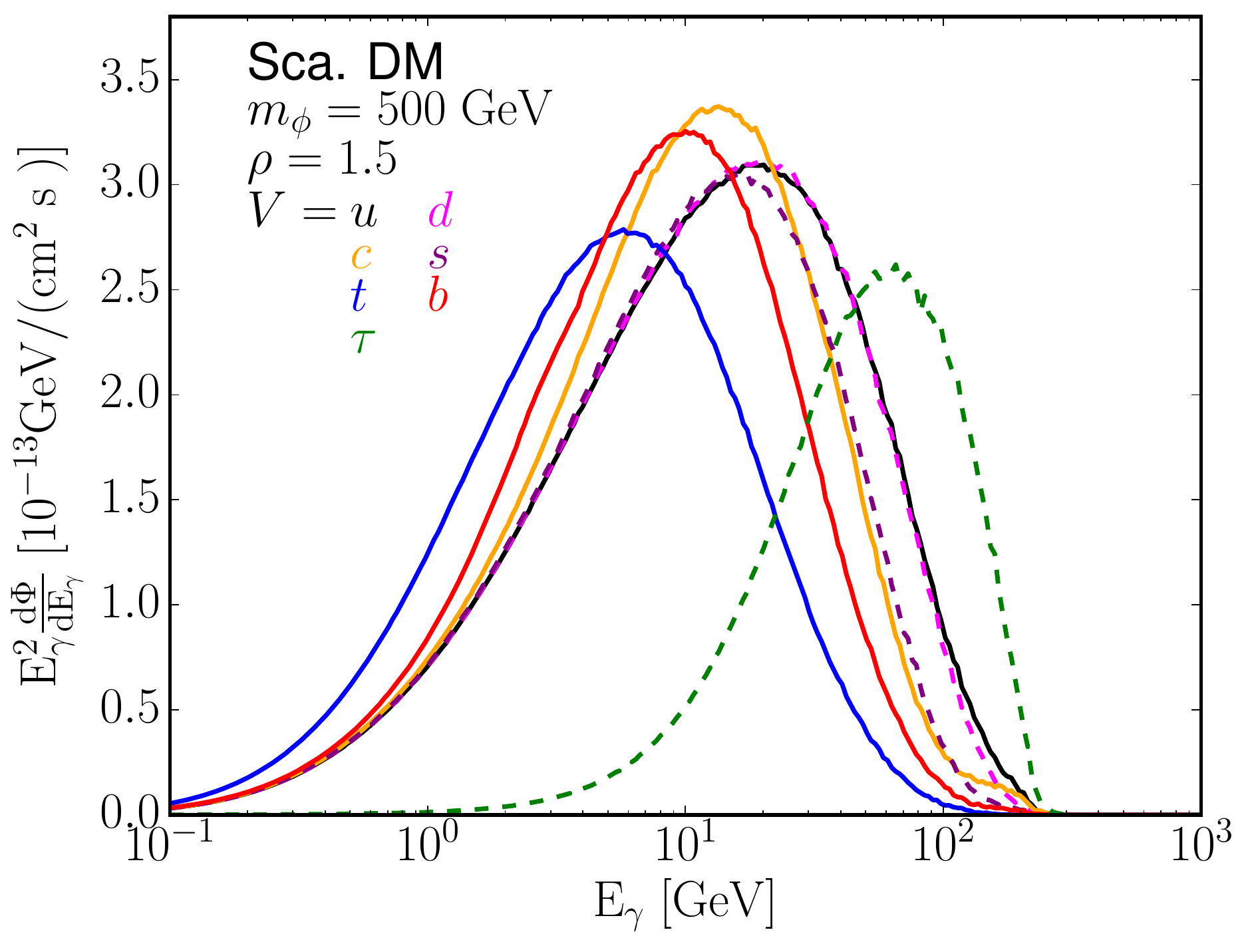}
\hfill
\includegraphics[width=0.48\textwidth]{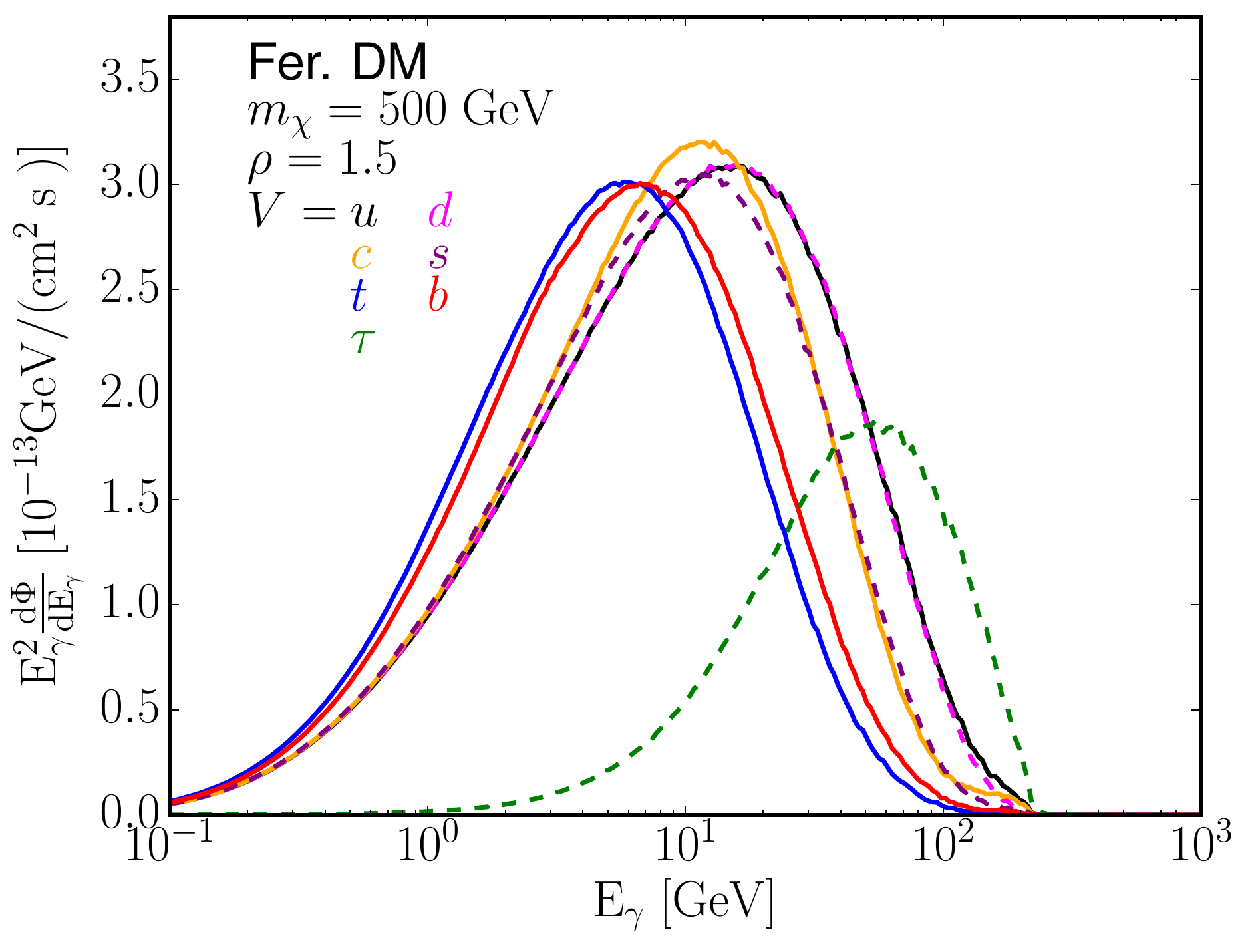}
\caption{Differential flux per annihilation for scalar (fermion) dark matter perators containing quarks and $\tau$
in the left (right) panel.
The dark matter mass is set to 500 GeV and $\rho=1.5$.}
\label{fig:qflux}
\end{figure}

For quarks and tau, cosmic gamma ray searches impose the most stringent constraints. Therefore,
we will focus on the Fermi-LAT limits and the CTA projected sensitivites.
In \figref{fig:qflux} we show the differential gamma ray spectra for operators with scalar DM on the left
and fermion DM on the right.
As we can see, the spectra for operators with the lightest quarks $u$ and $d$ are almost identical.
In general, operators with quarks have  similar spectra and thus similar limits on the cross section. 
Therefore we will show the limits from Fermi-LAT and projected sensitivities of CTA only for $u$, $b$ and $t$, 
to avoid over-crowdedness in the figures.  The limits for $u$ and $b$ quarks ($\tau$) exclude the thermal relic cross sections for $m_{\text{DM}} \lesssim$~100~GeV (50~GeV).

\begin{figure}[htp]
\centering
\includegraphics[width=0.48\textwidth]{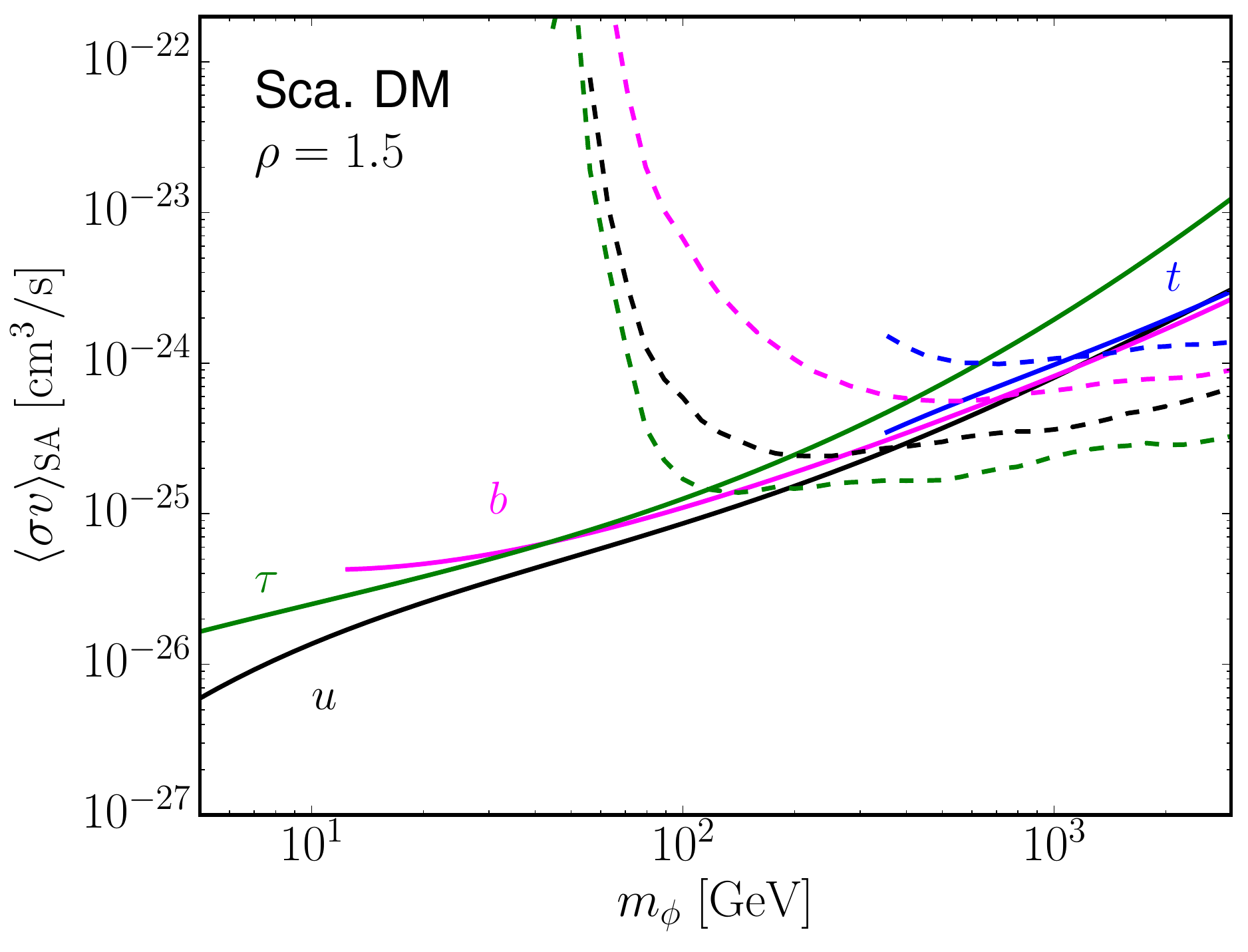}
\hfill
\includegraphics[width=0.48\textwidth]{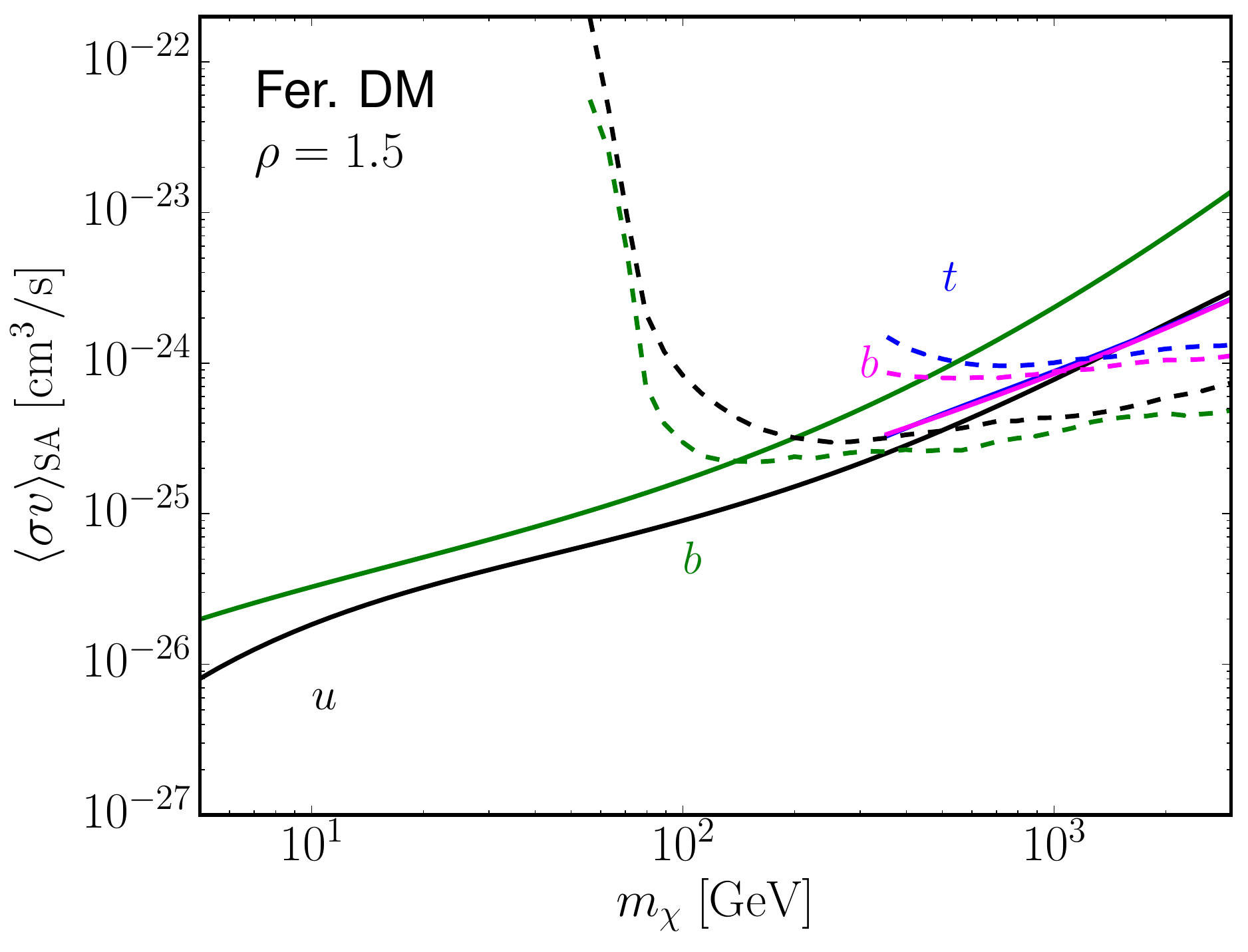}
\caption{Limits from Fermi-LAT observations and future sensitivies of CTA on the semi-annihilation cross sections 
of scalar (fermion) dark matter operators with selected quark flavors in solid  and dashed lines in the
left (right) panel.
The results for $u$, $b$, $t$ and $\tau$ are drawn in black, magenta, blue and green. 
The mass ratio between the dark partner and the DM is set to 1.5.}
\label{fig:qcrox}
\end{figure}

\begin{figure}[htp]
\centering
\includegraphics[width=0.48\textwidth]{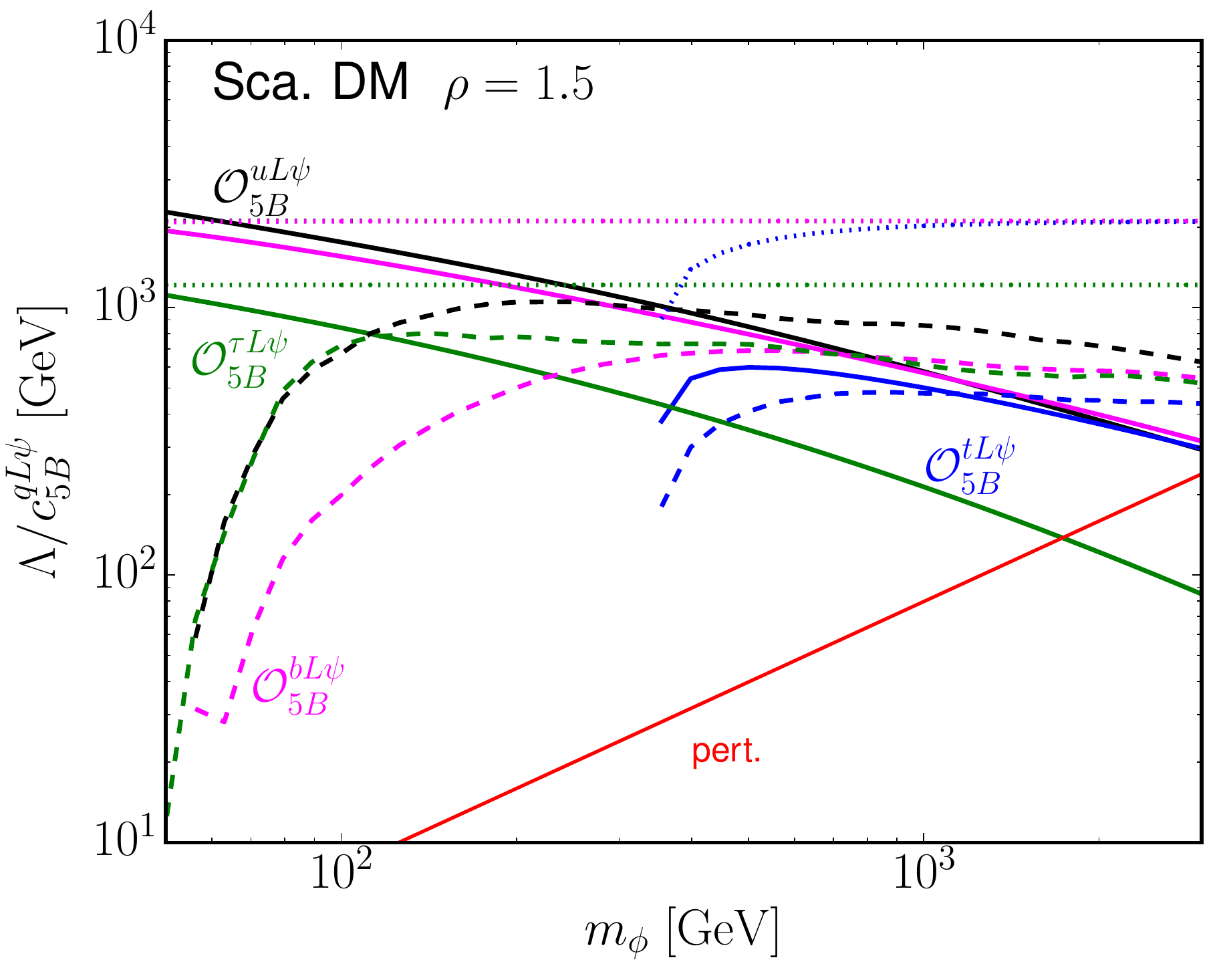}
\hfill
\includegraphics[width=0.48\textwidth]{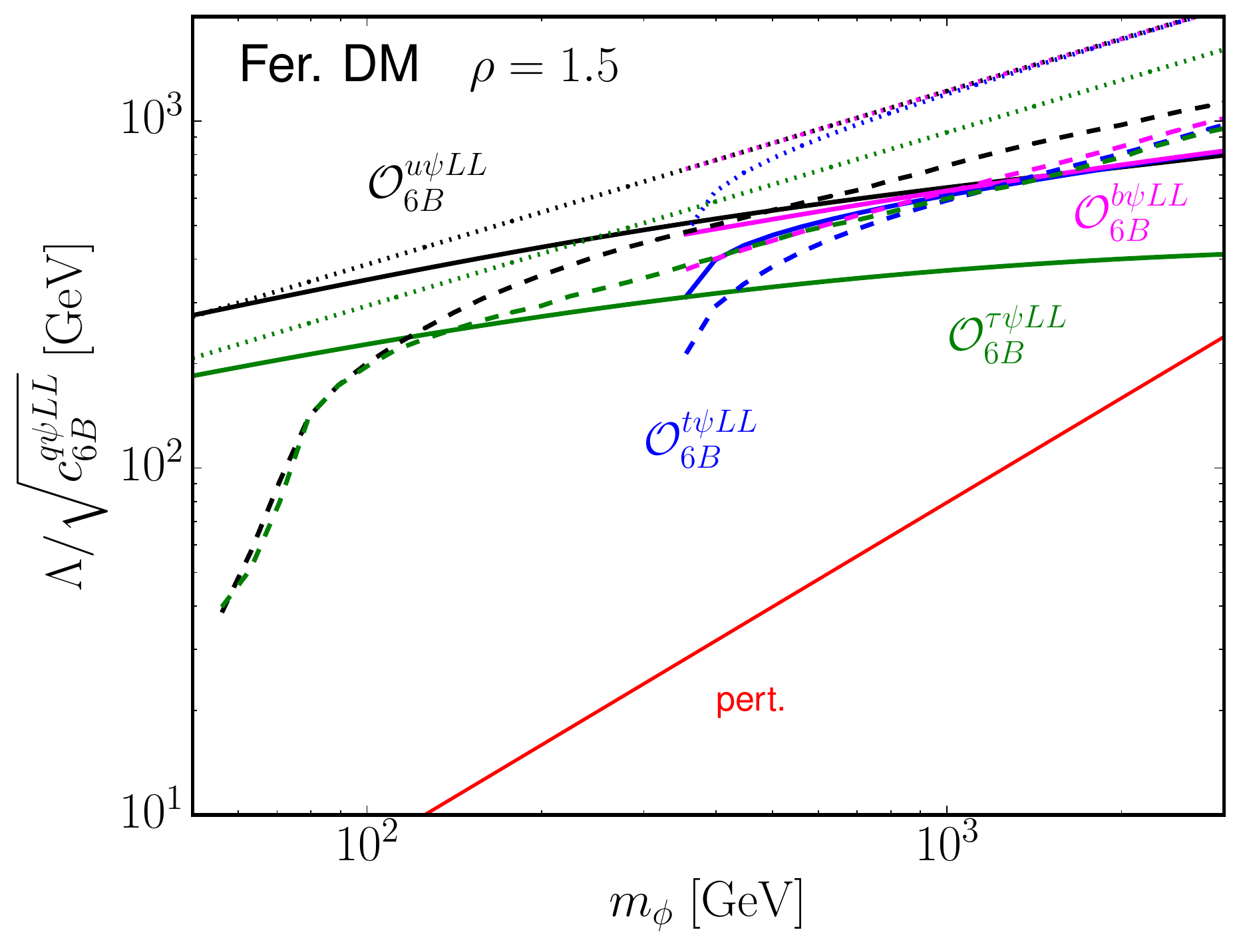}
\caption{Limits from Fermi-LAT and sensitivities from CTA on the coefficients of scalar (fermion)
dark matter operators with selected quark flavor in solid and dashed line respectively in the left (right) panel. 
The results for $u$, $b$, $t$ and $\tau$ are drawn in black, magenta, blue and green. 
The mass ratio between the dark partner and the DM is set to 1.5.
The dotted lines denote the thermal relic cross sections. 
The perturbativity limit is shown as the red solid line. }
\label{fig:qlimit}
\end{figure}

\section{Conclusions}\label{sec:conc}

As constraints from direct and collider searches become more severe, semi-annihilating dark matter becomes more attractive due to its resistance to these bounds.  It therefore behooves us to explore the range of possible SA phenomenology: to understand what limits exist, what signals can be searched for, and what models still survive.  In this work, we have made an initial effort to answer these questions by applying tools familiar from model-independent studies of annihilating DM to the SA paradigm.  By assuming that mediator particles connecting the DM to the SM are heavy, we can construct an effective theory with non-renormalisable operators as the leading coupling between the two.  Indeed, SA typically requires these mediators to carry dark sector charges, so that they \emph{must} be more massive than the DM.  The most phenomenologically relevant processes are given by those operators with lowest dimension.  The number of possible operators up to a given dimension is finite, making a study of the general theory space tractable.

Our results fall into three main categories.  The first is the enumeration of SA operators where the only additional dark sector fields are DM, as given in \tabsref{tab:ScaDM} through~\ref{tab:MixDMUnb} in \secref{sec:DMonly}.  The possible model space is tightly constrained; since DM must be colourless and neutral, only neutral SM fields can be produced in $2\to 2$ SA.  We also found that photon and gluon final states require at least three dark matter fields, making signals likely suppressed by DM fractional abundances but raising the possibility of observing multiple gamma-ray lines.  With one exception, all EW symmetric phase operators we found involve five or more fields.  They therefore also lead to $2\to 3$ SA processes, but these will be subleading for $m \lesssim 3$~TeV.  If we impose that DM consist of only a single field, only two possibilities exist within our assumptions, given by the operators $\op_{5U}^{H}$ of \tabref{tab:ScaDM} and $\op_{7U}^{LL}$ of \tabref{tab:FerDM}.  The former is closely related to the minimal model of scalar SADM, while the latter represents a new model not discussed in previous literature.

Our second set of results is the enumeration of models with dark partners, additional light unstable states in the dark sector which may carry SM gauge charges.  The relevant operators here are listed in \tabsref{tab:SdmBro} through~\ref{tab:MixdmUnb} in \secref{sec:DP}.  The addition of the dark partners substantially modifies the model space.  All SM fields can now appear as final states, so that the SA phenomenology itself is greatly expanded.  Even for processes that can appear in DM-only models, the dark partner operators often have lower dimension, leading to enhanced cross sections.  The dark partners can be produced at colliders, opening up a new range of search channels.  Since the dark partners must be light enough to be produced in SA, this also constrains the DM mass.  Most importantly, the required instability of the dark partners demands the EFT be extended with at least one additional operator to mediate that decay.  This introduces an additional model-dependence into the phenomenology, with the choice of decay mode affecting both collider limits and indirect detection signals.

Finally, in \secref{sec:ID} we gave the current and near future constraints on these effective operators from searches in gamma rays, positrons, and neutrinos.  To be concrete, we considered only operators consistent with a unique DM state plus possible dark partners, and assume that particle saturates the observed cosmological abundance.  We considered limits from observation of gamma rays, positrons, and neutrinos, as well as modifications to the CMB anisotropies.  The lack of any unambiguous observation translates to lower bounds on the UV cut-off $\Lambda$.  We expressed the limits in terms of operators in the EW broken phase, to ease the application of these limits to other models of SADM.  We also compared the limits we found to the operator coefficients necessary to produce the observed DM relic abundance from SA alone.  For SA to gauge bosons, Higgses, and neutrinos, the current and near-future limits are insufficient to exclude the thermal relic cross sections, except possibly near threshold.  In contrast, pure SA to charged leptons or quarks is either excluded or nearly so unless the DM mass $m \gtrsim \order{100}$~GeV, with the exact limit depending on the operator and final state.

While constraints on SADM are important to understand, the ultimate hope is that we can make a discovery and further distinguish these models from those of annihilating dark matter.  In this direction, models with dark partners are particularly promising.  In addition to the prospect of producing dark partners directly at colliders, their decays can leave direct imprints on cosmic ray spectra.  This is clearly shown in \emph{e.g.} \figref{fig:nuspec} for the case of neutrinos.  In the cases we have studied here, direct production of visible states through SA would be discovered first, and the decay products then serve as confirmation of the model.  However, the reverse is also possible when the dark partner is relatively heavy, $m_\dkp \sim 2m$.  Alternatively, correlating SA and annihilation signals could also serve this role.  This looks most promising when they appear in different channels, \emph{e.g.} SA into neutrinos or positrons and Higgs-portal annihilation to gamma-rays.

\section*{Acknowledgements}

This work was supported by IBS under the project code, IBS-R018-D1.  YC was supported by the Australian Research Council.

\appendix

\section{Tensor Symmetry Properties}\label{app:TensSym}

Let $n^{ij}$ be an arbitrary tensor with $N^2$ components,  $i,j = 1, \ldots, N$.  As is well-known, we can construct two tensors with definite symmetry properties:
\begin{align}
	s^{ij} & \equiv \frac{1}{2} (n^{ij} + n^{ji}) \qquad \text{with } \frac{1}{2} N (N +1) \text{ components,} \\
	a^{ij} & \equiv \frac{1}{2} (n^{ij} - n^{ji}) \qquad \text{with } \frac{1}{2} N (N -1) \text{ components.}
\end{align}
Now let us consider $n^{ijk}$ an arbitrary tensor with $N^3$ components.  We can construct tensors of definite symmetry properties as follows:
\begin{align}
	s^{ijk} & \equiv \frac{1}{6} (n^{ijk} + n^{ikj} + n^{kij} + n^{kji} + n^{jki} + n^{jik}) && \text{with } \frac{1}{6} N (N +1) (N+2) \text{ components,} \\
	a^{ijk} & \equiv \frac{1}{6} (n^{ijk} - n^{ikj} + n^{kij} - n^{kji} + n^{jki} - n^{jik}) && \text{with } \frac{1}{6} N (N -1) (N - 2) \text{ components,} \\
	x^{ijk} & \equiv \frac{1}{6} (n^{ikj} + n^{kij} - n^{kji} - n^{jki}) && \text{with } \frac{1}{3} N (N^2 -1) \text{ components,} \\
	y^{ijk} & \equiv \frac{1}{6} (n^{ikj} - n^{kij} - n^{kji} + n^{jki}) && \text{with } \frac{1}{3} N (N^2 -1) \text{ components.}
\end{align}
These objects are linearly independent and have $N^3$ components, and hence form a complete basis for three-index tensors.  In particular, there is the inversion formula
\begin{equation}
	n^{ijk} = s^{ijk} + a^{ijk} + x^{ikj} - x^{kji} + y^{ikj} - y^{kji} \,.
\end{equation}
In addition to the familiar completely (anti)-symmetric matrices $a$ and $s$, the matrices $x$ and $y$ have the symmetry properties
\begin{align}
	x^{ijk} &= - x^{jik} \,, & x^{ijk} + x^{jki} + x^{kij} &= 0 \,, \\
	y^{ijk} &= y^{jik} \,, & y^{ijk} + y^{jki} + y^{kij} &= 0 \,.
\end{align}
When there are repeated indices, we have the following symmetry rules for $x$ and $y$:
\begin{align}
  x^{iii} & = 0 \,, & x^{ijj} & = - x^{jij} \,, & x^{jji} & = 0 \,, \\
  y^{iii} & = 0 \,, & y^{ijj} & = y^{jij} = - \frac{1}{2} \, y^{jji} \,.
\end{align}
Let $\mathcal{S}_{ijk}$ ($\mathcal{A}_{ijk}$) be an arbitrary matrix (anti)-symmetric under $i \leftrightarrow j$.  The non-trivial contractions of these matrices with $x$ and $y$ satisfy
\begin{equation}
	y^{ikj} \mathcal{S}_{ijk} = - \frac{1}{2} \, y^{ijk} \mathcal{S}_{ijk} \,, \qquad x^{ikj} \mathcal{A}_{ijk} = \frac{1}{2} \, x^{ijk} \mathcal{A}_{ijk} \,.
\end{equation}

\section{Additional Operators at Dimension 7 and 8}\label{app:HigherD}

In this appendix in \tabsref{tab:d8MixDMonly}--\ref{tab:d8MixDM} we give some additional dimension 7 and 8 operators in the EW unbroken phase to supplement those already listed in \secsref{sec:DMonly} and \ref{sec:DP}.  These terms fall into two categories.  Some are the leading operators needed to generate broken phase operators of dimension 6 and below, where the broken phase operators are not the dominant contributions to any SA process.  The remainder generate broken phase operators individually that are only produced in groups by the terms listed previously.  For operators including dark partners, we do not list decay operators for brevity, but they are easily constructed following the procedures discussed in \secref{sec:DP}.

For theories without dark partners, the only case where we have new operators are when there are both scalar and fermion DM.  We find eight such operators, all at dimension 8, which we list in \tabref{tab:d8MixDMonly}.  The first four lines are the leading generators of four dimension 6 broken phase operators.  The last two lines are needed produce the broken phase couplings to the photon and $Z$ field strength tensors with arbitrary coefficients.  (Strictly, only $\brtilde{\op}_{8U}^W$ plus the dimension-6 term of \tabref{tab:MixDMUnb} are required, but we list both dimension-8 terms.)

For theories with dark partners, we list operators for scalar DM, fermion DM, and both in \tabsref{tab:d8ScaDM}, \ref{tab:d8FerDM}, and~\ref{tab:d8MixDM}, respectively.  For scalar DM, we find 6 (12) relevant terms for scalar (fermion) dark partners, all at dimension 7.  None of these are the leading contributions to any broken phase operators, but rather allow us to generate certain broken phase operators uniquely.  We find four dimension-7 operators with scalar dark partners that are the leading generators of terms in \tabref{tab:FdmBro}.  There are an additional 17 (48) operators with scalar (fermion) dark partners that allow us to generate broken phase terms individually.  Lastly there are four dimension-7 operators with fermion dark partners that are the leading generators of terms in \tabref{tab:MixdmBro}; plus a further 12 (17) operators with scalar (fermion) dark partners that allow us to generate broken phase terms individually.  

\begin{table}
	\centering
	\begin{tabular}{|c|c|c|}
		\hline
		Operator & Definition & Broken Phase \\
		\hline
		$\op_{8U}^{ZsS}$ & $s^{ij} \bigl( \dms \, \overleftrightarrow{\partial_\mu} (\bar{\dmf}_i^c \dmf_k) \bigr) \bigl(i H^\dagger \Dlr{\mu} H\bigr)$ & $- \frac{v m_Z}{\Lambda^2} \, \op_{6B}^{ZsS}$ \\
		$\op_{8U}^{ZsP}$ & $s^{ij} \bigl( \dms \, \dlr{\mu} (\bar{\dmf}_i^c \gamma^5 \dmf_k) \bigr) \bigl(i H^\dagger \Dlr{\mu} H\bigr)$ & $- \frac{v m_Z}{\Lambda^2} \, \op_{6B}^{ZsP}$ \\
		$\op_{8U}^{ZaS}$ & $a^{ij} (\bar{\dmf}_i^c \partial^\mu \dmf_j) \, \dms \, \bigl(i H^\dagger \Dlr{\mu} H\bigr)$ & $- \frac{v m_Z}{\Lambda^2} \, \op_{6B}^{ZaS}$ \\
		$\op_{8U}^{ZaP}$ & $a^{ij} (\bar{\dmf}_i^c \gamma^5 \partial^\mu \dmf_j) \, \dms \, \bigl(i H^\dagger \Dlr{\mu} H\bigr)$ & $- \frac{v m_Z}{\Lambda^2} \, \op_{6B}^{ZaP}$ \\
		$\brtilde{\op}{}_{8U}^{B}$ & $a^{ij} \, \bar{\dmf}_i^c \sigma^{\mu\nu} \dmf_j \, \dms \, \brtilde{B}_{\mu\nu} (H^\dagger H)$ & $\frac{v^2}{2 \Lambda^2} (c_W \brtilde{\op}{}_{6B}^{\gamma} - s_W \brtilde{\op}{}_{6B}^{Z})$ \\
		$\brtilde{\op}{}_{8U}^{W}$ & $a^{ij} \, \bar{\dmf}_i^c \sigma^{\mu\nu} \dmf_j \, \dms \, (H^\dagger \brtilde{W}_{\mu\nu}^a \sigma^a H)$ & $\frac{v^2}{2 \Lambda^2} (s_W \brtilde{\op}{}_{6B}^{\gamma} + c_W \brtilde{\op}{}_{6B}^{Z})$ \\
		\hline
	\end{tabular}
	\caption{Dimension 8 operators for scalar and fermion DM with no dark partners.}\label{tab:d8MixDMonly}
\end{table}

\begin{table}
	\centering
	\begin{tabular}{|c|c|c|c|}
		\hline
		Operator & Definition & $\dps$/$\dpf$ & Broken Phase \\
		\hline
		\hline
		$\op_{7U}^{Z\dps}$ & $s^{ij} \, \dms_i \dms_j (\partial_\mu \dps) \, \bigl(i H^\dagger \Dlr{\mu} H\bigr) $ & (1, 1, 0) & $- \frac{v m_Z}{\Lambda^2} \, \op_{5B}^{Z\dps}$ \\
		$\op_{7U}^{Z\dms}$ & $a^{ij} \, \dps \, \dms_i (\partial_\mu \dms_j) \, \bigl(i H^\dagger \Dlr{\mu} H\bigr) $ & (1, 1, 0) & $- \frac{v m_Z}{\Lambda^2} \, \op_{5B}^{Z\dms}$ \\
		$\op_{7U}^{W\dps}$ & $s^{ij} \, \dms_i \dms_j (D_\mu \dps) \, \bigl(i H^\dagger \Dlr{\mu} \tilde{H} \bigr) $ & (1, 1, 1) & $\frac{\sqrt{2}\,v m_W}{\Lambda^2} \, \op_{5B}^{W\dps}$ \\
		$\op_{7U}^{W\dms}$ & $a^{ij} \, \dps \, \dms_i (\partial_\mu \dms_j) \, \bigl(i H^\dagger \Dlr{\mu} \tilde{H} \bigr) $ & (1, 1, 1) & $\frac{\sqrt{2}\,v m_W}{\Lambda^2} \, \op_{5B}^{W\dps}$ \\
		$\op_{7U}^{H \partial D}$ & $a^{ij} \, \dms_i \, (\partial^\mu \dms_j) \, \bigl( (H^\dagger H) \dlr{\mu} \dps)$ & (1, 1, 0) & $\frac{v}{\Lambda} \, \op_{6B}^{h\partial\dps}$ \\
		$\op_{7U}^{H D^2}$ & $s^{ij} \, \dms_i \dms_j \, (\partial_\mu \dps) \bigl( \partial^\mu (H^\dagger H) \bigr)$ & (1, 1, 0) & $\frac{v}{\Lambda} \, \op_{6B}^{\partial h\dps}$ \\
		\hline
		\hline
		$\op_{7U}^{F_d\dpf}$ & $a^{ij} \dms_i \partial_\mu \dms_j \, \bigl((F^\dagger H) \bar{\sigma}^\mu \dpfL\bigr)$ & ($R_F$, 1, $Y_F - \frac{1}{2}$) & $\frac{v}{\sqrt{2}\Lambda} \, \op_{6B}^{F_dL\dpf}$ \\
		$\op_{7U}^{F_u\dpf}$ & $a^{ij} \dms_i \partial_\mu \dms_j \, \bigl((F^\dagger \tilde{H}) \bar{\sigma}^\mu \dpfL\bigr)$ & ($R_F$, 1, $Y_F + \frac{1}{2}$) & $\frac{v}{\sqrt{2}\Lambda} \, \op_{6B}^{F_uL\dpf}$ \\
		\hline
	\end{tabular}
	\caption{Dimension 7 operators for scalar DM with scalar dark partners $\dps$ (top) and fermion dark partners $\dpf = (\dpfL, \dpfR)^T$ (bottom).  In the latter case, $F = (F_u, F_d)$ is an $SU(2)$-doublet fermion in the gauge representation $(R_F, 2, Y_F$).}\label{tab:d8ScaDM}
\end{table}

\begin{table}
	\centering
	\begin{tabular}{|c|c|c|c|}
		\hline
		Operator & Definition & $\dps$/$\dpf$ & Broken Phase \\
		\hline
		\hline
		$\op_{7U}^{H\dps V}$ & $a^{ij} \, \bar{\dmf}_i^c \gamma^\mu \dmf_j \, \bigl( \dps \dlr{\mu} (H^\dagger H) \bigr)$ & (1, 1, 0) & $\frac{v}{\Lambda} \, \op_{6B}^{h\dps V}$ \\
		$\op_{7U}^{H\dps A}$ & $s^{ij} \, \bar{\dmf}_i^c \gamma^\mu \gamma^5 \dmf_j \, \bigl( \dps \dlr{\mu} (H^\dagger H) \bigr)$ & (1, 1, 0) & $\frac{v}{\Lambda} \, \op_{6B}^{h\dps A}$ \\
		$\op_{7U}^{Z\dps V}$ & $a^{ij} \, \bar{\dmf}_i^c \gamma^\mu \dmf_j \, \dps \bigl( i H^\dagger \Dlr{\mu} H \bigr)$ & (1, 1, 0) & $- \frac{v m_Z}{\Lambda^2} \, \op_{5B}^{Z\dps V}$ \\
		$\op_{7U}^{Z\dps A}$ & $s^{ij} \, \bar{\dmf}_i^c \gamma^\mu \gamma^5 \dmf_j \, \dps \bigl( i H^\dagger \Dlr{\mu} H \bigr)$ & (1, 1, 0) & $- \frac{v m_Z}{\Lambda^2} \, \op_{5B}^{Z\dps A}$ \\
		$\op_{7U}^{W\dps V}$ & $a^{ij} \, \bar{\dmf}_i^c \gamma^\mu \dmf_j \, \dps \bigl( i H^\dagger \Dlr{\mu} \tilde{H} \bigr)$ & (1, 1, 1) & $\frac{\sqrt{2}\,v m_W}{\Lambda^2} \, \op_{5B}^{W\dps V}$ \\
		$\op_{7U}^{W\dps A}$ & $s^{ij} \, \bar{\dmf}_i^c \gamma^\mu \gamma^5 \dmf_j \, \dps \bigl( i H^\dagger \Dlr{\mu} \tilde{H} \bigr)$ & (1, 1, 1) & $\frac{\sqrt{2}\,v m_W}{\Lambda^2} \, \op_{5B}^{W\dps A}$ \\
		$\op_{7U}^{\V\dps sS}$ & $s^{ij} \, (i D^\mu H^\dagger) \bigl( \dps \Dlr{\mu} (\bar{\dmf}_i^c \dmf_j) \bigr)$ & (1, 2, $\frac{1}{2}$) & $- \frac{m_Z}{\sqrt{2}\Lambda} \, \op_{6B}^{Z\dps sS} + \frac{m_W}{\Lambda} \, \op_{6B}^{W\dps sS}$ \\
		$\op_{7U}^{\V\dps sP}$ & $s^{ij} \, (i D^\mu H^\dagger) \bigl( \dps \Dlr{\mu} (\bar{\dmf}_i^c \gamma^5 \dmf_j) \bigr)$ & (1, 2, $\frac{1}{2}$) & $- \frac{m_Z}{\sqrt{2}\Lambda} \, \op_{6B}^{Z\dps sP} + \frac{m_W}{\Lambda} \, \op_{6B}^{W\dps sP}$ \\
		$\op_{7U}^{\V\dps aS}$ & $a^{ij} \, (\bar{\dmf}_i^c \partial_\mu \dmf_j) \, \bigl( i H^\dagger \Dlr{\mu} \dps\bigr)$ & (1, 2, $\frac{1}{2}$) & $- \frac{\sqrt{2}\,m_Z}{\Lambda} \, \op_{6B}^{Z\dps aS} + \frac{2 m_W}{\Lambda} \, \op_{6B}^{W\dps aS}$ \\
		$\op_{7U}^{\V\dps aP}$ & $a^{ij} \, (\bar{\dmf}_i^c \gamma^5 \partial_\mu \dmf_j) \, \bigl( i H^\dagger \Dlr{\mu} \dps\bigr)$ & (1, 2, $\frac{1}{2}$) & $- \frac{\sqrt{2}\,m_Z}{\Lambda} \, \op_{6B}^{Z\dps aP} + \frac{2 m_W}{\Lambda} \, \op_{6B}^{W\dps aP}$ \\
		\hline
		$\op_{8U}^{Z\dps sS}$ & $s^{ij} \, \bigl( \dps \dlr{\mu} (\bar{\dmf}_i^c \dmf_j) \bigr) \bigl(i H^\dagger \Dlr{\mu} H\bigr)$ & (1, 1, 0) & $- \frac{v m_Z}{\Lambda^2} \, \op_{6B}^{Z\dps sS}$ \\
		$\op_{8U}^{Z\dps sP}$ & $s^{ij} \, \bigl( \dps \dlr{\mu} (\bar{\dmf}_i^c \gamma^5 \dmf_j) \bigr) \bigl(i H^\dagger \Dlr{\mu} H\bigr)$ & (1, 1, 0) & $- \frac{v m_Z}{\Lambda^2} \, \op_{6B}^{Z\dps sP}$ \\
		$\op_{8U}^{Z\dps aS}$ & $a^{ij} \, (\bar{\dmf}_i^c \partial_\mu \dmf_j) \, \dps \bigl(i H^\dagger \Dlr{\mu} H\bigr)$ & (1, 1, 0) & $- \frac{v m_Z}{\Lambda^2} \, \op_{6B}^{Z\dps aS}$ \\
		$\op_{8U}^{Z\dps aP}$ & $a^{ij} \, (\bar{\dmf}_i^c \gamma^5 \partial_\mu \dmf_j) \, \dps \bigl(i H^\dagger \Dlr{\mu} H\bigr)$ & (1, 1, 0) & $- \frac{v m_Z}{\Lambda^2} \, \op_{6B}^{Z\dps aP}$ \\
		$\op_{8U}^{W\dps sS}$ & $s^{ij} \bigl( \dps \dlr{\mu} (\bar{\dmf}_i^c \dmf_j) \bigr) \, \bigl(i H^\dagger \Dlr{\mu} \tilde{H}\bigr)$ & (1, 1, 1) & $\frac{\sqrt{2}\,v m_W}{\Lambda^2} \, \op_{6B}^{W\dps sS}$ \\
		$\op_{8U}^{W\dps sP}$ & $s^{ij} \, \bigl( \dps \dlr{\mu} (\bar{\dmf}_i^c \gamma^5 \dmf_j) \bigr) \bigl(i H^\dagger \Dlr{\mu} \tilde{H}\bigr)$ & (1, 1, 1) & $\frac{\sqrt{2}\,v m_W}{\Lambda^2} \, \op_{6B}^{W\dps sS}$ \\
		$\op_{8U}^{W\dps aS}$ & $a^{ij} \, \bar{\dmf}_i^c \partial_\mu \dmf_j \, \dps \bigl(i H^\dagger \Dlr{\mu} \tilde{H}\bigr)$ & (1, 1, 1) & $\frac{\sqrt{2}\,v m_W}{\Lambda^2} \, \op_{6B}^{W\dps sS}$ \\
		$\op_{8U}^{W\dps aP}$ & $a^{ij} \, \bar{\dmf}_i^c \gamma^5 \partial_\mu \dmf_j \, \dps \bigl(i H^\dagger \Dlr{\mu} \tilde{H}\bigr)$ & (1, 1, 1) & $\frac{\sqrt{2}\,v m_W}{\Lambda^2} \, \op_{6B}^{W\dps sS}$ \\
		$\brtilde{\op}{}_{8U}^{B\dps}$ & $a^{ij} \, \bar{\dmf}_i^c \sigma^{\mu\nu} \dmf_j \, \dps \, \brtilde{B}_{\mu\nu} (H^\dagger H)$ & (1, 1, 0) & $\frac{v^2}{2 \Lambda^2} \, \bigl( c_W \op_{6B}^{\gamma\dps} - s_W \op_{6B}^{Z\dps} \bigr)$ \\
		$\brtilde{\op}{}_{8U}^{Z\dps}$ & $a^{ij} \, \bar{\dmf}_i^c \sigma^{\mu\nu} \dmf_j \, \dps \, (H^\dagger \brtilde{W}^a_{\mu\nu} \sigma^a H)$ & (1, 1, 0) & $\frac{v^2}{2 \Lambda^2} \, \bigl( s_W \op_{6B}^{\gamma\dps} + c_W \op_{6B}^{Z\dps} \bigr)$ \\
		$\brtilde{\op}{}_{8U}^{W\dps}$ & $a^{ij} \, \bar{\dmf}_i^c \sigma^{\mu\nu} \dmf_j \, \dps \, (H^\dagger \brtilde{W}^a_{\mu\nu} \sigma^a \tilde{H})$ & (1, 1, 1) & $\frac{v^2}{\sqrt{2}\Lambda^2} \, \op_{6B}^{W\dps}$ \\
		\hline
		\hline
		$\op_{7U}^{F_d\dpf L}$ & $s^{ij}\, (\dmfL_i \dmfL_j) \, \bigl((F^\dagger H) \dpfR\bigr)$ & ($R_F$, 1, $Y_F - \frac{1}{2}$) & $\frac{v}{\sqrt{2}\Lambda} \, \op_{6B}^{F_d\dpf L}$ \\
		$\op_{7U}^{F_d\dpf R}$ & $s^{ij} \, (\bar{\xi}_i^\dagger \bar{\xi}_j^\dagger) \, \bigl((F^\dagger H) \dpfR\bigr)$ & ($R_F$, 1, $Y_F - \frac{1}{2}$) & $\frac{v}{\sqrt{2}\Lambda} \, \op_{6B}^{F_d\dpf R}$ \\
		$\op_{7U}^{F_d\dmf L}$ & $(s^{ij} + a^{ij}) \, (\dmfL_i \dpfL) \, \bigl((F^\dagger H) \bar{\xi}_j^\dagger\bigr)$ & ($R_F$, 1, $Y_F - \frac{1}{2}$) & $\frac{v}{\sqrt{2}\Lambda} \, \op_{6B}^{F_d\dmf L}$ \\
		$\op_{7U}^{F_d\dmf R}$ & $(s^{ij} + a^{ij}) \, (\bar{\xi}_i^\dagger \dpfR) \, \bigl((F^\dagger H) \bar{\xi}_j^\dagger\bigr)$ & ($R_F$, 1, $Y_F - \frac{1}{2}$) & $\frac{v}{\sqrt{2}\Lambda} \, \op_{6B}^{F_d\dmf R}$ \\
		$\op_{7U}^{F_u\dpf L}$ & $s^{ij}\, (\dmfL_i \dmfL_j) \, \bigl((F^\dagger \tilde{H}) \dpfR\bigr)$ & ($R_F$, 1, $Y_F + \frac{1}{2}$) & $\frac{v}{\sqrt{2}\Lambda} \, \op_{6B}^{F_u\dpf L}$ \\
		$\op_{7U}^{F_u\dpf R}$ & $s^{ij} \, (\bar{\xi}_i^\dagger \bar{\xi}_j^\dagger) \, \bigl((F^\dagger \tilde{H}) \dpfR\bigr)$ & ($R_F$, 1, $Y_F + \frac{1}{2}$) & $\frac{v}{\sqrt{2}\Lambda} \, \op_{6B}^{F_u\dpf R}$ \\
		$\op_{7U}^{F_u\dmf L}$ & $(s^{ij} + a^{ij}) \, (\dmfL_i \dpfL) \, \bigl((F^\dagger \tilde{H}) \bar{\xi}_j^\dagger\bigr)$ & ($R_F$, 1, $Y_F + \frac{1}{2}$) & $\frac{v}{\sqrt{2}\Lambda} \, \op_{6B}^{F_u\dmf L}$ \\
		$\op_{7U}^{F_u\dmf R}$ & $(s^{ij} + a^{ij}) \, (\bar{\xi}_i^\dagger \dpfR) \, \bigl((F^\dagger \tilde{H}) \bar{\xi}_j^\dagger\bigr)$ & ($R_F$, 1, $Y_F + \frac{1}{2}$) & $\frac{v}{\sqrt{2}\Lambda} \, \op_{6B}^{F_u\dmf R}$ \\
		\hline
	\end{tabular}
	\caption{Dimension 7 and 8 operators for fermion DM $\dmf = (\dmfL, \dmfR)^T$ with scalar dark partners $\dps$ (top) and fermion dark partners $\dpf = (\dpfL, \dpfR)^T$ (bottom).  In the latter case, $F = (F_u, F_d)$ is an $SU(2)$-doublet fermion in the gauge representation $(R_F, 2, Y_F$).}\label{tab:d8FerDM}
\end{table}

\begin{table}
	\centering
	\begin{tabular}{|c|c|c|c|}
		\hline
		Operator & Definition & $\dps$/$\dpf$ & Broken Phase \\
		\hline
		\hline
		$\op_{7U}^{F_d\dps}$ & $\bigl( \dms \Dlr{\mu} \dps \bigr) \, \bigl( (F^\dagger H) \bar{\sigma}^\mu \dmfL \bigr)$ & ($R_F$, 1, $Y_F - \frac{1}{2}$) & $\frac{v}{\sqrt{2}\Lambda} \, \op_{6B}^{F_dL\dps}$ \\
		$\op_{7U}^{F_u\dps}$ & $\bigl( \dms \Dlr{\mu} \dps \bigr) \, \bigl( (F^\dagger \tilde{H}) \bar{\sigma}^\mu \dmfL \bigr)$ & ($R_F$, 1, $Y_F + \frac{1}{2}$) & $\frac{v}{\sqrt{2}\Lambda} \, \op_{6B}^{F_uL\dps}$ \\
		\hline
		\hline
		$\op_{7U}^{H\dpf V}$ & $\bar{\dmf}^c \gamma^\mu \dpf \, \bigl( \dms \dlr{\mu} (H^\dagger H) \bigr)$ & (1, 1, 0) & $\frac{v}{\Lambda} \, \op_{6B}^{h\dpf V}$ \\
		$\op_{7U}^{H\dpf A}$ & $\bar{\dmf}^c \gamma^\mu \gamma^5 \dpf \, \bigl( \dms \dlr{\mu} (H^\dagger H) \bigr)$ & (1, 1, 0) & $\frac{v}{\Lambda} \, \op_{6B}^{h\dpf A}$ \\
		$\op_{7U}^{Z\dpf V}$ & $\bar{\dmf}^c \gamma^\mu \dpf \, \dms \bigl( i H^\dagger \Dlr{\mu} H \bigr)$ & (1, 1, 0) & $- \frac{vm_Z}{\Lambda^2} \, \op_{5B}^{Z\dpf V}$ \\
		$\op_{7U}^{Z\dpf A}$ & $\bar{\dmf}^c \gamma^\mu \gamma^5 \dpf \, \dms \bigl( i H^\dagger \Dlr{\mu} H \bigr)$ & (1, 1, 0) & $- \frac{vm_Z}{\Lambda^2} \, \op_{5B}^{Z\dpf A}$ \\
		$\op_{7U}^{W\dpf V}$ & $\bar{\dmf}^c \gamma^\mu \dpf \, \dms \bigl( i H^\dagger \Dlr{\mu} \tilde{H} \bigr)$ & (1, 1, 1) & $\frac{\sqrt{2}\,vm_W}{\Lambda^2} \, \op_{5B}^{W\dpf V}$ \\
		$\op_{7U}^{W\dpf A}$ & $\bar{\dmf}^c \gamma^\mu \gamma^5 \dpf \, \dms \bigl( i H^\dagger \Dlr{\mu} \tilde{H} \bigr)$ & (1, 1, 1) & $\frac{\sqrt{2}\,vm_W}{\Lambda^2} \, \op_{5B}^{W\dpf A}$ \\
		$\op_{7U}^{\V\dpf sS}$ & $(i D^\mu H^\dagger) \bigl( \dms \Dlr{\mu} (\bar{\dmf}^c \dpf) \bigr)$ & (1, 2, $\frac{1}{2}$) & $- \frac{m_Z}{\sqrt{2}\Lambda} \, \op_{6B}^{Z\dpf sS} + \frac{m_W}{\Lambda} \, \op_{6B}^{Z\dpf sS}$ \\
		$\op_{7U}^{\V\dpf sP}$ & $(i D^\mu H^\dagger) \bigl( \dms \Dlr{\mu} (\bar{\dmf}^c \gamma^5 \dpf) \bigr)$ & (1, 2, $\frac{1}{2}$) & $- \frac{m_Z}{\sqrt{2}\Lambda} \, \op_{6B}^{Z\dpf sP} + \frac{m_W}{\Lambda} \, \op_{6B}^{Z\dpf sP}$ \\
		$\op_{7U}^{\V\dpf aS}$ & $\bigl(\bar{\dmf}^c \Dlr{\mu} \dpf\bigr) \bigl(i H^\dagger \Dlr{\mu} \dms\bigr)$ & (1, 2, $\frac{1}{2}$) & $- \frac{\sqrt{2}\,m_Z}{\Lambda} \, \op_{6B}^{Z\dpf aS} + \frac{2 m_W}{\Lambda} \, \op_{6B}^{W\dpf aS}$ \\
		$\op_{7U}^{\V\dpf aP}$ & $\bigl( \bar{\dmf}^c \gamma^5 \Dlr{\mu} \dpf \bigr) \bigl( i H^\dagger \Dlr{\mu} \dms \bigr)$ & (1, 2, $\frac{1}{2}$) & $- \frac{\sqrt{2}\,m_Z}{\Lambda} \, \op_{6B}^{Z\dpf aP} + \frac{2 m_W}{\Lambda} \, \op_{6B}^{W\dpf aP}$ \\
		\hline
		$\op_{8U}^{Z\dpf sS}$ & $\bigl( \dms \dlr{\mu} (\bar{\dmf}^c \dpf) \bigr) \bigl(i H^\dagger \Dlr{\mu} H\bigr)$ & (1, 1, 0) & $- \frac{vm_Z}{\Lambda^2} \, \op_{6B}^{Z\dpf sS}$ \\
		$\op_{8U}^{Z\dpf sP}$ & $\bigl( \dms \dlr{\mu} (\bar{\dmf}^c \gamma^5 \dpf) \bigr) \bigl(i H^\dagger \Dlr{\mu} H\bigr)$ & (1, 1, 0) & $- \frac{vm_Z}{\Lambda^2} \, \op_{6B}^{Z\dpf sP}$ \\
		$\op_{8U}^{Z\dpf aS}$ & $\bigl( \bar{\dmf}^c \dlr{\mu} \dpf \bigr) \, \dms \bigl(i H^\dagger \Dlr{\mu} H\bigr)$ & (1, 1, 0) & $- \frac{vm_Z}{\Lambda^2} \, \op_{6B}^{Z\dpf aS}$ \\
		$\op_{8U}^{Z\dpf aP}$ & $\bigl( \bar{\dmf}^c \gamma^5 \dlr{\mu} \dpf \bigr) \, \dms \bigl(i H^\dagger \Dlr{\mu} H\bigr)$ & (1, 1, 0) & $- \frac{vm_Z}{\Lambda^2} \, \op_{6B}^{Z\dpf aP}$ \\
		$\op_{8U}^{W\dpf sS}$ & $\bigl( \dms \Dlr{\mu} (\bar{\dmf}^c \dpf) \bigr) \bigl(i H^\dagger \Dlr{\mu} \tilde{H}\bigr)$ & (1, 1, 1) & $\frac{\sqrt{2}vm_W}{\Lambda^2} \, \op_{6B}^{W\dpf sS}$ \\
		$\op_{8U}^{W\dpf sP}$ & $\bigl( \dms \Dlr{\mu} (\bar{\dmf}^c \gamma^5 \dpf) \bigr) \bigl(i H^\dagger \Dlr{\mu} \tilde{H}\bigr)$ & (1, 1, 1) & $\frac{\sqrt{2}vm_W}{\Lambda^2} \, \op_{6B}^{W\dpf sP}$ \\
		$\op_{8U}^{W\dpf aS}$ & $\bigl( \bar{\dmf}^c \Dlr{\mu} \dpf \bigr) \, \dms \bigl(i H^\dagger \Dlr{\mu} \tilde{H}\bigr)$ & (1, 1, 1) & $\frac{\sqrt{2}vm_W}{\Lambda^2} \, \op_{6B}^{W\dpf aS}$ \\
		$\op_{8U}^{W\dpf aP}$ & $\bigl( \bar{\dmf}^c \gamma^5 \Dlr{\mu} \dpf \bigr) \, \dms \bigl(i H^\dagger \Dlr{\mu} \tilde{H}\bigr)$ & (1, 1, 1) & $\frac{\sqrt{2}vm_W}{\Lambda^2} \, \op_{6B}^{W\dpf aP}$ \\
		$\brtilde{\op}{}_{8U}^{B\dpf}$ & $\bar{\dmf}^c \sigma^{\mu\nu} \dpf \, \dms \, \brtilde{B}_{\mu\nu}(H^\dagger H)$ & (1, 1, 0) & $\frac{v^2}{2\Lambda^2} \bigl( c_W \brtilde{\op}{}_{6B}^{\gamma\dpf} - s_W \brtilde{\op}{}_{6B}^{Z\dpf} \bigr)$ \\
		$\brtilde{\op}{}_{8U}^{Z\dpf}$ & $\bar{\dmf}^c \sigma^{\mu\nu} \dpf \, \dms \, (H^\dagger \brtilde{W}^a_{\mu\nu} \sigma^a H)$ & (1, 1, 0) & $\frac{v^2}{2\Lambda^2} \bigl( s_W \brtilde{\op}{}_{6B}^{\gamma\dpf} + c_W \brtilde{\op}{}_{6B}^{Z\dpf} \bigr)$ \\
		$\brtilde{\op}{}_{8U}^{W\dpf}$ & $\bar{\dmf}^c \sigma^{\mu\nu} \dpf \, \dms \, (H^\dagger \brtilde{W}^a_{\mu\nu} \sigma^a \tilde{H})$ & (1, 1, 1) & $\frac{v^2}{\sqrt{2}\Lambda^2} \brtilde{\op}{}_{6B}^{W\dpf}$ \\
		\hline
	\end{tabular}
	\caption{Dimension 7 and 8 operators with both scalar and fermion DM with scalar dark partners $\dps$ (top) and fermion dark partners $\dpf = (\dpfL, \dpfR)^T$ (bottom).  In the former case, $F = (F_u, F_d)$ is an $SU(2)$-doublet fermion in the gauge representation $(R_F, 2, Y_F$).}\label{tab:d8MixDM}
\end{table}

\section{General Models}\label{app:DMwithEW}

In this section we relax the requirement that the DM be a pure gauge singlet.  We otherwise retain the assumptions outlined in \secref{sec:method}.  Since DM must be a singlet under the unbroken SM gauge group $SU(3)_C \times U(1)_{em}$, this only leads to new operators in the EW symmetric phase.  We continue to focus on operators with at least two DM fields, which can now take the general SM gauge representations $(1, I_{1,2}, Y_{1,2})$; the gauge representation of the third dark sector field is then fixed.  For operators with multiple SM fields, we give their possible SM gauge index contractions (but not of the dark sector fields).  When it is possible for all three dark sector multiplets to contain a charge- and colour-singlet, we denote them with $\dms$ and $\dmf$; we find that this is always possible when the visible sector particle is uncoloured.  For operators involving quarks and gluons, one of the dark sector fields must be unstable, and we use the dark partner fields $\dps$/$\dpf$.   For brevity, we will not give explicit decay operators $\op_{dec}$; these can easily be inferred from the results in \secref{sec:DP}.  Finally, we label these operators using the suffix `$dG$', where $d$ is the operator dimension and $G$ stands for `General'.

We give operators with three (two, one, zero) dark sector scalar fields in \tabref{tab:SSSGen} (\ref{tab:SSFGen}, \ref{tab:SFFGen}, \ref{tab:FFFGen}).  For the three-scalar case, all operators couple to the Higgs and can contain DM multiplets only.  For the other possibilities, we split the tables such that the upper section contains possible DM-only operators, and the lower section(s) operators that require dark partners.  Not counting different DM gauge assignments, for DM-only operators we find 8 (30, 14, 12) operators with three (two, one, zero) dark sector scalar fields, summing over lepton generations where appropriate.  For operators involving dark partners, we have 30 operators with two scalar DM (all with a fermion dark partner); 26 operators with two fermion DM (2 with a scalar dark partner); and 32 operators with scalar-fermion DM (2 with a fermion dark partner).

In the various models of \secsref{sec:DMonly} and~\ref{sec:DP}, we were concerned about the presence of Higgs portals (\modeqsref{eq:scaHP} and~\eqref{eq:ferHP}) and renormalisable SA couplings (\modeqsref{eq:phi3} and~\eqref{eq:ffsterm}).  The latter are in general not allowed, though this is operator-dependent.  The former can never be forbidden, leading to similar restrictions on the UV completions as before.  As discussed in \secref{sec:method}, allowing non-zero DM gauge interactions will open new processes that in general might be expected to dominate over the non-renormalisable operators listed in this section.  In particular, the relic density and cosmic ray searches will be affected by DM annihilation through gauge bosons; production at colliders will be enhanced; and, for DM with non-zero hypercharge, direct detection bounds can be very severe.  Ensuring that SA processes remain phenomenologically relevant is an open model-building challenge that is beyond the scope of this work.  However, we note that direct detection bounds can be avoided if there is a small mass splitting $\delta \gtrsim 500$~keV among the components of the DM multiplet~\cite{1608.02662}, which can be achieved through \emph{e.g.} appropriate couplings to the Higgs.  A larger mass splitting $\delta \gtrsim 0.05 \, m$ can suppress the contributions to the relic density, \emph{e.g.} through mixing with other dark sector fields.  We defer the question of whether viable models of this form exist to future work.

\begin{table}
	\centering
	\begin{tabular}{|c|c|l@{ }l@{ }l|}
		\hline
		Operator & Definition & \multicolumn{3}{|c|}{3rd DM Field Representation} \\
		\hline
		$\op_{4G}^H$ & $s^{ijk} \, \dms_i \dms_j \dms_k H^\dagger$ & (1, & $I_1 \otimes I_2 \otimes 2,$ & $\frac{1}{2} - Y_1 - Y_2)$ \\
		$\op_{5G}^{|H|^2_1}$ & $s^{ijk} \, \dms_i \dms_j \dms_k \, H^\dagger H$ & (1, & $I_1 \otimes I_2,$ & $- Y_1 - Y_2)$ \\
		$\op_{5G}^{|H|^2_3}$ & $s^{ijk} \, \dms_i \dms_j \dms_k \, H^\dagger \sigma^a H$ & (1, & $I_1 \otimes I_2 \otimes 3,$ & $- Y_1 - Y_2)$ \\
		$\op_{5G}^{H^2}$ & $s^{ijk} \, \dms_i \dms_j \dms_k \, H^\dagger \sigma^a \tilde{H}$ & (1, & $I_1 \otimes I_2 \otimes 3,$ & $1 - Y_1 - Y_2)$ \\
		$\op_{6G}^{H_2}$ & $s^{ijk} \, \dms_i \dms_j \dms_k \, H^\dagger (H^\dagger H)$ & (1, & $I_1 \otimes I_2 \otimes 2,$ & $\frac{1}{2} - Y_1 - Y_2)$ \\
		$\op_{6G}^{H_4}$ & $s^{ijk} \, \dms_i \dms_j \dms_k \, \tilde{H}^\dagger H^\dagger H^\dagger$ & (1, & $I_1 \otimes I_2 \otimes 4,$ & $\frac{1}{2} - Y_1 - Y_2)$ \\
		$\op_{6G}^{H^3}$ & $s^{ijk} \, \dms_i \dms_j \dms_k \, H^\dagger H^\dagger H^\dagger$ & (1, & $I_1 \otimes I_2 \otimes 4,$ & $\frac{3}{2} - Y_1 - Y_2)$ \\
		$\op_{6G}^{HD}$ & $\bigl( y^{ijk} + x^{ikj} \bigr) \, (D_\mu \dms_i) (D^\mu \dms_j) \dms_k H^\dagger$ & (1, & $I_1 \otimes I_2 \otimes 2,$ & $\frac{1}{2} - Y_1 - Y_2)$ \\
		\hline
	\end{tabular}
	\caption{General operators with three dark sector scalar fields.  In all these operators, it is possible for all three dark sector multiplets to contain a DM particle, so we use $\dms$ for all fields.}\label{tab:SSSGen}
\end{table}

\begin{table}
	\centering
	\begin{tabular}{|c|c|l@{ }l@{ }l|}
		\hline
		Operator & Definition & \multicolumn{3}{|c|}{3rd Field Representation} \\
		\hline
		\hline
		$\op_{5G}^{\bar{e}}$ & $s^{ij} \, \dms_i \dms_j \, \bar{e} \dmfL$ & (1, & $I_1 \otimes I_2,$ & $- 1 - Y_1 - Y_2)$ \\
		$\op_{5G}^{L}$ & $s^{ij} \, \dms_i \dms_j \, L^\dagger \dmfR$ & (1, & $I_1 \otimes I_2 \otimes 2,$ & $\frac{1}{2} - Y_1 - Y_2)$ \\
		$\op_{6G}^{\bar{e}H}$ & $s^{ij} \, \dms_i \dms_j \, H \, \bar{e} \dmfR$ & (1, & $I_1 \otimes I_2 \otimes 2,$ & $- \frac{3}{2} - Y_1 - Y_2)$ \\
		$\op_{6G}^{\bar{e}H^\dagger}$ & $s^{ij} \, \dms_i \dms_j \, H^\dagger \bar{e} \dmfR$ & (1, & $I_1 \otimes I_2 \otimes 2,$ & $- \frac{1}{2} - Y_1 - Y_2)$ \\
		$\op_{6G}^{LH_1}$ & $s^{ij} \, \dms_i \dms_j \, \bigl( (L^\dagger H) \dmfL \bigr)$ & (1, & $I_1 \otimes I_2,$ & $- 1 - Y_1 - Y_2)$ \\
		$\op_{6G}^{LH_1^\dagger}$ & $s^{ij} \, \dms_i \dms_j \, \bigl( (L^\dagger \tilde{H}) \dmfL \bigr)$ & (1, & $I_1 \otimes I_2,$ & $- Y_1 - Y_2)$ \\
		$\op_{6G}^{LH_3}$ & $s^{ij} \, \dms_i \dms_j \, \bigl( (L^\dagger \sigma^a H) \dmfL \bigr)$ & (1, & $I_1 \otimes I_2 \otimes 3,$ & $- 1 - Y_1 - Y_2)$ \\
		$\op_{6G}^{LH_3^\dagger}$ & $s^{ij} \, \dms_i \dms_j \, \bigl( (L^\dagger \sigma^a \tilde{H}) \dmfL \bigr)$ & (1, & $I_1 \otimes I_2 \otimes 3,$ & $- Y_1 - Y_2)$ \\
		$\op_{6G}^{\bar{e}V}$ & $a^{ij} \bigl( \dms_i \Dlr{\mu} \dms_j \bigr) \, \bar{e} \sigma^\mu \dmfR$ & (1, & $I_1 \otimes I_2,$ & $-1 - Y_1 - Y_2$) \\
		$\op_{6G}^{LV}$ & $a^{ij} \bigl( \dms_i \Dlr{\mu} \dms_j \bigr) \, L^\dagger \bar{\sigma}^\mu \dmfL$ & (1, & $I_1 \otimes I_2 \otimes 2,$ & $\frac{1}{2} - Y_1 - Y_2$) \\
		\hline
		\hline
		$\op_{5G}^{\bar{f}\dpf}$ & $s^{ij} \, \dms_i \dms_j \, \bar{f} \dpfL$ & (3, & $I_1 \otimes I_2,$ & $- Y_{\bar{f}} - Y_1 - Y_2)$ \\
		$\op_{5G}^{Q\dpf}$ & $s^{ij} \, \dms_i \dms_j \, Q^\dagger \dpfR$ & (3, & $I_1 \otimes I_2 \otimes 2,$ & $\frac{1}{6} - Y_1 - Y_2)$ \\
		$\op_{6G}^{\bar{f}H\dpf}$ & $s^{ij} \, \dms_i \dms_j \, H \, \bar{f} \dpfL$ & (3, & $I_1 \otimes I_2 \otimes 2,$ & $- \frac{1}{2} - Y_{\bar{f}} - Y_1 - Y_2)$ \\
		$\op_{6G}^{\bar{f}H^\dagger\dpf}$ & $s^{ij} \, \dms_i \dms_j \, H^\dagger \bar{f} \dpfL$ & (3, & $I_1 \otimes I_2 \otimes 2,$ & $\frac{1}{2} - Y_{\bar{f}} - Y_1 - Y_2)$ \\
		$\op_{6G}^{QH_1}$ & $s^{ij} \, \dms_i \dms_j \, \bigl( (Q^\dagger H) \dpfR \bigr)$ & (3, & $I_1 \otimes I_2,$ & $- \frac{1}{3} - Y_1 - Y_2)$ \\
		$\op_{6G}^{QH_1^\dagger\dpf}$ & $s^{ij} \, \dms_i \dms_j \, \bigl( (Q^\dagger \tilde{H}) \dpfR \bigr)$ & (3, & $I_1 \otimes I_2,$ & $\frac{2}{3} - Y_1 - Y_2)$ \\
		$\op_{6G}^{QH_3\dpf}$ & $s^{ij} \, \dms_i \dms_j \, \bigl( (Q^\dagger \sigma^a H) \dpfR \bigr)$ & (3, & $I_1 \otimes I_2 \otimes 3,$ & $- \frac{1}{3} - Y_1 - Y_2)$ \\
		$\op_{6G}^{QH_3^\dagger\dpf}$ & $s^{ij} \, \dms_i \dms_j \, \bigl( (Q^\dagger \sigma^a \tilde{H}) \dpfR \bigr)$ & (3, & $I_1 \otimes I_2 \otimes 3,$ & $\frac{2}{3} - Y_1 - Y_2)$ \\
		$\op_{6G}^{\bar{f}V\dpf}$ & $a^{ij} \bigl( \dms_i \Dlr{\mu} \dms_j \bigr) \, \bar{f} \sigma^\mu \dpfR$ & (3, & $I_1 \otimes I_2,$ & $- Y_{\bar{f}} - Y_1 - Y_2$) \\
		$\op_{6G}^{QV\dpf}$ & $a^{ij} \bigl( \dms_i \Dlr{\mu} \dms_j \bigr) \, Q^\dagger \bar{\sigma}^\mu \dpfL$ & (3, & $I_1 \otimes I_2 \otimes 2,$ & $\frac{1}{6} - Y_1 - Y_2$) \\
		\hline
		\hline
		$\op_{5G}^{\bar{f}\dps}$ & $\dms \, \dps \, \bar{f} \dmfL$ & (3, & $I_1 \otimes I_2,$ & $- Y_{\bar{f}} - Y_1 - Y_2)$ \\
		$\op_{5G}^{Q\dps}$ & $\dms \, \dps \, Q^\dagger \dmfR$ & (3, & $I_1 \otimes I_2 \otimes 2,$ & $\frac{1}{6} - Y_1 - Y_2)$ \\
		$\op_{6G}^{\bar{f}H\dps}$ & $\dms \, \dps \, H \, \bar{f} \dmfL$ & (3, & $I_1 \otimes I_2 \otimes 2,$ & $- \frac{1}{2} - Y_{\bar{f}} - Y_1 - Y_2)$ \\
		$\op_{6G}^{\bar{f}H^\dagger\dps}$ & $\dms \, \dps \, H^\dagger \bar{f} \dmfL$ & (3, & $I_1 \otimes I_2 \otimes 2,$ & $\frac{1}{2} - Y_{\bar{f}} - Y_1 - Y_2)$ \\
		$\op_{6G}^{QH_1\dps}$ & $\dms \, \dps \, \bigl( (Q^\dagger H) \dmfR \bigr)$ & (3, & $I_1 \otimes I_2,$ & $- \frac{1}{3} - Y_1 - Y_2)$ \\
		$\op_{6G}^{QH_1^\dagger\dps}$ & $\dms \, \dps \, \bigl( (Q^\dagger \tilde{H}) \dmfR \bigr)$ & (3, & $I_1 \otimes I_2,$ & $\frac{2}{3} - Y_1 - Y_2)$ \\
		$\op_{6G}^{QH_3\dps}$ & $\dms \, \dps \, \bigl( (Q^\dagger \sigma^a H) \dmfR \bigr)$ & (3, & $I_1 \otimes I_2 \otimes 3,$ & $- \frac{1}{3} - Y_1 - Y_2)$ \\
		$\op_{6G}^{QH_3^\dagger\dps}$ & $\dms \, \dps \, \bigl( (Q^\dagger \sigma^a \tilde{H}) \dmfR \bigr)$ & (3, & $I_1 \otimes I_2 \otimes 3,$ & $\frac{2}{3} - Y_1 - Y_2)$ \\
		$\op_{6G}^{\bar{f}V\dps}$ & $\bigl( \dms \Dlr{\mu} \dps \bigr) \, \bar{f} \sigma^\mu \dmfR$ & (3, & $I_1 \otimes I_2,$ & $- Y_{\bar{f}} - Y_1 - Y_2$) \\
		$\op_{6G}^{QV\dps}$ & $\bigl( \dms \Dlr{\mu} \dps \bigr) \, Q^\dagger \bar{\sigma}^\mu \dmfL$ & (3, & $I_1 \otimes I_2 \otimes 2,$ & $\frac{1}{6} - Y_1 - Y_2$) \\
		\hline
	\end{tabular}
	\caption{General operators with two scalar and one fermion dark sector field.  The operators in the upper section can all be DM, so we label the fields using $\dms$ and $\dmf = (\dmfL, \dmfR)^T$.  Operators in the lower two sections necessarily involve an unstable dark partner.  In the middle section, that dark partner is a fermion $\dpf = (\dpfL, \dpfR)^T$, while in the lower section it is a scalar $\dps$.  $\bar{f}$ represents the quark singlet fields $\{\bar{u}, \bar{d}\}$, with hypercharge $Y_{\bar{f}}$.}\label{tab:SSFGen}
\end{table}

\begin{table}
	\centering
	\begin{tabular}{|c|c|l@{ }l@{ }l|}
		\hline
		Operator & Definition & \multicolumn{3}{|c|}{3rd Field Representation} \\
		\hline
		$\op_{5G}^{HS}$ & $s^{ij} \, \bar{\dmf}_i^c \dmf_j \, \dms H^\dagger$ & (1, & $I_1 \otimes I_2 \otimes 2,$ & $\frac{1}{2} - Y_1 - Y_2)$ \\
		$\op_{5G}^{HP}$ & $s^{ij} \, \bar{\dmf}_i^c \gamma^5 \dmf_j \, \dms H^\dagger$ & (1, & $I_1 \otimes I_2 \otimes 2,$ & $\frac{1}{2} - Y_1 - Y_2)$ \\
		$\op_{6G}^{|H|^2_1S}$ & $s^{ij} \, \bar{\dmf}_i^c \dmf_j \, \dms \, H^\dagger H$ & (1, & $I_1 \otimes I_2,$ & $- Y_1 - Y_2)$ \\
		$\op_{6G}^{|H|^2_3S}$ & $s^{ij} \, \bar{\dmf}_i^c \dmf_j \, \dms \, H^\dagger \sigma^a H$ & (1, & $I_1 \otimes I_2 \otimes 3,$ & $- Y_1 - Y_2)$ \\
		$\op_{6G}^{H^2S}$ & $s^{ij} \, \bar{\dmf}_i^c \dmf_j \, \dms \, H^\dagger \sigma^a \tilde{H}$ & (1, & $I_1 \otimes I_2 \otimes 3,$ & $1 - Y_1 - Y_2)$ \\
		$\op_{6G}^{|H|^2_1P}$ & $s^{ij} \, \bar{\dmf}_i^c \gamma^5 \dmf_j \, \dms \, H^\dagger H$ & (1, & $I_1 \otimes I_2,$ & $- Y_1 - Y_2)$ \\
		$\op_{6G}^{|H|^2_3P}$ & $s^{ij} \, \bar{\dmf}_i^c \gamma^5 \dmf_j \, \dms \, H^\dagger \sigma^a H$ & (1, & $I_1 \otimes I_2 \otimes 3,$ & $- Y_1 - Y_2)$ \\
		$\op_{6G}^{H^2P}$ & $s^{ij} \, \bar{\dmf}_i^c \gamma^5 \dmf_j \, \dms \, H^\dagger \sigma^a \tilde{H}$ & (1, & $I_1 \otimes I_2 \otimes 3,$ & $1 - Y_1 - Y_2)$ \\
		$\op_{6G}^{HV}$ & $a^{ij} \, \bar{\dmf}_i^c \gamma^\mu \dmf_j \, \bigl( i H^\dagger \Dlr{\mu} \dms)$ & (1, & $I_1 \otimes I_2 \otimes 2,$ & $\frac{1}{2} - Y_1 - Y_2)$ \\
		$\op_{6G}^{HA}$ & $s^{ij} \, \bar{\dmf}_i^c \gamma^\mu \gamma^5 \dmf_j \, \bigl( i H^\dagger \Dlr{\mu} \dms)$ & (1, & $I_1 \otimes I_2 \otimes 2,$ & $\frac{1}{2} - Y_1 - Y_2)$ \\
		$\brtilde{\op}{}_{6G}^{B T}$ & $a^{ij} \, \bar{\dmf}_i^c \sigma^{\mu\nu} \gamma^5 \dmf_j \, \dms \, \brtilde{B}_{\mu\nu}$ & (1, & $I_1 \otimes I_2,$ & $- Y_1 - Y_2)$ \\
		$\brtilde{\op}{}_{6G}^{W T}$ & $a^{ij} \, \bar{\dmf}_i^c \sigma^{\mu\nu} \gamma^5 \dmf_j \, \dms \, \brtilde{W}^a_{\mu\nu}$ & (1, & $I_1 \otimes I_2 \otimes 3,$ & $- Y_1 - Y_2)$ \\
		\hline
		\hline
		$\brtilde{\op}{}_{6G}^{G\dps T}$ & $a^{ij} \, \bar{\dmf}_i^c \sigma^{\mu\nu} \gamma^5 \dmf_j \, \dps \, \brtilde{G}_{\mu\nu}$ & (8, & $I_1 \otimes I_2,$ & $- Y_1 - Y_2)$ \\
		$\brtilde{\op}{}_{6G}^{G\dpf T}$ & $\bar{\dmf}^c \sigma^{\mu\nu} \gamma^5 \dpf \, \dms \, \brtilde{G}_{\mu\nu}$ & (8, & $I_1 \otimes I_2,$ & $- Y_1 - Y_2)$ \\
		\hline
	\end{tabular}
	\caption{General operators with one scalar and two fermion dark sector fields.  The operators in the upper section can all be DM, so we label the fields using $\dms$ and $\dmf$.  Operators in the lower section necessarily involve an unstable dark partner $\dps$ or $\dpf$.}\label{tab:SFFGen}
\end{table}

\begin{table}
	\centering
	\begin{tabular}{|c|c|l@{ }l@{ }l|}
		\hline
		Operator & Definition & \multicolumn{3}{|c|}{3rd Field Representation} \\
		\hline
		\hline
		$\op_{6G}^{\bar{e}L}$ & $(y^{ijk} + x^{ikj}) \, (\dmfL_i \dmfL_j) \, (\bar{e} \dmfL_k)$ & (1, & $I_1 \otimes I_2,$ & $-1 - Y_1 - Y_2)$ \\
		$\op_{6G}^{\bar{e}R}$ & $n^{ijk} \, (\dmfR_i \dmfR_j) \, (\bar{e} \dmfL_k)$ & (1, & $I_1 \otimes I_2,$ & $-1 - Y_1 - Y_2)$ \\
		$\op_{6G}^{LL}$ & $n^{ijk} \, (\dmfL_i \dmfL_j) \, (L^\dagger \dmfR_k)$ & (1, & $I_1 \otimes I_2 \otimes 2,$ & $\frac{1}{2}  - Y_1 - Y_2)$ \\
		$\op_{6G}^{RL}$ & $(y^{ijk} + x^{ikj}) \, (\dmfR_i \dmfR_j) \, (L^\dagger \dmfR_k)$ & (1, & $I_1 \otimes I_2 \otimes 2,$ & $\frac{1}{2}  - Y_1 - Y_2)$ \\
		\hline
		\hline
		$\op_{6G}^{\bar{f}\dpf L}$ & $s^{ij} \, (\dmfL_i \dmfL_j) \, (\bar{f} \dpfL)$ & (3, & $I_1 \otimes I_2,$ & $- Y _{\bar{f}} - Y_1 - Y_2)$ \\
		$\op_{6G}^{\bar{f}\dpf R}$ & $s^{ij} \, (\dmfR_i \dmfR_j) \, (\bar{f} \dpfL)$ & (3, & $I_1 \otimes I_2,$ & $- Y _{\bar{f}} - Y_1 - Y_2)$ \\
		$\op_{6G}^{Q\dpf L}$ & $s^{ij} \, (\dmfL_i \dmfL_j) \, (Q^\dagger \dpfR)$ & (3, & $I_1 \otimes I_2 \otimes 2,$ & $\frac{1}{6} - Y_1 - Y_2)$ \\
		$\op_{6G}^{Q\dpf R}$ & $s^{ij} \, (\dmfR_i \dmfR_j) \, (Q^\dagger \dpfR)$ & (3, & $I_1 \otimes I_2 \otimes 2,$ & $\frac{1}{6} - Y_1 - Y_2)$ \\
		$\op_{6G}^{\bar{f}\dmf L}$ & $n^{ij} \, (\dmfL_i \dpfL) \, (\bar{f} \dmfL_j)$ & (3, & $I_1 \otimes I_2,$ & $- Y _{\bar{f}} - Y_1 - Y_2)$ \\
		$\op_{6G}^{\bar{f}\dmf R}$ & $n^{ij} \, (\dmfR_i \dpfR) \, (\bar{f} \dmfL_j)$ & (3, & $I_1 \otimes I_2,$ & $- Y _{\bar{f}} - Y_1 - Y_2)$ \\
		$\op_{6G}^{Q\dmf L}$ & $n^{ij} \, (\dmfL_i \dpfL) \, (Q^\dagger \dmfR_j)$ & (3, & $I_1 \otimes I_2 \otimes 2,$ & $\frac{1}{6} - Y_1 - Y_2)$ \\
		$\op_{6G}^{Q\dmf R}$ & $n^{ij} \, (\dmfR_i \dpfR) \, (Q^\dagger \dmfR_j)$ & (3, & $I_1 \otimes I_2 \otimes 2,$ & $\frac{1}{6} - Y_1 - Y_2)$ \\
		\hline
	\end{tabular}
	\caption{General operators with three fermion dark sector fields.  The operators in the upper section can all be DM, so we label the fields using $\dmf = (\dmfL, \dmfR)^T$.  Operators in the lower section necessarily involve an unstable dark partner $\dpf = (\dpfL, \dpfR)^T$.  $\bar{f}$ represents the quark singlet fields $\{\bar{u}, \bar{d}\}$, with hypercharge $Y_{\bar{f}}$.}\label{tab:FFFGen}
\end{table}

\section{$P$-wave suppression of $\dms_i\dms_j \to \dms_k^\dagger \gamma/g$ from effective operators}\label{app:pwave}

The matrix element for these processes must contain a photon or gluon polarisation tensor $\varepsilon_\mu (p_4)$, and the only other non-scalar objects are initial and final state momenta.  The most general form the matrix element can take is then 
\begin{equation}
	i \mathcal{M} = f_1 \, p_1 \cdot \varepsilon + f_2 \, p_2 \cdot \varepsilon + f_3 \, p_3 \cdot \varepsilon + f_4 \, \epsilon_{\mu\nu\rho\sigma} p_1^\mu p_2^\nu p_3^\rho \varepsilon^\sigma \,,
\end{equation}
where the $f_i$ are scalar functions of the particle masses and momenta.  There is trivially no term involving $p_4 \cdot \varepsilon$, and any other possible contraction with the antiysymmetric tensor either vanishes identically or is reducible to the term we have included using conservation of momentum.  In the case of a gluon final state there are additional colour factors that do not affect our argument.

If we assume that this process is generated by a four-field contact operator, we have the further restriction that the $f_i$ must be non-singular functions.  Since a local operator has only positive powers of dimension, the associated Feynman rule can contain only positive powers of momenta and negative powers of the UV scale $\Lambda$.  The $f_i$ cannot have poles or branch cuts for physical momenta.

Conservation of momentum allows us to eliminate the third term by replacing $p_3 = p_1 + p_2 - p_4$:
\begin{equation}
	i \mathcal{M} = (f_1 + f_3) \, p_1 \cdot \varepsilon + (f_2 + f_3) \, p_2 \cdot \varepsilon + f_4 \, \epsilon_{\mu\nu\rho\sigma} p_1^\mu p_2^\nu p_3^\rho \varepsilon^\sigma \,,
\end{equation}
The Ward Identity tells us that, for all particles on-shell, the matrix element vanishes if we replace $\varepsilon \to p_4$.  The last term vanishes identically using conservation of momentum:
\begin{equation}
	\epsilon_{\mu\nu\rho\sigma} p_1^\mu p_2^\nu p_3^\rho p_4^\sigma = \epsilon_{\mu\nu\rho\sigma} p_1^\mu p_2^\nu p_3^\rho (p_1 + p_2 - p_3)^\sigma = 0 \,.
\end{equation}
The first two terms give us the relation
\begin{equation}
  0 = (f_1 + f_3) \, p_1 \cdot p_4 + (f_2 + f_3) \, p_2 \cdot p_4 \bigr\rvert_{\text{On-Shell}} \,.
\end{equation}
So we may write the general matrix element when all fields are on-shell as
\begin{equation}
	i \mathcal{M} = F_1 (s, t) \, \biggl( \frac{p_1 \cdot \varepsilon}{p_1 \cdot p_4} - \frac{p_2 \cdot \varepsilon}{p_2 \cdot p_4} \biggr) + i F_2 (s, t) \, \epsilon_{\mu\nu\rho\sigma} p_1^\mu p_2^\nu p_3^\rho \varepsilon^\sigma \,,
\end{equation}
where $F_{1,2}$ inherit from the $f_i$ the absence of any poles at physical momenta.

Now let us consider this expression in terms of the components of the momenta.  Without loss of generality, we may take
\begin{align}
	p_1 & = (E_1, 0, 0, \phantom{-} p) \,, & p_3 & = (E_3, - k \sin \theta, 0, - k \cos\theta ) \,, \\
	p_2 & = (E_2, 0, 0, -p) \,, & p_4 & = (k, k \sin \theta, 0, k \cos\theta ) \,, \\
	&& \varepsilon & = (0, \cos\theta, \pm i, - \sin\theta ) \,.
\end{align}
We have allowed for the possibility that $m_1 \neq m_2 \neq m_3$.  A short calculation then gives
\begin{equation}
	i \mathcal{M} = F_1 (s, t) \, \frac{p \sin \theta}{k} \biggl( \frac{1}{E_1 - p \cos\theta} - \frac{1}{E_2 + p \cos\theta} \biggr) \mp F_2 (s, t) \, (E_1 + E_2) \, p k \sin\theta \,.
\end{equation}
We see that this vanishes as $p \to 0$ unless at least one of the form factors $F_{1,2}$ has a pole at that point.  Since as we discussed they do not, it follows that in the non-relativistic limit 
\begin{equation}
	\sigma v = \frac{1}{2s} \int d\Pi \sum \abs{\mathcal{M}}^2 \sim \frac{1}{2s} \, p^2 \int d\cos\theta (\cdots) \sim v^2 \,,
\end{equation} 
so that this cross section is always $p$-wave suppressed.

\bibliography{SArefs}{}

\providecommand{\href}[2]{#2}\begingroup\raggedright\begin{thebibliography}{10}

\bibitem{1304.4279}
{\scshape CDMS} collaboration, R.~Agnese et~al., \emph{{Silicon Detector Dark
  Matter Results from the Final Exposure of CDMS II}},
  \href{http://dx.doi.org/10.1103/PhysRevLett.111.251301}{\emph{Phys. Rev.
  Lett.} {\bf 111} (2013) 251301}, [\href{https://arxiv.org/abs/1304.4279}{{\tt
  1304.4279}}].

\bibitem{1510.07754}
{\scshape PICO} collaboration, C.~Amole et~al., \emph{{Dark matter search
  results from the PICO-60 CF$_3$I bubble chamber}},
  \href{http://dx.doi.org/10.1103/PhysRevD.93.052014}{\emph{Phys. Rev.} {\bf
  D93} (2016) 052014}, [\href{https://arxiv.org/abs/1510.07754}{{\tt
  1510.07754}}].

\bibitem{1602.03489}
{\scshape LUX} collaboration, D.~S. Akerib et~al., \emph{{Results on the
  Spin-Dependent Scattering of Weakly Interacting Massive Particles on Nucleons
  from the Run 3 Data of the LUX Experiment}},
  \href{http://dx.doi.org/10.1103/PhysRevLett.116.161302}{\emph{Phys. Rev.
  Lett.} {\bf 116} (2016) 161302},
  [\href{https://arxiv.org/abs/1602.03489}{{\tt 1602.03489}}].

\bibitem{1602.03781}
F.~Mayet et~al., \emph{{A review of the discovery reach of directional Dark
  Matter detection}},
  \href{http://dx.doi.org/10.1016/j.physrep.2016.02.007}{\emph{Phys. Rept.}
  {\bf 627} (2016) 1--49}, [\href{https://arxiv.org/abs/1602.03781}{{\tt
  1602.03781}}].

\bibitem{lux2016talk}
{\scshape LUX} collaboration, A.~Manalaysay et~al., ``Dark-matter results from
  332 new live days of lux data.'' 2016.

\bibitem{1607.07400}
{\scshape PandaX-II} collaboration, A.~Tan et~al., \emph{{Dark Matter Results
  from First 98.7 Days of Data from the PandaX-II Experiment}},
  \href{http://dx.doi.org/10.1103/PhysRevLett.117.121303}{\emph{Phys. Rev.
  Lett.} {\bf 117} (2016) 121303},
  [\href{https://arxiv.org/abs/1607.07400}{{\tt 1607.07400}}].

\bibitem{1604.07773}
{\scshape ATLAS} collaboration, M.~Aaboud et~al., \emph{{Search for new
  phenomena in final states with an energetic jet and large missing transverse
  momentum in $pp$ collisions at $\sqrt{s}=13$  TeV using the ATLAS
  detector}}, \href{http://dx.doi.org/10.1103/PhysRevD.94.032005}{\emph{Phys.
  Rev.} {\bf D94} (2016) 032005}, [\href{https://arxiv.org/abs/1604.07773}{{\tt
  1604.07773}}].

\bibitem{1611.03568}
{\scshape CMS} collaboration, A.~M. Sirunyan et~al., \emph{{Search for dijet
  resonances in proton-proton collisions at $\sqrt{s}=$ 13 TeV and constraints
  on dark matter and other models}}, {\emph{Submitted to: Phys. Lett. B} (2016)
  }, [\href{https://arxiv.org/abs/1611.03568}{{\tt 1611.03568}}].

\bibitem{1604.00014}
J.~M. Gaskins, \emph{{A review of indirect searches for particle dark matter}},
   \href{https://arxiv.org/abs/1604.00014}{{\tt 1604.00014}}.

\bibitem{1609.08091}
{\scshape HESS} collaboration, H.~Abdalla et~al., \emph{{H.E.S.S. Limits on
  Linelike Dark Matter Signatures in the 100 GeV to 2 TeV Energy Range Close
  to the Galactic Center}},
  \href{http://dx.doi.org/10.1103/PhysRevLett.117.151302,
  10.3204/PUBDB-2016-05876}{\emph{Phys. Rev. Lett.} {\bf 117} (2016) 151302},
  [\href{https://arxiv.org/abs/1609.08091}{{\tt 1609.08091}}].

\bibitem{1611.03184}
{\scshape DES, Fermi-LAT} collaboration, A.~Albert et~al., \emph{{Searching for
  Dark Matter Annihilation in Recently Discovered Milky Way Satellites with
  Fermi-LAT}},  \href{https://arxiv.org/abs/1611.03184}{{\tt 1611.03184}}.

\bibitem{0811.0172}
T.~Hambye, \emph{{Hidden vector dark matter}},
  \href{http://dx.doi.org/10.1088/1126-6708/2009/01/028}{\emph{JHEP} {\bf 01}
  (2009) 028}, [\href{https://arxiv.org/abs/0811.0172}{{\tt 0811.0172}}].

\bibitem{0907.1007}
T.~Hambye and M.~H.~G. Tytgat, \emph{{Confined hidden vector dark matter}},
  \href{http://dx.doi.org/10.1016/j.physletb.2009.11.050}{\emph{Phys. Lett.}
  {\bf B683} (2010) 39--41}, [\href{https://arxiv.org/abs/0907.1007}{{\tt
  0907.1007}}].

\bibitem{0912.4496}
C.~Arina, T.~Hambye, A.~Ibarra and C.~Weniger, \emph{{Intense Gamma-Ray Lines
  from Hidden Vector Dark Matter Decay}},
  \href{http://dx.doi.org/10.1088/1475-7516/2010/03/024}{\emph{JCAP} {\bf 1003}
  (2010) 024}, [\href{https://arxiv.org/abs/0912.4496}{{\tt 0912.4496}}].

\bibitem{1003.5912}
F.~D'Eramo and J.~Thaler, \emph{{Semi-annihilation of Dark Matter}},
  \href{http://dx.doi.org/10.1007/JHEP06(2010)109}{\emph{JHEP} {\bf 06} (2010)
  109}, [\href{https://arxiv.org/abs/1003.5912}{{\tt 1003.5912}}].

\bibitem{err1202.2962}
G.~Belanger, K.~Kannike, A.~Pukhov and M.~Raidal, \emph{{Impact of
  semi-annihilations on dark matter phenomenology - an example of $Z_N$
  symmetric scalar dark matter}},
  \href{http://dx.doi.org/10.1088/1475-7516/2012/04/010}{\emph{JCAP} {\bf 1204}
  (2012) 010}, [\href{https://arxiv.org/abs/1202.2962}{{\tt 1202.2962}}].

\bibitem{1210.7817}
F.~D'Eramo, M.~McCullough and J.~Thaler, \emph{{Multiple Gamma Lines from
  Semi-Annihilation}},
  \href{http://dx.doi.org/10.1088/1475-7516/2013/04/030}{\emph{JCAP} {\bf 1304}
  (2013) 030}, [\href{https://arxiv.org/abs/1210.7817}{{\tt 1210.7817}}].

\bibitem{1211.1014}
G.~Belanger, K.~Kannike, A.~Pukhov and M.~Raidal, \emph{{$Z_3$ Scalar Singlet
  Dark Matter}},
  \href{http://dx.doi.org/10.1088/1475-7516/2013/01/022}{\emph{JCAP} {\bf 1301}
  (2013) 022}, [\href{https://arxiv.org/abs/1211.1014}{{\tt 1211.1014}}].

\bibitem{1402.6449}
P.~Ko and Y.~Tang, \emph{{Self-interacting scalar dark matter with local $Z_3$
  symmetry}},
  \href{http://dx.doi.org/10.1088/1475-7516/2014/05/047}{\emph{JCAP} {\bf 1405}
  (2014) 047}, [\href{https://arxiv.org/abs/1402.6449}{{\tt 1402.6449}}].

\bibitem{1403.4960}
G.~Bélanger, K.~Kannike, A.~Pukhov and M.~Raidal, \emph{{Minimal
  semi-annihilating $\mathbb{Z}_N$ scalar dark matter}},
  \href{http://dx.doi.org/10.1088/1475-7516/2014/06/021}{\emph{JCAP} {\bf 1406}
  (2014) 021}, [\href{https://arxiv.org/abs/1403.4960}{{\tt 1403.4960}}].

\bibitem{1405.5870}
M.~Aoki and T.~Toma, \emph{{Impact of semi-annihilation of $\mathbb{Z}_3$
  symmetric dark matter with radiative neutrino masses}},
  \href{http://dx.doi.org/10.1088/1475-7516/2014/09/016}{\emph{JCAP} {\bf 1409}
  (2014) 016}, [\href{https://arxiv.org/abs/1405.5870}{{\tt 1405.5870}}].

\bibitem{1410.2246}
J.~Berger, Y.~Cui and Y.~Zhao, \emph{{Detecting Boosted Dark Matter from the
  Sun with Large Volume Neutrino Detectors}},
  \href{http://dx.doi.org/10.1088/1475-7516/2015/02/005}{\emph{JCAP} {\bf 1502}
  (2015) 005}, [\href{https://arxiv.org/abs/1410.2246}{{\tt 1410.2246}}].

\bibitem{1507.08295}
N.~Fonseca, L.~Necib and J.~Thaler, \emph{{Dark Matter, Shared Asymmetries, and
  Galactic Gamma Ray Signals}},
  \href{http://dx.doi.org/10.1088/1475-7516/2016/02/052}{\emph{JCAP} {\bf 1602}
  (2016) 052}, [\href{https://arxiv.org/abs/1507.08295}{{\tt 1507.08295}}].

\bibitem{1508.03031}
A.~Karam and K.~Tamvakis, \emph{{Dark matter and neutrino masses from a
  scale-invariant multi-Higgs portal}},
  \href{http://dx.doi.org/10.1103/PhysRevD.92.075010}{\emph{Phys. Rev.} {\bf
  D92} (2015) 075010}, [\href{https://arxiv.org/abs/1508.03031}{{\tt
  1508.03031}}].

\bibitem{1509.08481}
Y.~Cai and A.~P. Spray, \emph{{Fermionic Semi-Annihilating Dark Matter}},
  \href{http://dx.doi.org/10.1007/JHEP01(2016)087}{\emph{JHEP} {\bf 01} (2016)
  087}, [\href{https://arxiv.org/abs/1509.08481}{{\tt 1509.08481}}].

\bibitem{1510.02179}
A.~P. Spray and Y.~Cai, \emph{{Semi-Annihilating Wino-Like Dark Matter}},
  {\emph{PoS} {\bf PLANCK2015} (2015) 125},
  [\href{https://arxiv.org/abs/1510.02179}{{\tt 1510.02179}}].

\bibitem{1511.09247}
Y.~Cai and A.~P. Spray, \emph{{The galactic center excess from $ {\mathbb{Z}}_3
  $ scalar semi-annihilations}},
  \href{http://dx.doi.org/10.1007/JHEP06(2016)156}{\emph{JHEP} {\bf 06} (2016)
  156}, [\href{https://arxiv.org/abs/1511.09247}{{\tt 1511.09247}}].

\bibitem{1601.06355}
R.~Ding, Z.-L. Han, Y.~Liao and W.-P. Xie, \emph{{Radiative neutrino mass with
  $\mathbb ℤ_{3}$ dark matter: from relic density to LHC signatures}},
  \href{http://dx.doi.org/10.1007/JHEP05(2016)030}{\emph{JHEP} {\bf 05} (2016)
  030}, [\href{https://arxiv.org/abs/1601.06355}{{\tt 1601.06355}}].

\bibitem{1607.01001}
A.~Karam and K.~Tamvakis, \emph{{Dark Matter from a Classically Scale-Invariant
  $SU(3)_X$}}, \href{http://dx.doi.org/10.1103/PhysRevD.94.055004}{\emph{Phys.
  Rev.} {\bf D94} (2016) 055004}, [\href{https://arxiv.org/abs/1607.01001}{{\tt
  1607.01001}}].

\bibitem{1611.09360}
Y.~Cai and A.~Spray, \emph{{A Systematic Effective Operator Analysis of
  Semi-Annihilating Dark Matter}},
  \href{https://arxiv.org/abs/1611.09360}{{\tt 1611.09360}}.

\bibitem{Birkedal:2004xn}
A.~Birkedal, K.~Matchev and M.~Perelstein, \emph{{Dark matter at colliders: A
  Model independent approach}},
  \href{http://dx.doi.org/10.1103/PhysRevD.70.077701}{\emph{Phys. Rev.} {\bf
  D70} (2004) 077701}, [\href{https://arxiv.org/abs/hep-ph/0403004}{{\tt
  hep-ph/0403004}}].

\bibitem{Feng:2005gj}
J.~L. Feng, S.~Su and F.~Takayama, \emph{{Lower limit on dark matter production
  at the large hadron collider}},
  \href{http://dx.doi.org/10.1103/PhysRevLett.96.151802}{\emph{Phys. Rev.
  Lett.} {\bf 96} (2006) 151802},
  [\href{https://arxiv.org/abs/hep-ph/0503117}{{\tt hep-ph/0503117}}].

\bibitem{Birkedal:2005ep}
A.~Birkedal, K.~T. Matchev, M.~Perelstein and A.~Spray, \emph{{Robust gamma ray
  signature of WIMP dark matter}},
  \href{https://arxiv.org/abs/hep-ph/0507194}{{\tt hep-ph/0507194}}.

\bibitem{1603.08002}
A.~De~Simone and T.~Jacques, \emph{{Simplified models vs. effective field
  theory approaches in dark matter searches}},
  \href{http://dx.doi.org/10.1140/epjc/s10052-016-4208-4}{\emph{Eur. Phys. J.}
  {\bf C76} (2016) 367}, [\href{https://arxiv.org/abs/1603.08002}{{\tt
  1603.08002}}].

\bibitem{0808.3384}
M.~Beltran, D.~Hooper, E.~W. Kolb and Z.~C. Krusberg, \emph{{Deducing the
  nature of dark matter from direct and indirect detection experiments in the
  absence of collider signatures of new physics}},
  \href{http://dx.doi.org/10.1103/PhysRevD.80.043509}{\emph{Phys. Rev.} {\bf
  D80} (2009) 043509}, [\href{https://arxiv.org/abs/0808.3384}{{\tt
  0808.3384}}].

\bibitem{0912.4511}
Q.-H. Cao, C.-R. Chen, C.~S. Li and H.~Zhang, \emph{{Effective Dark Matter
  Model: Relic density, CDMS II, Fermi LAT and LHC}},
  \href{http://dx.doi.org/10.1007/JHEP08(2011)018}{\emph{JHEP} {\bf 08} (2011)
  018}, [\href{https://arxiv.org/abs/0912.4511}{{\tt 0912.4511}}].

\bibitem{1008.1783}
J.~Goodman, M.~Ibe, A.~Rajaraman, W.~Shepherd, T.~M.~P. Tait and H.-B. Yu,
  \emph{{Constraints on Dark Matter from Colliders}},
  \href{http://dx.doi.org/10.1103/PhysRevD.82.116010}{\emph{Phys. Rev.} {\bf
  D82} (2010) 116010}, [\href{https://arxiv.org/abs/1008.1783}{{\tt
  1008.1783}}].

\bibitem{1111.5331}
A.~Friedland, M.~L. Graesser, I.~M. Shoemaker and L.~Vecchi, \emph{{Probing
  Nonstandard Standard Model Backgrounds with LHC Monojets}},
  \href{http://dx.doi.org/10.1016/j.physletb.2012.06.078}{\emph{Phys. Lett.}
  {\bf B714} (2012) 267--275}, [\href{https://arxiv.org/abs/1111.5331}{{\tt
  1111.5331}}].

\bibitem{1308.6799}
O.~Buchmueller, M.~J. Dolan and C.~McCabe, \emph{{Beyond Effective Field Theory
  for Dark Matter Searches at the LHC}},
  \href{http://dx.doi.org/10.1007/JHEP01(2014)025}{\emph{JHEP} {\bf 01} (2014)
  025}, [\href{https://arxiv.org/abs/1308.6799}{{\tt 1308.6799}}].

\bibitem{1506.03116}
J.~Abdallah et~al., \emph{{Simplified Models for Dark Matter Searches at the
  LHC}}, \href{http://dx.doi.org/10.1016/j.dark.2015.08.001}{\emph{Phys. Dark
  Univ.} {\bf 9-10} (2015) 8--23},
  [\href{https://arxiv.org/abs/1506.03116}{{\tt 1506.03116}}].

\bibitem{1510.03434}
M.~J. Baker et~al., \emph{{The Coannihilation Codex}},
  \href{http://dx.doi.org/10.1007/JHEP12(2015)120}{\emph{JHEP} {\bf 12} (2015)
  120}, [\href{https://arxiv.org/abs/1510.03434}{{\tt 1510.03434}}].

\bibitem{1512.00476}
N.~F. Bell, Y.~Cai and R.~K. Leane, \emph{{Mono-W Dark Matter Signals at the
  LHC: Simplified Model Analysis}},
  \href{http://dx.doi.org/10.1088/1475-7516/2016/01/051}{\emph{JCAP} {\bf 1601}
  (2016) 051}, [\href{https://arxiv.org/abs/1512.00476}{{\tt 1512.00476}}].

\bibitem{1603.01366}
A.~J. Brennan, M.~F. McDonald, J.~Gramling and T.~D. Jacques, \emph{{Collide
  and Conquer: Constraints on Simplified Dark Matter Models using Mono-X
  Collider Searches}},
  \href{http://dx.doi.org/10.1007/JHEP05(2016)112}{\emph{JHEP} {\bf 05} (2016)
  112}, [\href{https://arxiv.org/abs/1603.01366}{{\tt 1603.01366}}].

\bibitem{1112.5457}
I.~M. Shoemaker and L.~Vecchi, \emph{{Unitarity and Monojet Bounds on Models
  for DAMA, CoGeNT, and CRESST-II}},
  \href{http://dx.doi.org/10.1103/PhysRevD.86.015023}{\emph{Phys. Rev.} {\bf
  D86} (2012) 015023}, [\href{https://arxiv.org/abs/1112.5457}{{\tt
  1112.5457}}].

\bibitem{0905.0956}
X.~Calmet and S.~K. Majee, \emph{{Effective Theory for Dark Matter and a New
  Force in the Dark Matter Sector}},
  \href{http://dx.doi.org/10.1016/j.physletb.2009.07.049}{\emph{Phys. Lett.}
  {\bf B679} (2009) 267--269}, [\href{https://arxiv.org/abs/0905.0956}{{\tt
  0905.0956}}].

\bibitem{1407.5492}
P.~Ko and Y.~Tang, \emph{{Galactic center $\gamma$-ray excess in hidden sector
  DM models with dark gauge symmetries: local $Z_{3}$ symmetry as an example}},
  \href{http://dx.doi.org/10.1088/1475-7516/2015/01/023}{\emph{JCAP} {\bf 1501}
  (2015) 023}, [\href{https://arxiv.org/abs/1407.5492}{{\tt 1407.5492}}].

\bibitem{1407.6588}
S.~Baek, P.~Ko and W.-I. Park, \emph{{Local $Z_2$ scalar dark matter model
  confronting galactic ${\mathrm GeV}$-scale $\gamma$-ray}},
  \href{http://dx.doi.org/10.1016/j.physletb.2015.06.002}{\emph{Phys. Lett.}
  {\bf B747} (2015) 255--259}, [\href{https://arxiv.org/abs/1407.6588}{{\tt
  1407.6588}}].

\bibitem{Lin:2000qq}
W.~B. Lin, D.~H. Huang, X.~Zhang and R.~H. Brandenberger, \emph{{Nonthermal
  production of WIMPs and the subgalactic structure of the universe}},
  \href{http://dx.doi.org/10.1103/PhysRevLett.86.954}{\emph{Phys. Rev. Lett.}
  {\bf 86} (2001) 954}, [\href{https://arxiv.org/abs/astro-ph/0009003}{{\tt
  astro-ph/0009003}}].

\bibitem{Politzer:1980me}
H.~D. Politzer, \emph{{Power Corrections at Short Distances}},
  \href{http://dx.doi.org/10.1016/0550-3213(80)90172-8}{\emph{Nucl. Phys.} {\bf
  B172} (1980) 349--382}.

\bibitem{KlubergStern:1975hc}
H.~Kluberg-Stern and J.~B. Zuber, \emph{{Renormalization of Nonabelian Gauge
  Theories in a Background Field Gauge. 2. Gauge Invariant Operators}},
  \href{http://dx.doi.org/10.1103/PhysRevD.12.3159}{\emph{Phys. Rev.} {\bf D12}
  (1975) 3159--3180}.

\bibitem{GrosseKnetter:1993td}
C.~Grosse-Knetter, \emph{{Effective Lagrangians with higher derivatives and
  equations of motion}},
  \href{http://dx.doi.org/10.1103/PhysRevD.49.6709}{\emph{Phys. Rev.} {\bf D49}
  (1994) 6709--6719}, [\href{https://arxiv.org/abs/hep-ph/9306321}{{\tt
  hep-ph/9306321}}].

\bibitem{Arzt:1993gz}
C.~Arzt, \emph{{Reduced effective Lagrangians}},
  \href{http://dx.doi.org/10.1016/0370-2693(94)01419-D}{\emph{Phys. Lett.} {\bf
  B342} (1995) 189--195}, [\href{https://arxiv.org/abs/hep-ph/9304230}{{\tt
  hep-ph/9304230}}].

\bibitem{Simma:1993ky}
H.~Simma, \emph{{Equations of motion for effective Lagrangians and penguins in
  rare B decays}}, \href{http://dx.doi.org/10.1007/BF01641888}{\emph{Z. Phys.}
  {\bf C61} (1994) 67--82}, [\href{https://arxiv.org/abs/hep-ph/9307274}{{\tt
  hep-ph/9307274}}].

\bibitem{Wudka:1994ny}
J.~Wudka, \emph{{Electroweak effective Lagrangians}},
  \href{http://dx.doi.org/10.1142/S0217751X94000959}{\emph{Int. J. Mod. Phys.}
  {\bf A9} (1994) 2301--2362},
  [\href{https://arxiv.org/abs/hep-ph/9406205}{{\tt hep-ph/9406205}}].

\bibitem{1412.0520}
M.~Duch, B.~Grzadkowski and J.~Wudka, \emph{{Classification of effective
  operators for interactions between the Standard Model and dark matter}},
  \href{http://dx.doi.org/10.1007/JHEP05(2015)116}{\emph{JHEP} {\bf 05} (2015)
  116}, [\href{https://arxiv.org/abs/1412.0520}{{\tt 1412.0520}}].

\bibitem{Kim:2006af}
Y.~G. Kim and K.~Y. Lee, \emph{{The Minimal model of fermionic dark matter}},
  \href{http://dx.doi.org/10.1103/PhysRevD.75.115012}{\emph{Phys. Rev.} {\bf
  D75} (2007) 115012}, [\href{https://arxiv.org/abs/hep-ph/0611069}{{\tt
  hep-ph/0611069}}].

\bibitem{Kawasaki:2004qu}
M.~Kawasaki, K.~Kohri and T.~Moroi, \emph{{Big-Bang nucleosynthesis and
  hadronic decay of long-lived massive particles}},
  \href{http://dx.doi.org/10.1103/PhysRevD.71.083502}{\emph{Phys. Rev.} {\bf
  D71} (2005) 083502}, [\href{https://arxiv.org/abs/astro-ph/0408426}{{\tt
  astro-ph/0408426}}].

\bibitem{Jedamzik:2006xz}
K.~Jedamzik, \emph{{Big bang nucleosynthesis constraints on hadronically and
  electromagnetically decaying relic neutral particles}},
  \href{http://dx.doi.org/10.1103/PhysRevD.74.103509}{\emph{Phys. Rev.} {\bf
  D74} (2006) 103509}, [\href{https://arxiv.org/abs/hep-ph/0604251}{{\tt
  hep-ph/0604251}}].

\bibitem{1608.07648}
D.~S. Akerib et~al., \emph{{Results from a search for dark matter in the
  complete LUX exposure}},  \href{https://arxiv.org/abs/1608.07648}{{\tt
  1608.07648}}.

\bibitem{1403.5294}
{\scshape ATLAS} collaboration, G.~Aad et~al., \emph{{Search for direct
  production of charginos, neutralinos and sleptons in final states with two
  leptons and missing transverse momentum in $pp$ collisions at $\sqrt{s} =$ 8
  TeV with the ATLAS detector}},
  \href{http://dx.doi.org/10.1007/JHEP05(2014)071}{\emph{JHEP} {\bf 05} (2014)
  071}, [\href{https://arxiv.org/abs/1403.5294}{{\tt 1403.5294}}].

\bibitem{1405.7570}
{\scshape CMS} collaboration, V.~Khachatryan et~al., \emph{{Searches for
  electroweak production of charginos, neutralinos, and sleptons decaying to
  leptons and W, Z, and Higgs bosons in pp collisions at 8 TeV}},
  \href{http://dx.doi.org/10.1140/epjc/s10052-014-3036-7}{\emph{Eur. Phys. J.}
  {\bf C74} (2014) 3036}, [\href{https://arxiv.org/abs/1405.7570}{{\tt
  1405.7570}}].

\bibitem{ATLAS:2016kts}
{\scshape ATLAS} collaboration, T.~A. collaboration, \emph{{Further searches
  for squarks and gluinos in final states with jets and missing transverse
  momentum at $\sqrt{s}$ =13 TeV with the ATLAS detector}}, .

\bibitem{CMS-err}
{\scshape CMS} collaboration, V.~Khachatryan et~al., \emph{{A search for new
  phenomena in pp collisions at $\sqrt(s)$ = 13 TeV in final states with
  missing transverse momentum and at least one jet using the alphaT variable}},
   \href{https://arxiv.org/abs/1611.00338}{{\tt 1611.00338}}.

\bibitem{ATLAS-CONF-2016-096}
{\scshape ATLAS} collaboration, T.~A. collaboration, \emph{{Search for
  supersymmetry with two and three leptons and missing transverse momentum in
  the final state at $\sqrt{s}$=13 TeV with the ATLAS detector}}, .

\bibitem{ATLAS:2016tsc}
{\scshape ATLAS} collaboration, T.~A. collaboration, \emph{{Search for Dark
  Matter production associated with bottom quarks with 13.3 fb−1 of pp
  collisions at √s = 13 TeV with the ATLAS detector at the LHC}}, .

\bibitem{1609.06555}
M.~Kakizaki, A.~Santa and O.~Seto, \emph{{Phenomenological signatures of mixed
  complex scalar WIMP dark matter}},
  \href{https://arxiv.org/abs/1609.06555}{{\tt 1609.06555}}.

\bibitem{0710.3820}
T.~Sjostrand, S.~Mrenna and P.~Z. Skands, \emph{{A Brief Introduction to PYTHIA
  8.1}}, \href{http://dx.doi.org/10.1016/j.cpc.2008.01.036}{\emph{Comput. Phys.
  Commun.} {\bf 178} (2008) 852--867},
  [\href{https://arxiv.org/abs/0710.3820}{{\tt 0710.3820}}].

\bibitem{0807.4730}
C.~Evoli, D.~Gaggero, D.~Grasso and L.~Maccione, \emph{{Cosmic-Ray Nuclei,
  Antiprotons and Gamma-rays in the Galaxy: a New Diffusion Model}},
  \href{http://dx.doi.org/10.1088/1475-7516/2008/10/018,
  10.1088/1475-7516/2016/04/E01}{\emph{JCAP} {\bf 0810} (2008) 018},
  [\href{https://arxiv.org/abs/0807.4730}{{\tt 0807.4730}}].

\bibitem{1012.4515}
M.~Cirelli, G.~Corcella, A.~Hektor, G.~Hutsi, M.~Kadastik, P.~Panci et~al.,
  \emph{{PPPC 4 DM ID: A Poor Particle Physicist Cookbook for Dark Matter
  Indirect Detection}}, \href{http://dx.doi.org/10.1088/1475-7516/2012/10/E01,
  10.1088/1475-7516/2011/03/051}{\emph{JCAP} {\bf 1103} (2011) 051},
  [\href{https://arxiv.org/abs/1012.4515}{{\tt 1012.4515}}].

\bibitem{1503.02641}
{\scshape Fermi-LAT} collaboration, M.~Ackermann et~al., \emph{{Searching for
  Dark Matter Annihilation from Milky Way Dwarf Spheroidal Galaxies with Six
  Years of Fermi Large Area Telescope Data}},
  \href{http://dx.doi.org/10.1103/PhysRevLett.115.231301}{\emph{Phys. Rev.
  Lett.} {\bf 115} (2015) 231301},
  [\href{https://arxiv.org/abs/1503.02641}{{\tt 1503.02641}}].

\bibitem{Navarro:1995iw}
J.~F. Navarro, C.~S. Frenk and S.~D.~M. White, \emph{{The Structure of cold
  dark matter halos}}, \href{http://dx.doi.org/10.1086/177173}{\emph{Astrophys.
  J.} {\bf 462} (1996) 563--575},
  [\href{https://arxiv.org/abs/astro-ph/9508025}{{\tt astro-ph/9508025}}].

\bibitem{1312.1535}
J.~Aleksić et~al., \emph{{Optimized dark matter searches in deep observations
  of Segue 1 with MAGIC}},
  \href{http://dx.doi.org/10.1088/1475-7516/2014/02/008}{\emph{JCAP} {\bf 1402}
  (2014) 008}, [\href{https://arxiv.org/abs/1312.1535}{{\tt 1312.1535}}].

\bibitem{1601.06590}
{\scshape Fermi-LAT, MAGIC} collaboration, M.~L. Ahnen et~al., \emph{{Limits to
  dark matter annihilation cross-section from a combined analysis of MAGIC and
  Fermi-LAT observations of dwarf satellite galaxies}},
  \href{http://dx.doi.org/10.1088/1475-7516/2016/02/039}{\emph{JCAP} {\bf 1602}
  (2016) 039}, [\href{https://arxiv.org/abs/1601.06590}{{\tt 1601.06590}}].

\bibitem{1408.4131}
H.~Silverwood, C.~Weniger, P.~Scott and G.~Bertone, \emph{{A realistic
  assessment of the CTA sensitivity to dark matter annihilation}},
  \href{http://dx.doi.org/10.1088/1475-7516/2015/03/055}{\emph{JCAP} {\bf 1503}
  (2015) 055}, [\href{https://arxiv.org/abs/1408.4131}{{\tt 1408.4131}}].

\bibitem{Navarro:2003ew}
J.~F. Navarro, E.~Hayashi, C.~Power, A.~Jenkins, C.~S. Frenk, S.~D.~M. White
  et~al., \emph{{The Inner structure of Lambda-CDM halos 3: Universality and
  asymptotic slopes}},
  \href{http://dx.doi.org/10.1111/j.1365-2966.2004.07586.x}{\emph{Mon. Not.
  Roy. Astron. Soc.} {\bf 349} (2004) 1039},
  [\href{https://arxiv.org/abs/astro-ph/0311231}{{\tt astro-ph/0311231}}].

\bibitem{Graham:2005xx}
A.~W. Graham, D.~Merritt, B.~Moore, J.~Diemand and B.~Terzic, \emph{{Empirical
  models for Dark Matter Halos. I. Nonparametric Construction of Density
  Profiles and Comparison with Parametric Models}},
  \href{http://dx.doi.org/10.1086/508988}{\emph{Astron. J.} {\bf 132} (2006)
  2685--2700}, [\href{https://arxiv.org/abs/astro-ph/0509417}{{\tt
  astro-ph/0509417}}].

\bibitem{Aguilar:2014mma}
{\scshape AMS} collaboration, M.~Aguilar et~al., \emph{{Electron and Positron
  Fluxes in Primary Cosmic Rays Measured with the Alpha Magnetic Spectrometer
  on the International Space Station}},
  \href{http://dx.doi.org/10.1103/PhysRevLett.113.121102}{\emph{Phys. Rev.
  Lett.} {\bf 113} (2014) 121102}.

\bibitem{Accardo:2014lma}
{\scshape AMS} collaboration, L.~Accardo et~al., \emph{{High Statistics
  Measurement of the Positron Fraction in Primary Cosmic Rays of 0.5–500 GeV
  with the Alpha Magnetic Spectrometer on the International Space Station}},
  \href{http://dx.doi.org/10.1103/PhysRevLett.113.121101}{\emph{Phys. Rev.
  Lett.} {\bf 113} (2014) 121101}.

\bibitem{1308.0133}
{\scshape PAMELA} collaboration, O.~Adriani et~al., \emph{{Cosmic-Ray Positron
  Energy Spectrum Measured by PAMELA}},
  \href{http://dx.doi.org/10.1103/PhysRevLett.111.081102}{\emph{Phys. Rev.
  Lett.} {\bf 111} (2013) 081102}, [\href{https://arxiv.org/abs/1308.0133}{{\tt
  1308.0133}}].

\bibitem{DuVernois:2001bb}
M.~A. DuVernois et~al., \emph{{Cosmic ray electrons and positrons from 1-GeV to
  100-GeV: Measurements with HEAT and their interpretation}},
  \href{http://dx.doi.org/10.1086/322324}{\emph{Astrophys. J.} {\bf 559} (2001)
  296--303}.

\bibitem{1306.3983}
L.~Bergstrom, T.~Bringmann, I.~Cholis, D.~Hooper and C.~Weniger, \emph{{New
  limits on dark matter annihilation from AMS cosmic ray positron data}},
  \href{http://dx.doi.org/10.1103/PhysRevLett.111.171101}{\emph{Phys. Rev.
  Lett.} {\bf 111} (2013) 171101}, [\href{https://arxiv.org/abs/1306.3983}{{\tt
  1306.3983}}].

\bibitem{1309.2570}
A.~Ibarra, A.~S. Lamperstorfer and J.~Silk, \emph{{Dark matter annihilations
  and decays after the AMS-02 positron measurements}},
  \href{http://dx.doi.org/10.1103/PhysRevD.89.063539}{\emph{Phys. Rev.} {\bf
  D89} (2014) 063539}, [\href{https://arxiv.org/abs/1309.2570}{{\tt
  1309.2570}}].

\bibitem{1602.05966}
F.~S. Queiroz, C.~E. Yaguna and C.~Weniger, \emph{{Gamma-ray Limits on Neutrino
  Lines}}, \href{http://dx.doi.org/10.1088/1475-7516/2016/05/050}{\emph{JCAP}
  {\bf 1605} (2016) 050}, [\href{https://arxiv.org/abs/1602.05966}{{\tt
  1602.05966}}].

\bibitem{1505.07259}
{\scshape IceCube} collaboration, M.~G. Aartsen et~al., \emph{{Search for Dark
  Matter Annihilation in the Galactic Center with IceCube-79}},
  \href{http://dx.doi.org/10.1140/epjc/s10052-015-3713-1}{\emph{Eur. Phys. J.}
  {\bf C75} (2015) 492}, [\href{https://arxiv.org/abs/1505.07259}{{\tt
  1505.07259}}].

\bibitem{1510.07999}
{\scshape Super-Kamiokande} collaboration, K.~Frankiewicz, \emph{{Searching for
  Dark Matter Annihilation into Neutrinos with Super-Kamiokande}},  in
  \emph{{Proceedings, Meeting of the APS Division of Particles and Fields (DPF
  2015): Ann Arbor, Michigan, USA, 4-8 Aug 2015}}, 2015.
\newblock \href{https://arxiv.org/abs/1510.07999}{{\tt 1510.07999}}.

\bibitem{1505.04866}
{\scshape ANTARES} collaboration, S.~Adrian-Martinez et~al., \emph{{Search of
  Dark Matter Annihilation in the Galactic Centre using the ANTARES Neutrino
  Telescope}},
  \href{http://dx.doi.org/10.1088/1475-7516/2015/10/068}{\emph{JCAP} {\bf 1510}
  (2015) 068}, [\href{https://arxiv.org/abs/1505.04866}{{\tt 1505.04866}}].

\bibitem{1409.5439}
M.~C. Gonzalez-Garcia, M.~Maltoni and T.~Schwetz, \emph{{Updated fit to three
  neutrino mixing: status of leptonic CP violation}},
  \href{http://dx.doi.org/10.1007/JHEP11(2014)052}{\emph{JHEP} {\bf 11} (2014)
  052}, [\href{https://arxiv.org/abs/1409.5439}{{\tt 1409.5439}}].

\bibitem{1502.01589}
{\scshape Planck} collaboration, P.~A.~R. Ade et~al., \emph{{Planck 2015
  results. XIII. Cosmological parameters}},
  \href{http://dx.doi.org/10.1051/0004-6361/201525830}{\emph{Astron.
  Astrophys.} {\bf 594} (2016) A13},
  [\href{https://arxiv.org/abs/1502.01589}{{\tt 1502.01589}}].

\bibitem{1506.03811}
T.~R. Slatyer, \emph{{Indirect dark matter signatures in the cosmic dark ages.
  I. Generalizing the bound on s-wave dark matter annihilation from Planck
  results}}, \href{http://dx.doi.org/10.1103/PhysRevD.93.023527}{\emph{Phys.
  Rev.} {\bf D93} (2016) 023527}, [\href{https://arxiv.org/abs/1506.03811}{{\tt
  1506.03811}}].

\bibitem{1608.02662}
J.~Bramante, P.~J. Fox, G.~D. Kribs and A.~Martin, \emph{{The Inelastic
  Frontier: Discovering Dark Matter at High Recoil Energy}}, {\emph{Submitted
  to: Phys. Rev. D} (2016) }, [\href{https://arxiv.org/abs/1608.02662}{{\tt
  1608.02662}}].

\end{thebibliography}\endgroup
\bibliographystyle{JHEP}

\end{document}